\documentclass[trackchanges,twocolumn]{aastex701}

\usepackage{booktabs}
\usepackage{breqn}
\usepackage{enumerate}

\newcommand\neii{[Ne~\textsc{ii}]}
\newcommand\neiii{[Ne~\textsc{iii}]}
\newcommand\nev{[Ne~\textsc{v}]}
\newcommand\oiv{[O~\textsc{iv}]}
\newcommand\oiii{[O~\textsc{iii}]}
\newcommand\oii{[O~\textsc{ii}]}
\newcommand\cii{[C~\textsc{ii}]}
\newcommand\ciii{C~\textsc{iii}]}
\newcommand\civ{C~\textsc{iv}}
\newcommand\siv{[S~\textsc{iv}]}
\newcommand\siii{[S~\textsc{iii}]}
\newcommand\sii{[S~\textsc{ii}]}
\newcommand\feii{[Fe~\textsc{ii}]}
\newcommand\arii{[Ar~\textsc{ii}]}
\newcommand\ariii{[Ar~\textsc{iii}]}
\newcommand\hb{H$\beta$}

\begin{document}

\title{The JWST and ALMA view of local dwarf Pox 186: Evidence of stripped binaries driving a hard radiation field and little cold molecular gas resulting in high star-formation efficiency}

\author[0000-0002-5320-2568]{Nimisha Kumari}
\affiliation{AURA for European Space Agency (ESA), ESA Office, Space Telescope Science Institute, 3700 San Martin Drive, Baltimore, MD, 21218, USA}
\email[show]{kumari@stsci.edu}

\author[0000-0001-8034-7802]{Renske Smit}
\affiliation{Astrophysics Research Institute, Liverpool John Moores University, Liverpool, L35 UG, UK}
\email{R.Smit@ljmu.ac.uk}

\author[0000-0003-2685-4488]{Claus Leitherer}
\affiliation{Space Telescope Science Institute, 3700 San Martin Drive, Baltimore, MD, 21218, USA}
\email{leitherer@stsci.edu}

\author[0000-0002-7093-1877]{J. \'Alvarez-M\'arquez}
\affiliation{Centro de Astrobiolog\'{\i}a (CAB), CSIC-INTA, Ctra. de Ajalvir km 4, Torrej\'on de Ardoz, E-28850, Madrid, Spain}
\email{jalvarez@cab.inta-csic.es}

\author[0000-0003-2000-3420]{Katherine Ormerod}
\affiliation{Astrophysics Research Institute, Liverpool John Moores University, Liverpool, L35 UG, UK}
\email{K.Ormerod@2023.ljmu.ac.uk}

\author[0000-0002-7595-121X]{Joris Witstok}
\affiliation{Cosmic Dawn Center (DAWN), Copenhagen, Denmark}
\email{joris.witstok@nbi.ku.dk}

\author[]{Suzanne Madden}
\affiliation{AIM, CEA, CNRS, Université Paris-Saclay, Université Paris Diderot, Sorbonne Paris Cité, 91191 Gif-sur-Yvette, France}
\email{uzanne.madden@cea.fr}

\author[0000-0001-8587-218X]{Matthew Hayes}
\affiliation{Stockholm University, Department of Astronomy and Oskar Klein Centre for Cosmoparticle Physics, AlbaNova University Centre,
SE-10691, Stockholm, Sweden}
\email{matthew.hayes@astro.su.se}

%\collaboration{all}{Pox186 collaboration}
 
\begin{abstract}

We present a JWST mid-infrared (MIR) view of Pox 186, one of the best local analogs of reionization-era galaxies based on its extreme properties in the ultraviolet, optical, and far-infrared (FIR). We extracted a high signal-to-noise spectrum of Pox 186 from the spatially resolved MIR data taken with the Medium Resolution Spectrometer (MRS) on the MIR Instrument (MIRI), and compared its MIR properties with those of the other dwarf galaxies in the Herschel Dwarf Galaxy Survey. We probe the radiation hardness of Pox 186 using the so-called softness diagram, based on the MIR line ratios \neiii/\neii~ and \siv/\siii, which suggest that Pox 186 hosts ionization sources with a hard radiation field. We probe the dominant ionization source in Pox 186 using diagnostic diagrams based on MIR line ratios. Our photoionization models provide evidence that stripped binary stars are the main cause of the extreme MIR line ratios. Similarly, a young starburst is needed to reproduce the MIR dust continuum. We probe the warm and cold molecular gas content using the MIR H$_2$ rotational lines (S(1)-S(7)) from JWST, and the archival sub-mm CO(2-1) line from ALMA, which indicate a high star-formation efficiency (SFE; $>$ 15 $\times$ 10$^{-9}$ yr$^{-1}$) and low gas depletion time comparable to that of the reionization-era galaxies. This analysis strongly suggests that star formation rather than active galactic nuclei drives the hard radiation field observed in Pox 186, and cautions against relying solely on the UV diagnostics to establish the nature of high-z galaxies.

\end{abstract}

\keywords{\uat{Galaxies}{573} --- \uat{Interstellar medium}{847} --- \uat{Dwarf irregular galaxies}{417} --- \uat{Infrared spectroscopy}{2285} --- \uat{Emission line galaxies}{459}}

\section{Introduction} 
\label{sec:intro}

\indent The launch of the James Webb Space Telescope \citep[JWST]{Gardner2023} has enabled the discovery and characterization of the first galaxies that reionized the Universe through rest-frame ultraviolet (UV) and optical observations \citep[e.g.,][]{Curtis-Lake2023, Stark2026, Alvarez-Marquez2025, Alvarez-Marquez2026}. Simultaneous observations with the Atacama Large Millimeter Array (ALMA) have allowed us to obtain a far-infrared view of these extreme objects \citep{Zavala2024, Carniani2025, Schouws2025, Schouws2026,  Witstok2026, Smit2026}. Still, several questions remain about the nature of these sources, which we can probe by examining their local analogs at wavelengths not yet accessible at high redshifts. 

\indent Identifying and characterizing the local analogs of high-redshift galaxies has been an active topic of research for decades \citep[e.g.,][]{Izotov1999, Leitherer1995, Kunth2000, Heckman2005, Borthakur2014, Bian2016, Amorin2017, Kumari2024, Jaskot2025}. Several criteria for finding local analogs have been proposed based on properties shared by the high-z galaxy populations, and they have been explored across multiple wavelengths, with a multitude of studies, especially in UV \citep[e.g.,][]{Berg2016, Senchyna2017, Mingozzi2022} and optical \citep[e.g.,][]{James2009, Kumari2018, Izotov2021}. However, UV and optical spectra of galaxies mainly provide information on ionized gas and stellar populations, and also suffer from dust attenuation. 

\indent The mid-infrared (MIR) part of the electromagnetic spectrum suffers relatively less from dust attenuation than UV and optical, and exhibits a plethora of spectral features which are useful in deriving information on other components of galaxies, such as molecular gas content via the H$_2$ rotational lines, dust via polyaromatic hydrocarbons (PAHs), in addition to ionized gas and stars via high ionization lines. MIR telescopes such as IRAS, ISO, and the Spitzer Space Telescope were significantly useful in studying these unique characteristics of galaxies \citep{Soifer1987, Sturm2000, Galliano2008, Hunt2010}.

\indent The MIR Instrument (MIRI) onboard JWST provides an ideal opportunity to perform detailed MIR studies of nearby galaxies owing to a higher sensitivity and spectral resolution compared to the last generation MIR instruments on the Spitzer Space Telescope. The Medium Resolution Spectrometer \citep[MRS,][]{Wells2015, Argyriou2023}, i.e., the integral field unit on MIRI, is particularly useful for spatially resolved spectroscopy, providing a MIR counterpart to the optical IFUs on the ground. Surveys such as PHANGS-JWST \citep{Williams2024}, MICONIC \citep{Hermosa2025}, GATOS \citep{Garcia-Bernete2024}, GOALS-JWST \citep{U2022, Rich2023} and CLASSSYIR \citep{Bergjwst2025} are specifically designed to take advantage of the MIRI/MRS Spectroscopy and help reveal the physical properties of different populations of local galaxies, including typical star-forming spiral galaxies, active galactic nuclei (AGNs), quasars, and dwarfs. In addition to these surveys, MIRI/MRS spectroscopy of single sources has also led to ground-breaking results, such as the discovery of molecular gas in the extremely metal-poor galaxy Leo P \citep{Telford2024} and the discovery of an AGN-diagnostic \nev~ line in a dwarf galaxy SBS0335-052E \citep{Mingozzi2025}.

\indent In this paper, we present the JWST mid-infrared view of Pox 186, which is potentially the best local analog of reionization-era galaxies based on the multi-wavelength study from \citet{Kumari2024} and shows the largest \oiii 88$\mu$m/\cii 157 $\mu$m ratio ($>$10) of any galaxy observed with Herschel and largest equivalent widths (EW) of \ciii 1907, 1909 ($\sim$36\AA) observed in the local Universe, along with extreme values of \oiii 5007, and \hb\ EW. %\citet{Kumari2024} also demonstrated that local analogs of reionization-era galaxies can be identified on the basis of their FIR line ratios, and that the galaxies exhibiting extreme FIR \oiii/\cii~ may also exhibit extreme UV and optical properties. This paper aims to identify MIR diagnostics for potential local analogs of high-redshift galaxies. 
The paper is organized as follows. Section \ref{sec:data} introduces the target Pox 186 in the context of its resemblance to the reionization-era galaxies. We provide an overview of the MIRI/MRS data of Pox 186, along with the reduction details. Here, we also describe the archival ALMA data used to probe the molecular gas from sub-mm observations of Pox 186, along with the MIR Spitzer data for other dwarf galaxies from the Herschel Dwarf Galaxy Survey \citep[HDGS,][]{Madden2013, Cormier2015}, which are compared with Pox 186 throughout the work. Section \ref{sec:results} describes our main results on probing the radiation hardness via MIR line ratios, MIR spectral decomposition via the  \texttt{CAFE} software to constrain the PAH properties, analysis of the warm and hot molecular gas content by modeling of the MIR H$_2$ rotational lines via \texttt{PDRTPY}, and constraining the cold-molecular gas content via the sub-mm CO(2-1) observations with ALMA. Section \ref{sec:summary} summarizes our main results. In the rest of the paper, we use \siv, \neii, \neiii, \siii~ and \oiv~ to denote the MIR lines \siv10.5$\mu$m, \neii12.8$\mu$m, \neiii15.6$\mu$m, \siii18.7$\mu$m, and \oiv25.8$\mu$m, respectively, unless stated otherwise. Similarly, we use \oiii~ and \cii~ to denote the FIR lines \oiii88$\mu$m and \cii157$\mu$m, respectively, unless mentioned otherwise.

\begin{figure}
    \centering
    \includegraphics[width=0.45\textwidth]{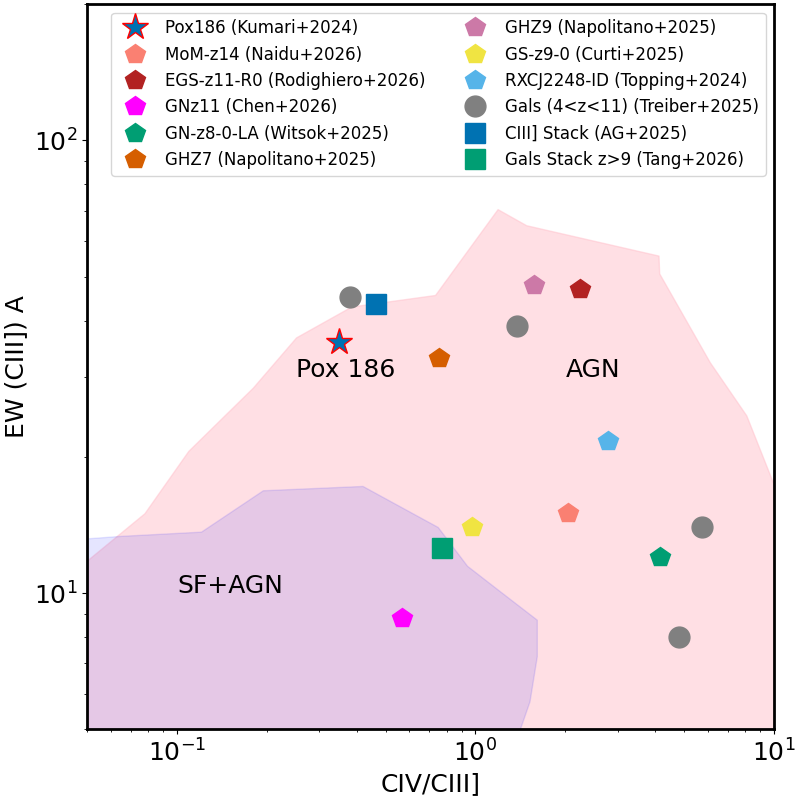}
    \caption{Equivalent widths of \ciii 1907,1909 versus \civ 1548, 1550/\ciii 1907,1909 line ratios of Pox 186, and individual galaxies at high redshift (z$\gtrsim$4) with confirmed JWST detections from literature including \citet{Topping2024, Curti2025, Napolitano2025, Treiber2025, Naidu2026, Chen2026, Rodighiero2026}, composite spectra of strong \ciii~ emitters and galaxies at redshift z$>$9 from \citet{Arevalo-Gonzalez2026} and \citet{Tang2026}, respectively. We also indicate the parameter space covered by the photoionization models from \citet{Nakajima2022}, where the shaded pink region indicates the upper envelope of the AGN models, while the purple shaded region denotes the upper envelope of the SF+AGN models.}
    \label{fig:intro_fig}
\end{figure}

\begin{figure}
    \centering
    \includegraphics[width=0.45\textwidth]{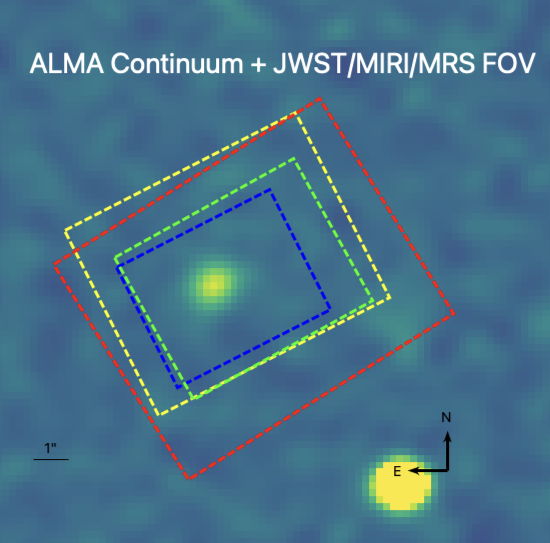}
    \caption{Footprint of MIRI/MRS in Channel 1 (Blue), Channel 2 (Green), Channel 3 (Yellow), and Channel 4 (Red) on an ALMA Continuum image.}
    \label{fig:alma_mrs_fov}
\end{figure}

\begin{figure*}
\centering
    \includegraphics[width=\textwidth]{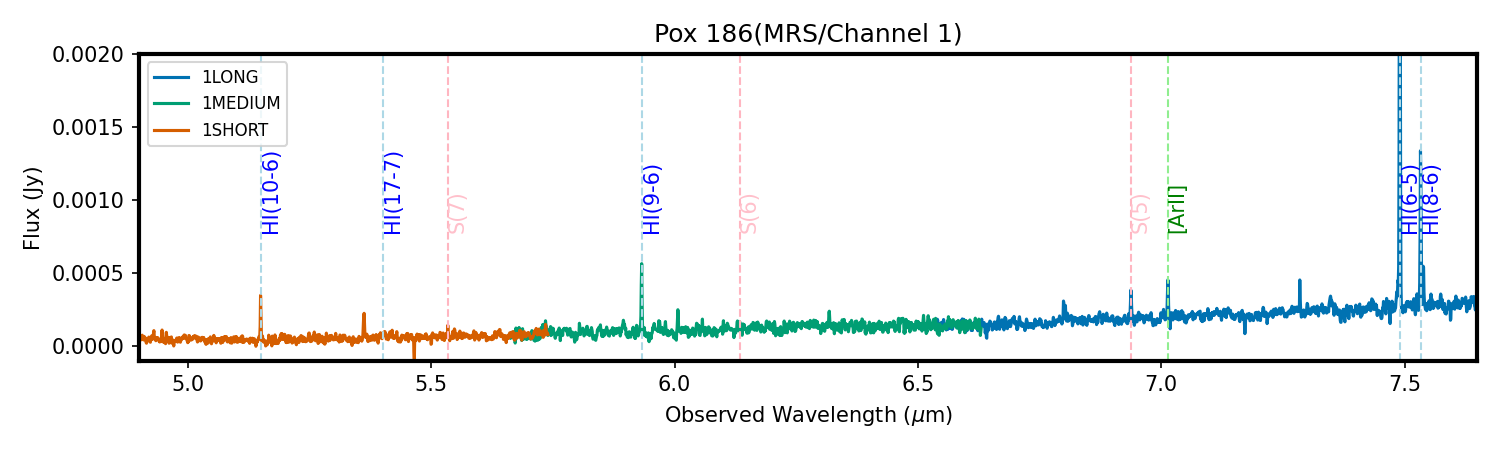}
    \includegraphics[width=\textwidth]{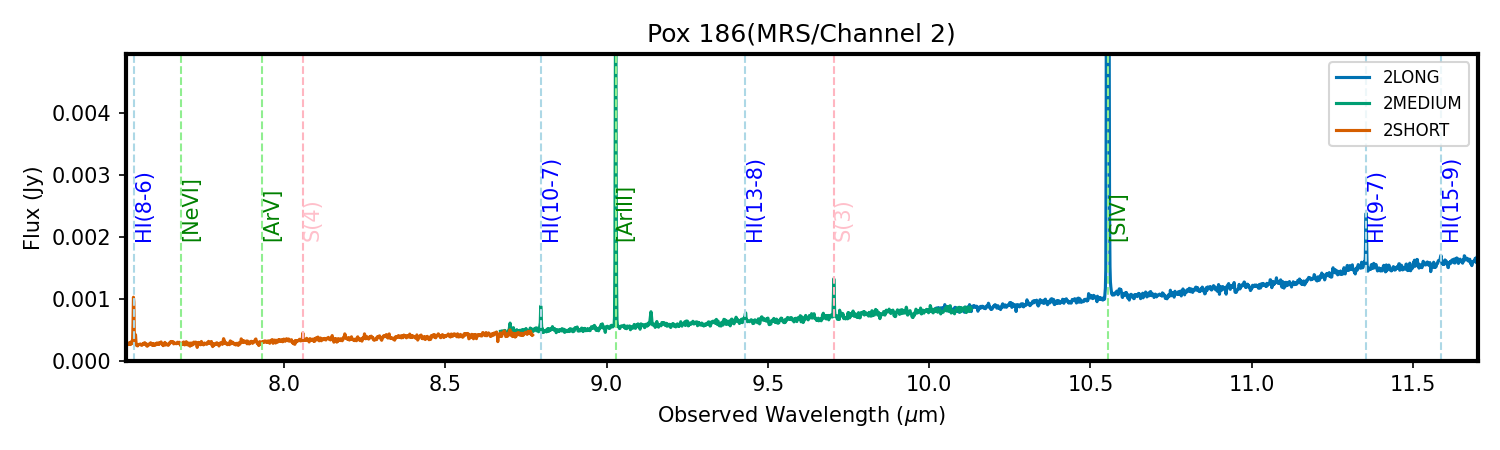}
    \includegraphics[width=\textwidth]{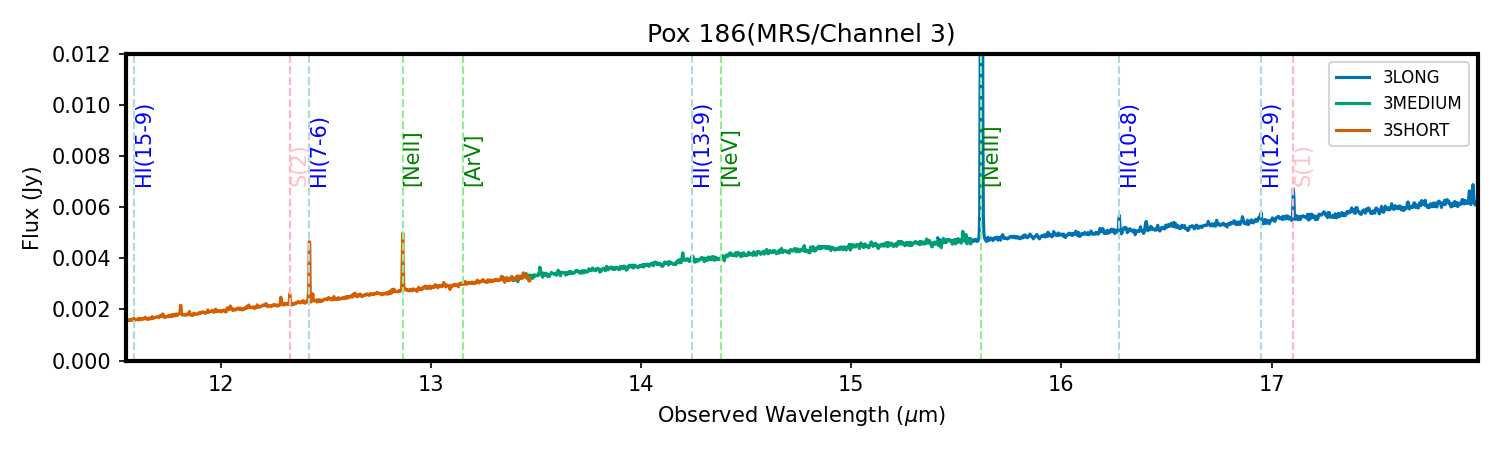}
    \includegraphics[width=\textwidth]{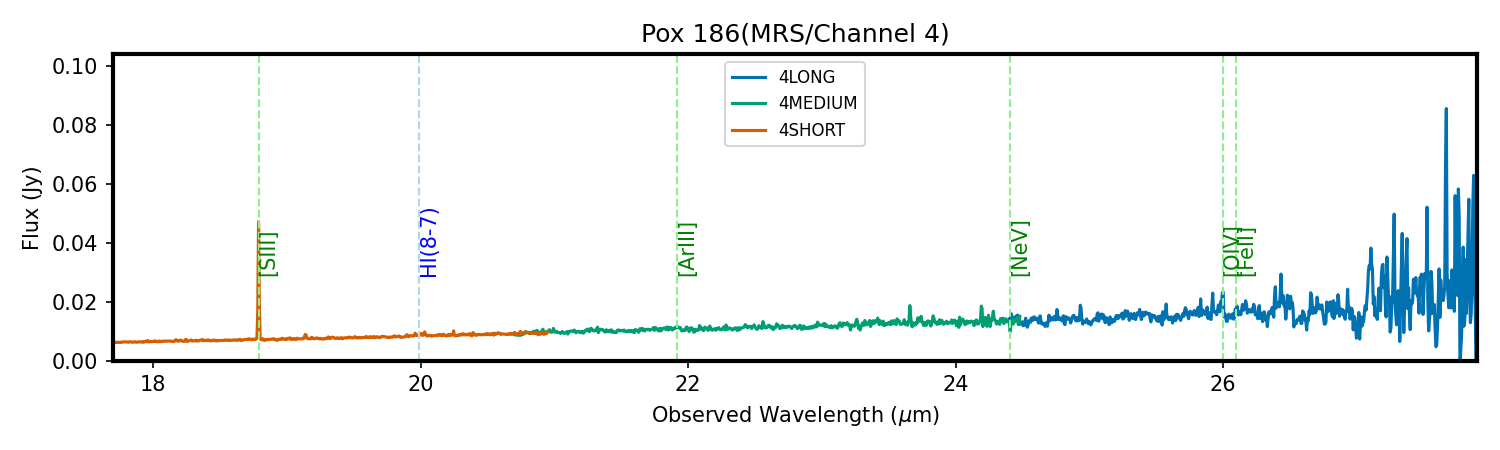}
    \caption{Integrated spectra of Pox 186 extracted from an aperture of extraction size of 1$\times$FWHM in Channels 1, 2, 3, and 4 within the 1st, 2nd, 3rd, and 4th panels (top-to-bottom), respectively, and in all three gratings, short A (brown), medium B (green), and long C (blue). The locations of fine-structure lines (light green), rotational H$_2$ lines (light pink), and H I recombination lines (light blue) are shown as dashed vertical lines. Not all marked lines are detected. The fluxes of detected lines are presented in the Appendix.}%All marked lines are not detected.}
    \label{fig:mrs_spectra}
\end{figure*}

\section{Data}
\label{sec:data}

\subsection{Pox 186 as a reionization era analog}
\label{sec:pox186}
\indent Pox 186, initially discovered by \citet{Kunth1981}, was considered an extreme blue compact dwarf galaxy \citep{Kunth1988} mainly because of its extremely compact size $\sim3\arcsec$ which appeared more like a giant H \textsc{ii} region like 30 Doradus than a galaxy. \citet{Madden2013} reports Pox 186 to have a stellar mass of 0.12$\times$10$^8$ M$_{\odot}$, a star-formation rate of 0.04 M$_{\odot}$yr$^{-1}$, placing this galaxy above the local main-sequence galaxies, and a neutral gas mass of $<$0.02$\times$10$^8$M$_{\odot}$, indicative of a high star-formation efficiency as speculated by \citet{Kunth1988}. It was later introduced as one of the best local analogs of reionization-era galaxies by \citet{Kumari2024} because of the largest FIR \oiii/\cii\ line ratio in the local Universe, indicative of a highly porous medium \citep{Inoue2016, Harikane2020, Katz2022} from which Lyman Continuum can escape to ionize the neutral medium, as should be the case for the sources which reionized the Universe. The photoionization modeling of the FIR data of Pox 186 obtained from Herschel \citep{Cormier2015, Madden2013} also indicates a very high Lyman Continuum escape fraction of 40\% \citep{Ramambason2022}, making it an analog of galaxies in the Epoch of Reionization (EoR). 

\indent Recently, a large population of high-redshift galaxies has been discovered via JWST, which strengthens the case that Pox 186 resembles very early galaxies. In terms of morphology, Pox 186 is similar to the really compact sources seen in the z$>$10 \citep[e.g.,][]{Roberts-Borsani2026} that shows larger equivalent widths of \ciii\ and \civ. In Figure \ref{fig:intro_fig}, we find that the \ciii\ equivalent width and \civ/\ciii\ line ratio of Pox 186 are similar to those of galaxies observed by JWST in the redshift range of $\sim$4-14. The high \ciii\ equivalent width of Pox 186 places it among galaxies such as GHZ9 (purple pentagon) at z=10.145 \citep{Napolitano2025}, EGS-z11-R0 (red pentagon) at z=11.45 \citep{Rodighiero2026} and  Mom-z-14 (light-orange pentagon) at z=14.44 \citep{Naidu2026}, for which the ionizing spectra are hard and various diagnostics are being used to investigate whether their ionizing source is dominated by star-formation or active galactic nuclei \citep[AGN][]{Kovacs2024, Napolitano2025, Naidu2026, Fabian2026}. For cases such as GNz-11 (Figure \ref{fig:intro_fig}, magenta pentagon), there is a dichotomy between star-formation \citep{Alvarez-Marquez2025, Chen2026} and AGN \citep{Maiolino2024} as for the dominant source of ionization. Similarly, the very red UV slope of Pox 186 \citep[$\sim$-0.36]{Kumari2024} raises the question of whether they could be analogs of little red dots \citep[LRDs;]{deGraff2025} and red monsters \citep{Rodighiero2026}, which are being discovered at high redshifts, and their power source is not yet known. The photoionization models of \citet{Nakajima2022} place Pox 186 and the majority of JWST-confirmed high-redshift galaxies in the UV parameter space dominated by the AGN (Figure \ref{fig:intro_fig}, red-shaded region), though there is no clear evidence of the presence of an AGN within Pox 186 \citep{Kumari2024}. We note that the stellar population models used for the interpretation of the high-z galaxies via photoionization models are based on the observations of the local Universe, particularly of resolved stars, with metallicities, significantly higher than those of the primordial Universe. To interpret these high-z sources correctly, we must improve the stellar population synthesis and evolution models by performing detailed multi-wavelength analysis of the closest analogs, such as Pox 186. Hence, in this work, we focus on the MIR data of Pox 186 to further probe the reionization-era galaxies for which MIR data are not yet available.

\subsection{MIRI/MRS Observations and Reduction}
\label{sec:mrs}

\indent Pox 186 was observed with the MIRI/MRS on June 14, 2025, via the GO program PID 6083 (PI Kumari/co-PI Smit). Observations were taken using all three gratings (Short A, Medium B, and Long C) and all 4 channels, thus sampling the entire wavelength range accessible by MIRI ($\sim$5-28 $\mu$m).  We used a 4-point dither optimized for extended sources. We performed these observations using the default FASTR1 readout pattern, with varying exposure times for each grating based on the predicted strength of the emission features it covered. Pox 186 is compact enough to be covered by the field-of-view of MIRI/MRS in all channels (Figure \ref{fig:alma_mrs_fov}). We also obtained background observations away from the source using all gratings and across all channels, with the exposure time matching that of the source. %A set of MIRI imaging data was also taken in F560W, F770W, and F1000W filters.

\indent We used the JWST pipeline version v1.21.0 \citep{Bushouse2025}, with the corresponding Calibration Reference Data System (CRDS) context \texttt{jwst\_1505.pmap}. We ran the MIRI/MRS data through Stage 1 with the default parameters. To run Stage 2, we enabled additional steps for residual cosmic-ray shower correction and 2D residual fringe correction for both on-source and background data. For running Stage 3, we adopted two different procedures for Channels 1/2 and Channels 3/4 to recover the highest signal-to-noise in all channels. In Channels 1/2, noise is dominated by the uncertainties in the calibration reference files. For the dataset in Channels 1/2, we wrote a custom routine outside the standard JWST pipeline to estimate the median of Stage 2 background exposures within a spectral setting (short, medium, long) and subtract it from the Stage 2 science exposures within the same spectral setting \citep{Alvarez-Marquez2024}. We then ran Stage 3 on these Stage 2 science exposures without further master background subtraction because of the intermediate background subtraction done after Stage 2. In Channels 3/4, uncertainties are dominated by the Poissonian noise \citep{Alvarez-Marquez2023}, for which the standard STScI pipeline is appropriate. Hence, for Channels 3/4, we ran Stage 3 on the direct products from Stage 2, and enabled master background subtraction.     

%We removed the detector imprints and background from the Stage 2 output using a custom routine outside the standard JWST pipeline. We ran Stage 3 on the background-subtracted Stage 2 on-source data. An inspection of the 3D data cube revealed that the source is enclosed within a 0.5 arcsec radius at longer wavelengths ($\sim$25$\mu$m). %The full-width half maximum (FWHM) of the MIRI/MRS point spread function is $\sim$1 arcsec at $\sim$25$\mu$m. 

We extracted the 1D spectrum assuming a point-source geometry and an extraction size of 1$\times$FWHM, while applying an additional 1D residual fringe correction. The JWST pipeline applies a wavelength-dependent aperture correction to the extracted 1D spectrum. Since the source is unresolved, we also experimented with a smaller aperture size of 0.5$\times$FWHM in an attempt to detect any faint \nev~ lines and ensure the highest signal-to-noise for other lines. However, our experiments revealed that a smaller extraction aperture led to flux offsets at higher wavelengths, likely due to issues with aperture corrections at these wavelengths. Hence, we decided to use the 1D spectrum extracted from a 1$\times$ FWHM aperture (see Figure \ref{fig:mrs_spectra}). We corrected the 1D spectrum for the foreground Galactic extinction \citep[E(B-V)=0.0385;][]{Schlafly2011} by using the dust extinction curve from \citet{Gordon2023}. We use this spectrum for the rest of the analysis.

\subsection{ALMA CO Observations}
\label{sec:alma}
\indent Interferometric observations of Pox 186 were taken with ALMA (\#2018.1.00434.S) within four blocks between March 21-23, 2019, and in four spectral windows, one centered at 229.642GHz and bandwidth of 1.875GHz, and other three centered at 227.651GHz, 212.710GHz and 214.702GHz with bandwidths of 2GHz each. %SPWNAM01= 'X343336007#ALMA_RB_06#BB_1#SW-01'
For this analysis, we used archival data products reduced with the standard data-reduction pipeline (version CASA54-P1-B). The archival continuum map was obtained by combining all spectral windows centered around 221 GHz and a bandwidth of 7.5 GHz in Band 6, and shows a clear detection centered on Pox 186. We converted the data cube originally in GHz to the kinematic local frame of rest (LSRK) velocity (km s$^{-1}$) using the software \texttt{carta} to inspect the sub-mm CO(2-1) line and probe the cold molecular gas. Using \texttt{carta}, we also extract a 1D spectrum of Pox 186 using a circular aperture of 1\arcsec radius centered on the source, covering the entire galaxy in the ALMA data cube. %Figure \ref{fig:alma_CO} shows the 1D extracted spectrum of Pox 186 in the LSRK frame.   

\subsection{Spitzer and Herschel line fluxes for HDGS}
\label{sec:HDGS}

We compare Pox 186 with other dwarf galaxies within HDGS using their mid- and far-infrared line fluxes published in \citet{Cormier2015}. In particular, we use the \siv, \neii, \neiii, \siii~ and \oiv~ MIR lines and \oiii~ and \cii~ FIR lines.  We note that the HDGS MIR data were obtained with Spitzer at both high and low resolutions. However, we use only the high-resolution line fluxes because a major portion of our line flux analysis relies on using the MIR \oiv\ data, while the low-resolution Spitzer data only provide upper limits on MIR \oiv~ line fluxes for the majority of galaxies. The HDGS MIR line fluxes are not corrected for dust extinction; however, we discuss the impact of dust correction in the line ratio analysis below.

\begin{figure}
\centering
\includegraphics[width = 0.45\textwidth]{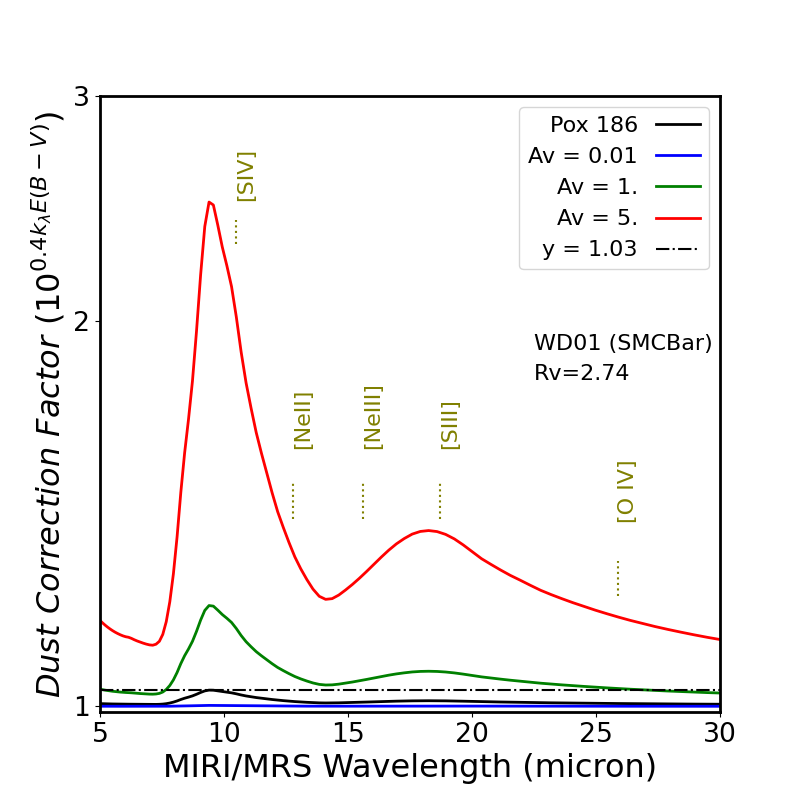}
\caption{Impact of dust correction on the spectral flux within the MIRI/MRS wavelength range when assuming an SMCBar-like dust-grain model from \citet{Weingartner2001}. The dotted horizontal line (y=1.03) shows the maximum dust-correction factor ($\sim$3\%) within the MIRI/MRS wavelength range.}
\label{fig:dust_WD01}
\end{figure}

\section{Results \& Discussion}
\label{sec:results}

\subsection{Negligible Dust-redenning}
\label{sec:emlines}

We estimated the observed line fluxes in the potential mid-infrared emission lines within the MRS wavelength range by fitting a Gaussian profile after subtracting a local linear continuum in the wavelength window enclosing the line. Tables \ref{tab:line} and  \ref{tab:H2_line} present the observed line fluxes and surface brightnesses, respectively. We chose not to correct the line fluxes for the intrinsic dust within Pox 186, because the dust content in this galaxy is too low to make an impact on the current analysis; however, we discuss the impact of dust correction in relevant sections. We estimated the dust correction factor within the MIRI/MRS spectral range using the dust grain model from \citet{Weingartner2001} for the Small Magellanic Cloud, as this is the only dust-extinction model which is valid in the mid-infrared range for a low-metallicity galaxy such as Pox 186. Figure \ref{fig:dust_WD01} shows that dust correction factor changes by only 3\% for the color excess reported for Pox 186 \citep[E(B-V)=0.058;][]{Kumari2024}. The maximum dust correction factor is lower than the flux uncertainty on the HI recombination line such as Hu$\alpha$ line and redenning correction to the line fluxes by using the observed MIR line ratio, e.g., Hu$\alpha$/Hu$\beta$, would still be in the noise of the current data set. 
This figure also shows the dust-correction factor for A$_v$ = 0.01 mag  (blue curve) corresponding to dust-less galaxy,  A$_v$ = 1 mag (green curve) above which galaxies are considered very dusty and, A$_v$ = 5 (red curve) corresponding to the typical A$_v$ of extremely dusty low-mass dwarfs galaxies ($\sim$ 10$^{7.3}$ M$_{\odot}$) determined from JWST observations \citep{Bisigello2023}, for which the redenning-correction would increase the flux in certain emission lines by more than a factor of 2.  

\begin{figure}
    \centering
    \includegraphics[width=0.55\textwidth]{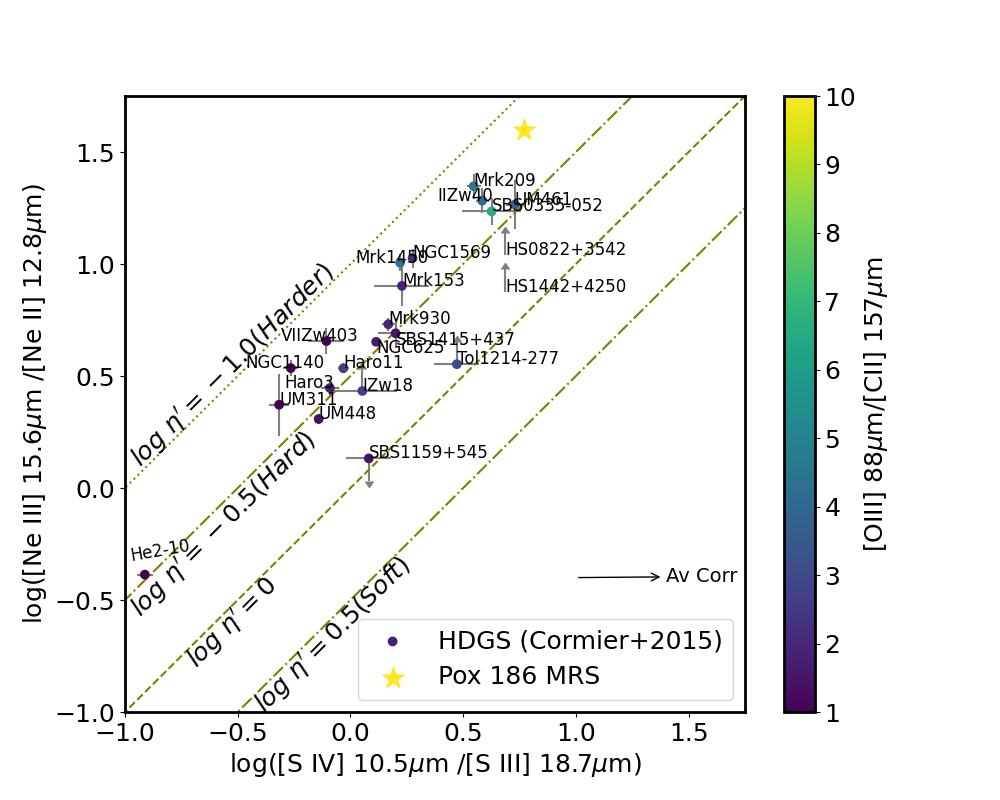}
\caption{MIR-Softness diagram showing Pox 186 from MIRI/MRS data along with the dwarf galaxies from \citet{Cormier2015} based on Spitzer high-resolution data, which are color-coded with respect to their FIR \oiii/\cii~ line ratios. The diagonal lines correspond to constant values of log $\eta\prime$ = -1 (dotted), - 0.5 (dash-dot), 0 (dashed), and 0.5 (dash-dot), with lower log $\eta\prime$ indicating a harder radiation field. The arrow in the bottom-right shows that the \siv/\siii~ line ratios will increase when the line fluxes are corrected for dust-extinction in case of large A$_v$ as demonstrated in Figure \ref{fig:dust_WD01}. }
    \label{fig:softness}
\end{figure}

\begin{figure}
    \centering
    \includegraphics[width=0.4\textwidth]{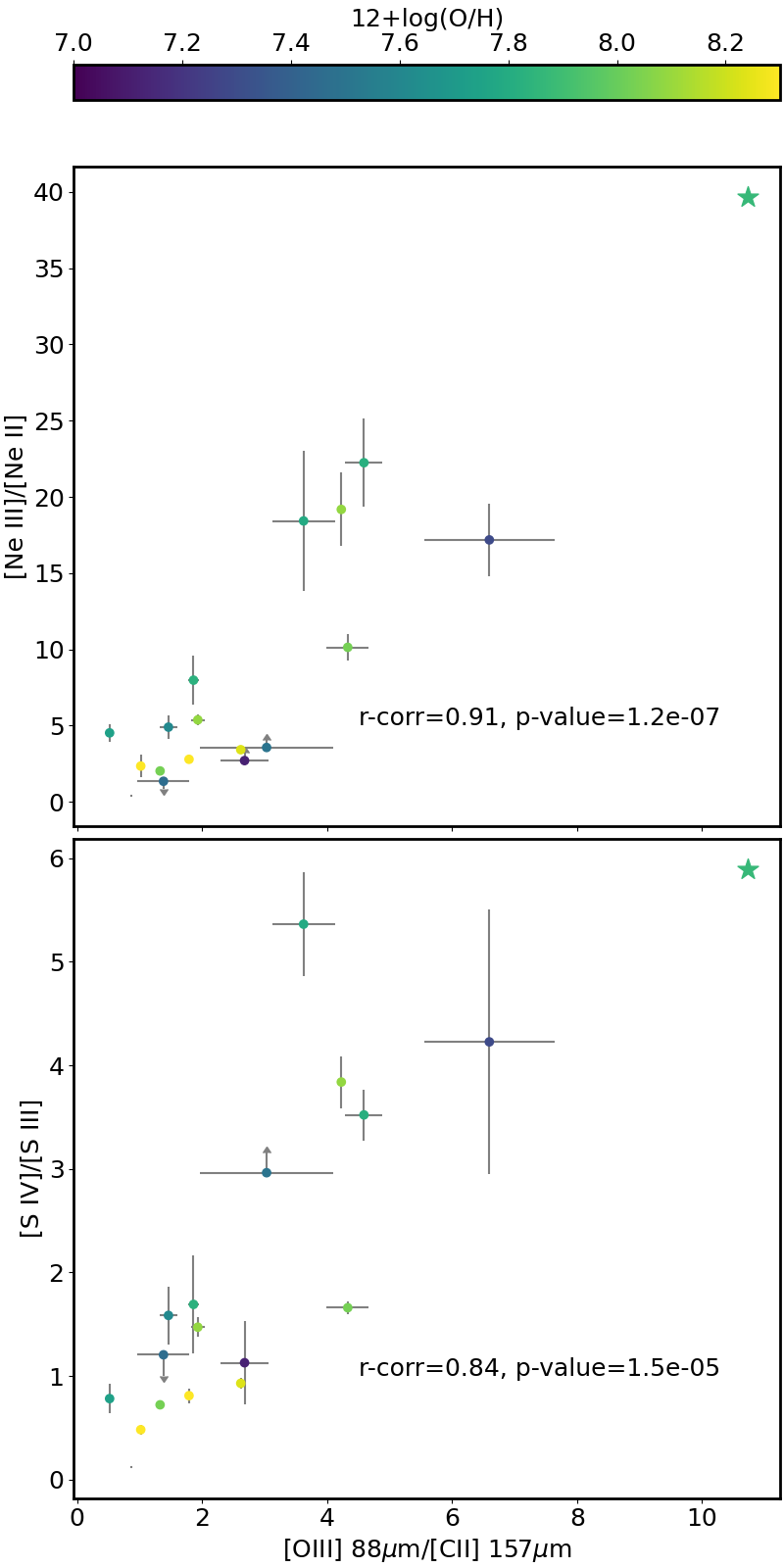}
    \caption{Variation of FIR \oiii/\cii\ with respect to MIR \neiii/\neii\ (upper-panel) and \siv/\siii~ (lower-panel) for Pox 186 (star) and other HDGS galaxies (dots). The data points are color-coded with respect to the gas-phase metallicities taken from \citet{Kumari2024} for Pox 186, and \citet{Madden2013} for other HDGS galaxies. The Pearson correlation coefficient along with the p-value for \oiii/\cii~ versus \neiii/\neii~ and \siv/\siii~ are reported in the bottom-right corners of the upper and lower panels, respectively.}
    \label{fig:o3c2_logU_Z}
\end{figure}

\begin{figure*}
    \centering    
    \includegraphics[width=0.49\textwidth]{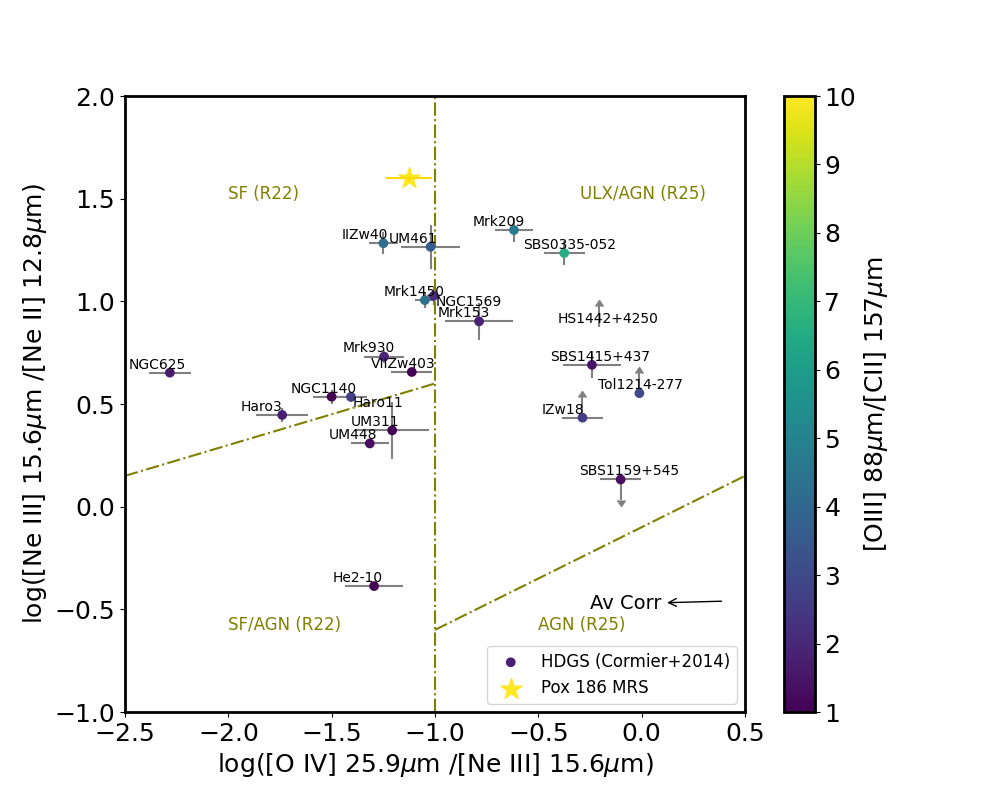}    
    \includegraphics[width=0.49\textwidth]{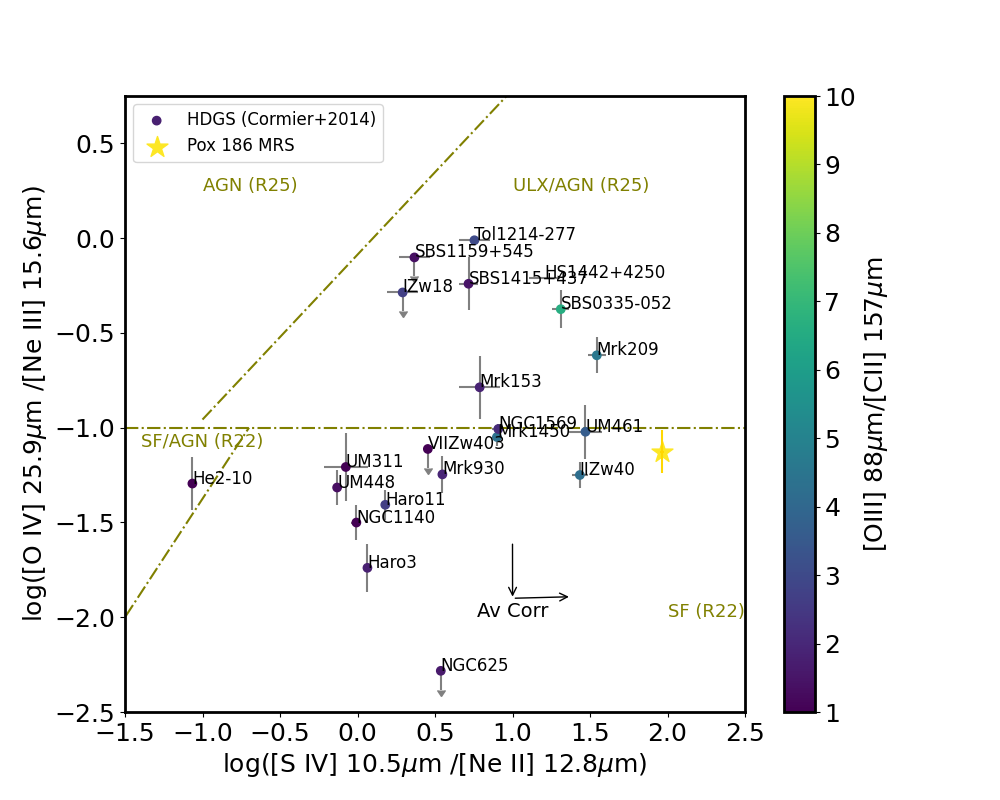}
    \caption{MIR line ratio diagnostic diagrams (left-hand panel: \neiii/\neii~ versus \oiv/\neiii~ and right-hand panel: \oiv/\neiii~ versus \siv/\neii) to distinguish the dominant ionizing source, showing Pox 186 (red star) from MIRI/MRS data along with the dwarf galaxies from \citet{Cormier2015} based on Spitzer high-resolution data, which are color-coded with respect to their FIR \oiii/\cii ~line ratios. On each panel, the dash-dot lines provides demarcation for line ratios occupied by different galaxy populations, star-forming, composite (SF/AGN), ULX/AGN and AGN and are taken from \citet{Richardson2022}, and \citet{Richardson2025}. The arrows in the bottom-right show that the increase or decrease of the corresponding line ratios will decrease when the line fluxes are corrected for dust extinction in the case of a large A$_v$ as demonstrated in Figure \ref{fig:dust_WD01}.}
    \label{fig:Ne3Ne2_O4Ne3}
\end{figure*}

\subsection{Hard radiation field in Pox 186}
\label{sec:hardness}
%\subsection{Radiation Hardness and source of ionization}
\indent Radiation hardness can be explored by using the softness diagram, which is based on the ratios of line fluxes of elements in successive ionization states, and has been explored in optical \citep{Perez-Montero2009a, Kumari2021} based on \oiii/\oii~ and \siii/\sii~ line ratios and MIR \citep{Perez-Montero2025} based on \neiii/\neii~ and \siv/\siii~ line ratios. Radiation hardness can be assessed by the theoretical softness parameter $\eta = \frac {X^i/X^{i+1}} {Y^i/Y^{i+1}}$, where X and Y are different elements, with X$^i$ and X$^{i+1}$ denoting the abundances of the $ith$ and $(i+1)th$ ionization states of element X, respectively. $\eta$ can be studied via the observational quantity $\eta\prime$ based on the line flux ratios of successive ionic species. Here we consider the MIR $\eta\prime = \frac{[Ne \textsc{ii}]/[Ne \textsc{iii}]}{[S \textsc{iii}]/[S \textsc{iv}]}$. Figure \ref{fig:softness} shows the MIR softness diagram with line ratios of Pox 186 from JWST data, and other dwarf galaxies from HDGS based on Spitzer line fluxes, where data points are color-coded with respect to the FIR \oiii/\cii~ line ratio. On Figure \ref{fig:softness}, we also show diagonal lines corresponding to log $\eta\prime$ = -1, -0.5, 0 and 0.5, where log$\eta\prime$ = 0 corresponds to \neiii/\neii = \siv/\siii. The typical SDSS galaxies in the nearby Universe show log$\eta\prime$=0 \citep{Kumari2021}, and a lower $\eta\prime$ indicates a harder radiation field. We find Pox 186 lies much higher (log $\eta\prime$ = -0.82$\pm$0.01) on this softness diagram, indicating the hardest radiation field compared to all HDGS dwarf galaxies, but still comparable to a few HDGS galaxies (Mrk 209, Mrk 1450, NGC 1140) with log $\eta\prime\sim$ -0.79 (see Table \ref{tab:hdgs}). All HDGS galaxies lie above the log$\eta\prime$=0 diagonal line, with the exception of SBS 1159+545, which also exhibits a relatively low value of the FIR \oiii/\cii~ line ratio.  In Figure \ref{fig:o3c2_logU_Z}, we investigate the variation of \oiii/\cii with respect to the two line ratios used in the calculation of the softness parameter, i.e., \neiii/\neii~(upper panel)  and \siv/\siii~(lower panel), and color-coded with respect to the gas-phase metallicity. We find a strong correlation between \oiii/\cii~ and the two line ratios sensitive to the ionization parameter, but no strong correlation with respect to metallicity or with respect to log$\eta\prime$ when we considered the entire available HDGS sample. Given that galaxies such as Pox 186 and SBS 1159+545 do indicate that the \oiii/\cii~ line ratio might be correlated to the softness parameter, and as such the radiation hardness, it would be useful to investigate this potential correlation further via a hardness diagnostic other than log$\eta\prime$.   %This could point to the FIR \oiii/\cii~ line ratio being a diagnostic for radiation hardness. 

\indent Figure \ref{fig:Ne3Ne2_O4Ne3} shows the MIR line ratio diagnostic diagrams which are proposed to distinguish the dominating source of ionization for galaxies, based on \neiii/\neii~ versus \oiv/\neiii~ (left-hand panel)  and \oiv/\neiii\ versus \siv/\neii~ (right-hand panel) line ratios. We have also indicated line ratios exhibited by different ionization sources by using the demarcation lines (dashed olive lines) from \citet{Richardson2022} and \citet{Richardson2025}. In both panels, Pox 186 line ratios (red star) obtained from JWST/MIRI data consistently lie within the star-forming region of these diagnostic diagrams and close to the demarcation line (log(\oiv/\neii) = -1.0) separating SF and ULX/AGN regions from \citet{Richardson2025}. However, the models from \citet{Richardson2025} are unable to produce the line ratios exhibited by Pox 186 (as shown in Figure \ref{fig:models} of Appendix \ref{app:models}) for the values of ionization parameter (log U $\sim$ --2) and metallicity (0.15Z$\odot$) estimated from the multi-wavelength observations. We further investigate photoionization models to reproduce the MIR emission line ratios of Pox 186 in Section \ref{sec:binaries}. %We note that the models in \citet{Richardson2025} consider a Kroupa initial mass function with an upper-mass limit of 100$M_{\odot}$. It would be worth exploring the same models with a higher upper-mass limit, e.g., 300$M_{\odot}$ considered in \citet{Kumari2024} to check whether the MIR line ratios of Pox 186 can be reproduced.} %  However, none of the model grids from \citet{Richardson2025}, either those of SF or AGN/ULX, overlap with the line ratios exhibited by Pox 186 in the two diagnostic diagrams, because of large values of \neiii/\neii~ and \siv/\neii~ of Pox 186. 

\indent Figure \ref{fig:Ne3Ne2_O4Ne3} also shows MIR line ratios of other HDGS dwarf galaxies obtained using Spitzer, color-coded with respect to their Herschel FIR \oiii/\cii~ line ratio, which consistently show that none of these dwarfs are unambiguously dominated by AGNs solely, but might have contributions from ULXs wherever \oiv/\neiii~ are too large to be explained by photoionization from stars. A future study of the broadband X-ray (2-10 keV) for these galaxies would help determine the dominant source of ionization. We also find that galaxies such as UM 311 and UM 448 are classified as composites via \oiv/\neiii~ versus \neiii/\neii~ diagram, and as SF via \oiv/\neiii~ versus \siv\neii~ diagram, showing that these diagnostics cannot explicitly distinguish between SF and Composites.

\indent \citet{Groves2008}  provide linear relations between \neiii/\neii~ and \siv/\neii~ line ratios for different populations of galaxies. We investigate these relations for Pox 186 and other HDGS dwarf galaxies in Figure \ref{fig:Ne3Ne2_S4Ne2} (Appendix \ref{app:groves}), and find that this sample satisfies both the starburst and AGN relations, and does not satisfy the relation provided for blue compact dwarfs.  

\begin{figure*}
    \includegraphics[width=0.3\textwidth]{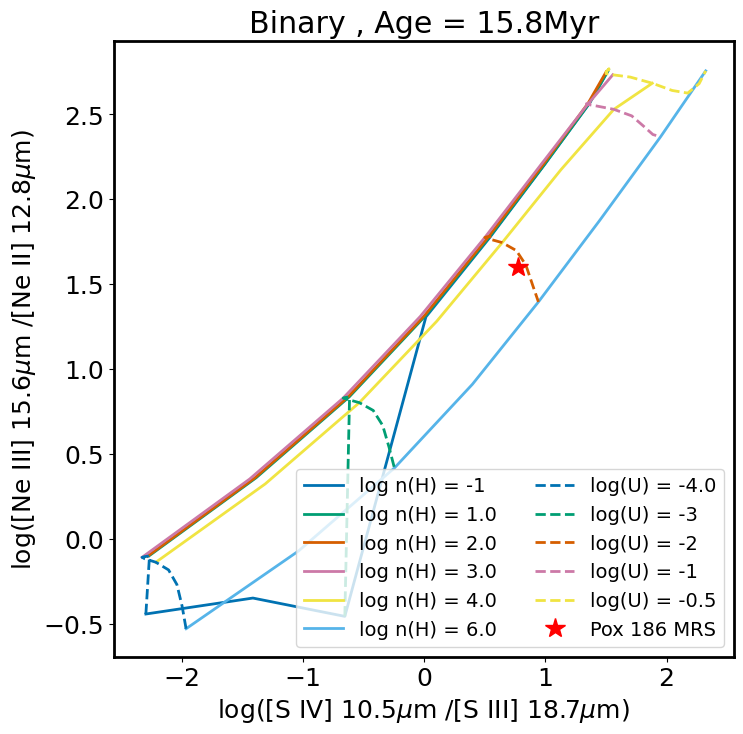}
    \includegraphics[width=0.29\textwidth]{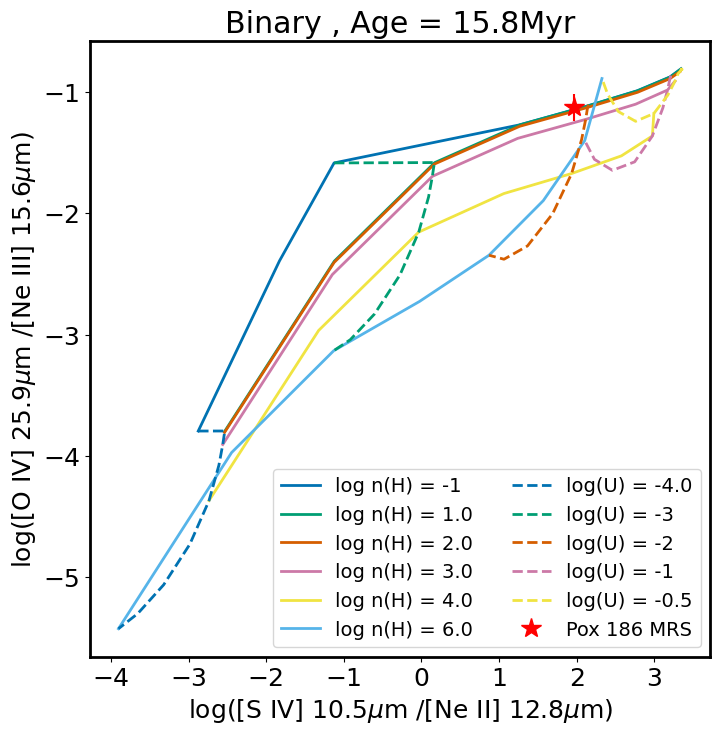}
    \includegraphics[width=0.3\textwidth]{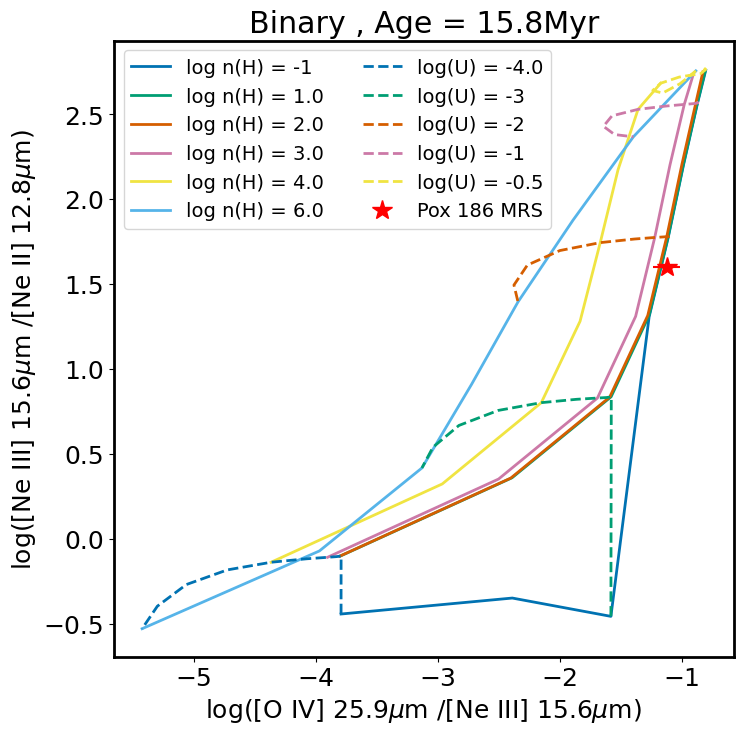}
    \caption{The cloudy photoionization models which best explain the MIR emission line ratios of Pox 186 (red star) as shown in left-hand panel: \neiii/\neii~ versus \siv/\siii, middle panel: \oiv/\neiii versus \siv/\neii, and right hand panel: \neiii/\neii versus \oiv/\neiii. The models correspond to the binary star populations with ages of $\sim$15.8 Myr, with solid colored curves showing different values of ionization parameters (log U = --0.5 - --4.0) and dashed colored curves showing the neutral gas densities (log n(H)/cm$^{-3}$ = --1 - 6).}
    \label{fig:cloudy}
\end{figure*}

\subsection{Impact of stripped binary stars}
\label{sec:binaries}

\indent  The high MIR line ratios of Pox 186 found in this work, along with its high EW(\ciii) found in \citet{Kumari2024} indicate that the radiation field is hard. Though the models from \citet{Richardson2025} do not reproduce the line ratios of Pox 186, they do indicate that the line ratios correspond to star-forming galaxies rather than AGNs. We will also show later in Section \ref{sec:PAH} that the MIR dust continuum is reproduced by a spectral energy distribution (SED) corresponding to a young star-forming galaxy. The hard radiation field for stellar sources can be obtained once their outer H-rich layers are stripped to expose the hot helium cores \citep{Gotberg2017}, for which envelope stripping via transfer of material from one star to its companion in binaries is the most viable and efficient mechanism \citep{Gotberg2019} for low-metallicity systems \citep{Hovis-Afflerbach2025}, though other mechanisms such as radiation pressure \citep{SmithNathan2014} may lead to single star stripping, and rapid stellar rotation \citep{Grasha2021} may also enhance the helium abundance on the stellar surface.

 \indent We investigate the extreme MIR emission line ratios of Pox 186 and explore the impact of binary stars using \texttt{cloudy} photoionization models, assuming plane-parallel geometry with a constant-density law and irradiation by an SED generated by \texttt{bpass}. We assumed a broken power-law initial mass function with slopes of $\alpha$=-1.3 for stars with 0.1-0.5 solar masses, and $\alpha$=-2.35 for stars with 0.5-300 M$\odot$, and composed of the binary star population. We ran these models for an age range of 1-100 Myr, neutral gas density varying from 10$^{-1}$ to 10$^6$ cm$^{-3}$, and log U varying from -0.5 to -4. We also ran the same set of models for a single-star population; however, the MIR line ratios of Pox 186 could be produced only by models with binary star populations of age $\sim$15.8 Myr,  as shown in Figure \ref{fig:cloudy}. 

\indent In the MIR softness diagram shown in Figure \ref{fig:cloudy} (left-hand panel), we find that the \neiii/\neii~ and \siv/\siii~ line ratios of Pox 186 correspond to a constant value of log U = --2 (red dashed curve) as shown by the photoionization model grids. This result itself adds confidence to our choice of model parameters, because, as we discuss below, we obtain different values of log U when we use different prescriptions found in the literature. Both \neiii/\neii~ and \siv/\siii~ line ratios, i.e., the quantities on both axes of this diagram in left-hand panel of Figure \ref{fig:cloudy} are sensitive to the ionization parameters, so 
we estimate ionization parameter from these MIR line ratios using the calibrations from \citet{Kewley2019b} based on the density and pressure models from \citet{Kewley2019a} which are derived by combining \texttt{Starburst99} \citep{Leitherer1999} and \texttt{MAPPINGSv5.1} \citep{Sutherland2018} photoionization models. Using these calibrations, we find log U = -1.549 $\pm$ 0.011 from MIR \siv/\siii~ line ratio, and log U = -2.097 $\pm$ 0.015 from  MIR \neiii/\neii\ line ratio from the data presented in this work, and log U =-2.661$\pm$0.014 from the optical emission line ratio \siii/\sii\ presented in \citet{Kumari2024}, hence differing by $\sim$1 dex. By modelling the optical data of Pox 186 via \texttt{HCM-Teff} \citep{Perez-Montero2019}, \citet{Kumari2024} reports log U = -2.4$\pm$0.4. Though, on the one hand, different values of log U from different ionic species might be indicative of different ionization zones being probed, we note that the ionization parameter is sensitive to the shape of the radiation field, which is shaped by the source itself. All the ionization parameter recipes considered here from the literature assume single stars rather than binaries. This highlights the need to consider binary star populations for ionization parameter recipes.

\indent We also find that the photoionization models incorporating binary stars are able to consistently reproduce other MIR emission line ratios of Pox 186, including \oiv/\neiii~ and \siv/\neii~ (middle panel) and \oiv/\neiii~ and \neiii/\neii~ (right-hand panel) for log U closer to -2, as for the line ratios considered in the softness diagram. However, we find that the MIR line ratios of Pox 186 shown in these figures may correspond to a wide range of gas densities, 10$^{1-6}$ cm$^{-3}$, indicative of density stratification. This is similar to the case of a low-metallicity dwarf galaxy IC 10, which exhibits a wide range of high densities closer to the ionizing cluster but constant density in the outer layers \citep{Polles2023}.

\subsection{Stable Polycyclic Aromatic Hydrocarbons (PAH)}
\label{sec:PAH}

\begin{figure*}
\centering
\includegraphics[width=\textwidth]{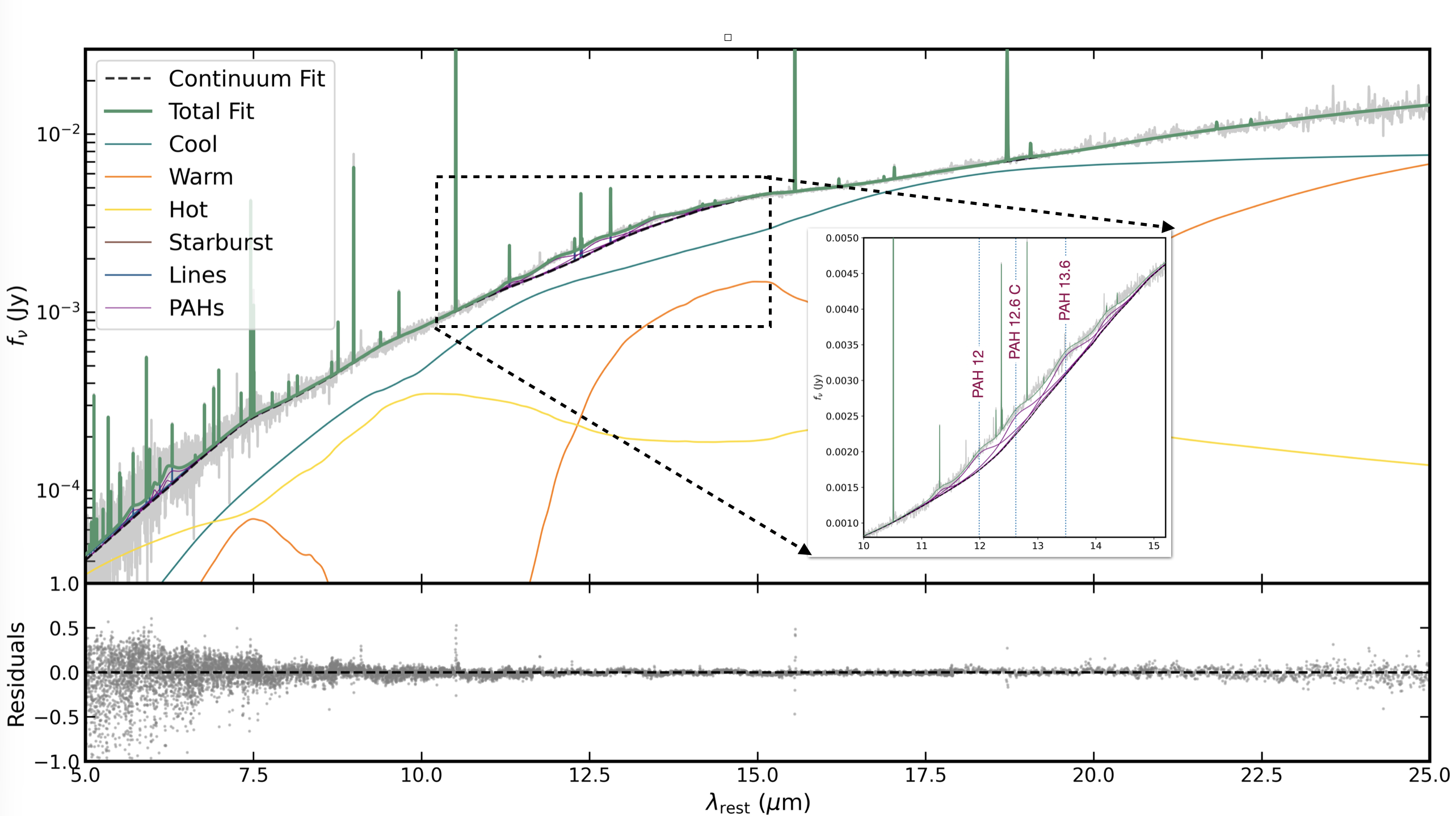}
\caption{Spectral decomposition of the MIRI/MRS spectrum of Pox 186 (light grey) using the software \texttt{cafe} showing the direct continuum (black dashed line), reprocessed continua from cool dust (light blue line), warm dust (orange line), and hot dust (yellow line), and PAH features (purple profiles). The overall fit is shown in dark green. Other spectral components such as the starburst component and emission lines, though included in \texttt{CAFE} fitting, are not obviously visible in the above figure. The normalized residuals are shown in the bottom panel. The inset in the upper panel shows the zoomed-in view of the three PAH features detected with S/N$>$3 and includes PAH 12 $\mu$m, PAH 12.6 Complex and PAH 13.6$\mu$m.}
\label{fig:cafe}
\end{figure*}

PAH features are characteristic of the mid-infrared spectrum of the interstellar medium of galaxies, and the equivalent widths and the flux ratios of certain PAH features have been explored previously to probe the radiation hardness and the nature of the ionizing source. We, therefore, perform a spectral decomposition via the software \texttt{CAFE} \citep[v2026.5.20;][]{cafe2025} and fit components related to the direct continuum, reprocessed continua from cold, warm, and hot dust, PAH features, emission lines, and absorption features from amorphous graphitic and silicate grains. Figure \ref{fig:cafe} shows the best-fit model to the data. 

\indent For modeling the dust continuum, we experimented with different SEDs such as the interstellar radiation field, AGNs, and starburst models generated from \texttt{starburst99} at 3 different ages, 2 Myr, 10 Myr, and 100 Myr, to illuminate and heat different components of the dust continuum. In principle, it would be preferable to use an SED consisting of the binary stars to model the dust continuum, because there is a strong indication of the presence of stripped binaries within Pox 186 as shown in Section \ref{sec:binaries}. However, the impact of binary stars on the dust grains and the related parameters and in turn on the dust continuum are less explored, which prevents us to use a custom SED dominated by binary stellar populations. We find that the starburst at 2 Myr best reproduces the MIR continuum, specifically at $\sim7.5-8.5\mu$m. The temperatures of the cool and warm dust components are the same, $\sim$40 K, and that of the hot dust component is $\sim$100 K. 

\indent When modeling the PAH features, we used all PAH features which are listed in \citet{Smith2007} and \citet{Marshall2007}, and allowed their widths to vary within $20\%$. We detect only three PAH features at SNR$>$3, including PAH 12$\mu$m, PAH 13.6$\mu$m, and PAH 12.6 Complex, which are the most stable PAH features \citep{Allamandola1989}. We report their equivalent widths and observed strengths in Table \ref{tab:pah}. These PAH features were previously identified in high-metallicity starburst galaxies M82 and NGC 253 \citep{Sturm2000}, though the other MIR PAH features identified in these two galaxies remain undetected in Pox 186. The 12.6$\mu$m feature is also reported in the low-metallicity dwarf galaxy CGCG 007-025 \citep{Valle-Espinosa2026}. The non-detection of ionized PAH 7.7 $\mu$m and neutral PAH 11.3 $\mu$m, along with a high \neiii/\neii\ ratio, indicates that these PAH features are destroyed by a hard radiation field \citep{Madden2006, Hunt2010}. However, some works \citep{Lai2025, Smith2007} also suggest that the larger PAH grains are enhanced in the higher metallicity environment. For example, the low-metallicity blue compact dwarf galaxy II Zw 40 shows an elevated contribution of the smaller 3.3 $\mu$m PAH compared to the total PAH emission \citep{Lai2025}. To test these competing theories and to understand the impact of a hard radiation field on dust grains, JWST/NIRSpec will be particularly useful by providing near-infrared (NIR) observations of low-metallicity dwarf galaxies, including Pox 186. 

\subsection{Warm and Cold Molecular gas}
\label{sec:molecular}

\indent Star formation is fuelled by the neutral gas (atomic and molecular) in the star-forming galaxies \citep[e.g.,][]{Kennicutt1998, Kumari2020}, though molecular gas is thought to be the main driver of star formation \citep[e.g.,][]{Bigiel2008}. Sub-mm carbon monoxide (CO) lines are mainly used as a proxy for estimating the molecular gas content of nearby star-forming galaxies \citep{KennicuttEvans2012}. However, it has been difficult to detect CO in nearby dwarf galaxies \citep{Hunter2024}. In this section, we explore the molecular gas content of Pox 186 via the MIR H$_2$ rotational lines in JWST data and CO(2-1) from ALMA.

\begin{figure}
\centering 
\includegraphics[width=0.5\textwidth]{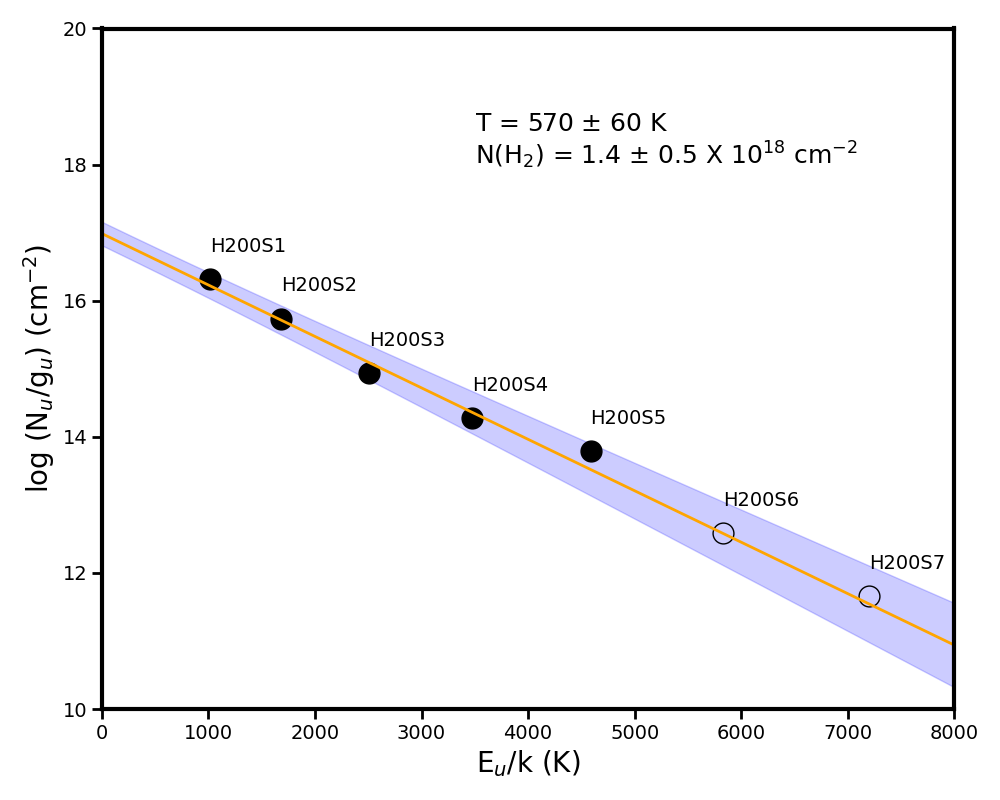}
\caption{Excitation diagram of the H$_2$ rotational lines  detected in the 1D MIRI/MRS spectrum of Pox 186. The best-fit (orange straight line) to the observed data (filled black circles) is obtained by using \texttt{pdrtpy}, and shows evidence of warm gas at $\sim$570 K corresponding to a column density of $\sim$1.4$\times$10$^{18}$ cm$^{-2}$. The $\pm$1$\sigma$ uncertainties on the best-fit line are indicated by the shaded blue region on either side of the best-fit line. The upper limits on the normalized upper-state column densities and energy levels for the undetected H$_2$(S6) and H$_2$ (S7) lines are shown by open circles, and are consistent with the best-fit 1-component model.}

\label{fig:pdrtpy}
\end{figure}

\begin{figure}
    \centering
    \includegraphics[width=0.5\textwidth]{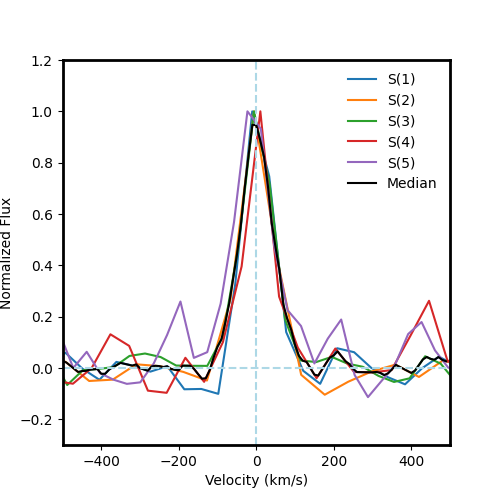}
    \caption{Continuum-subtracted line profiles of H$_2$ rotational lines (S(1)-S(5)) detected in the integrated spectrum of Pox 186, which are normalized by the peak flux. We also overplot the median of the normalized fluxes of all H$_2$ lines in black. We use this median profile to estimate the full width at half maximum ($\sim$94 km s$^{-1}$) without performing any line fitting.}
    \label{fig:H2_norm}
\end{figure}

\iffalse
\begin{table}
\caption{Molecular gas properties of Pox 186}
\resizebox{0.99\columnwidth}{!}
{%
\begin{tabular}{lr}
\toprule
Physical Parameters & Derived values \\
\midrule
%MIRI/MRS & \\
T(warm) (K) & 189  $\pm$  5 \\
T(hot) (K) & 741  $\pm$  3 \\
N(warm) (cm$^{-2}$) & 4.18E+19  $\pm$  6.0E+18 \\
N(hot) (cm$^{-2}$) & 1.28E+18  $\pm$  2.0E+16 \\
N(warm+hot) (cm$^{-2}$) & 4.31E+19  $\pm$  6.0E+18 \\
$\Sigma_M$ (warm) M$_{\odot}$pc$^{-2}$ & 0.67  $\pm$  0.097 \\
$\Sigma_M$ (hot) M$_{\odot}$pc$^{-2}$ & 0.0207  $\pm$  0.0003 \\
$\Sigma_M$(cold) (M$_{\odot}$pc$^{-2}$) & 0.159 $\pm$  0.010 \\

\bottomrule
\end{tabular}
}%
\label{tab:pdrtpy}
\end{table}
\fi

\begin{figure*}
    \centering
    \includegraphics[trim={0.5cm 41cm 2cm 4.5cm}, clip=true, width=\textwidth]{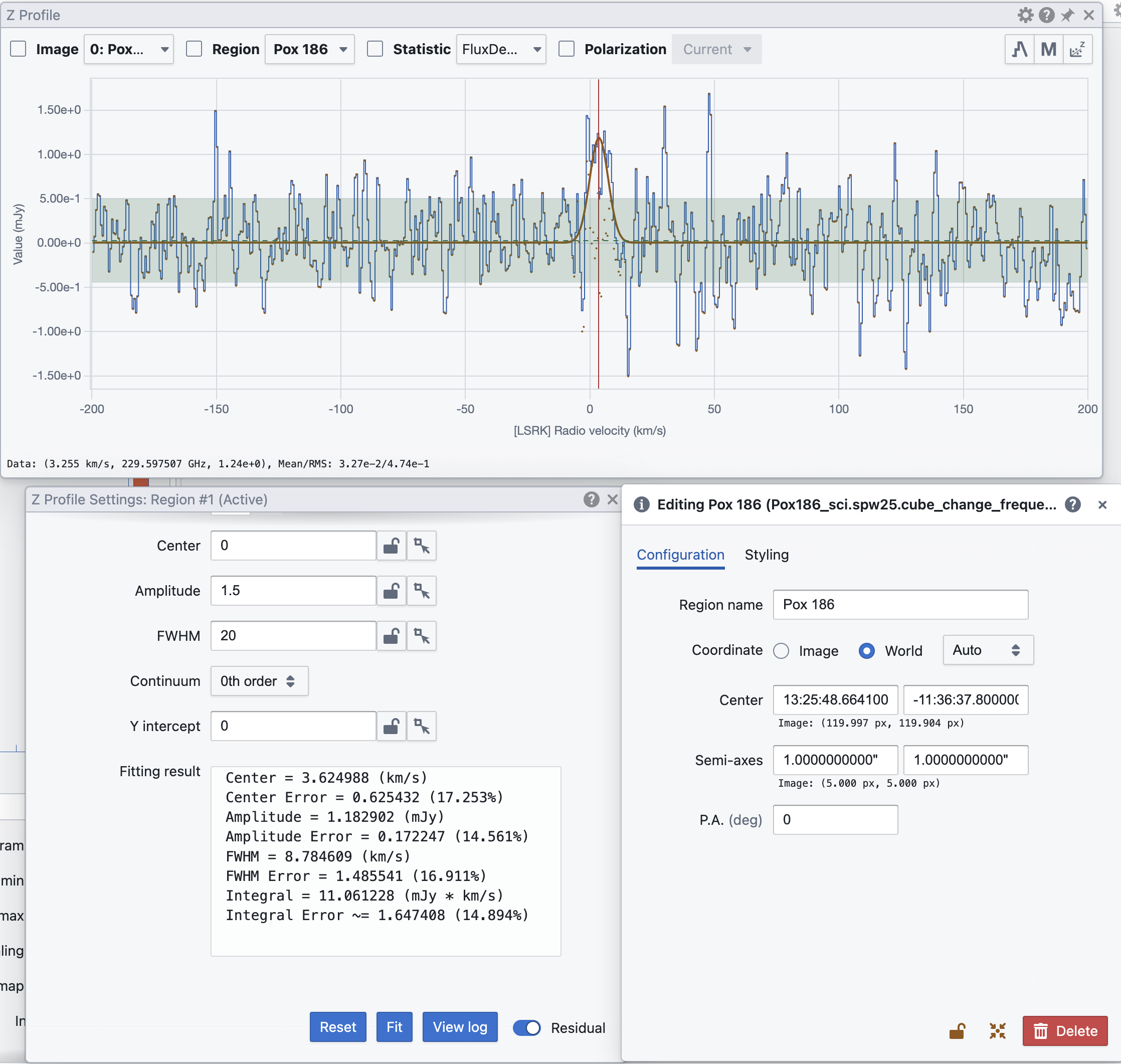}
    \caption{ALMA 1D spectrum (blue) of Pox 186 extracted from a circular aperture of radius  1\arcsec. Using \texttt{carta}, we fitted a Gaussian profile to the tentative CO (2-1) line with FWHM = 8.8$\pm$1.5 km s$^{-1}$ and amplitude 1.18 $\pm$ 0.17 mJy. The mean and rms are indicated at the bottom of the panel. %\textcolor{red}{May re-do this figure outside carta} 
    }
    \label{fig:alma_CO}
\end{figure*}

\indent \uline{MIRI/MRS:} The 1D spectrum of Pox 186 shows clear detection of H$_2$ rotational lines S(1)-S(5) (pink dashed vertical lines) (Figure \ref{fig:mrs_spectra}). Other transitions S(0), S(6), and S(7) are covered within the MIRI/MRS wavelength range, with S(0) being outside the credible wavelength range and the other two lines with no detection. %We do not include them in molecular gas analysis because the signal-to-noise on their fluxes is less than 3. 

\indent We analyze the molecular gas content of Pox 186 by modeling the line fluxes of H$_2$ rotational lines using the PhotoDissociation Region Toolbox -- Python \citep[\texttt{pdrtpy}, version 2.6.4;][]{Pound2023}, which allows us to study the variation of the upper stage column density (N$_u$\footnote{N$_u = 4\pi\frac{I}{A\Delta E}$, where I is the measured H$_2$ line intensity, A is the Einstein A coefficient and $\Delta E$ is the energy difference associated with the transition.}) for the observed H$_2$ line transition normalized by its statistical weight (g$_u$) with respect to the upper state energy of the transition (E$_u/k$ \footnote{$k$ is Boltzmann constant.}). Assuming local thermodynamic equilibrium (LTE) conditions and the ortho-to-para ratio of 3, we run \texttt{pdrtpy} on the surface brightness of MIR H$_2$ lines to fit a one-component gas model as well as a two-component (warm and hot) gas model. In the rest of the analysis, we focus on the results from a one-component model fit because a two-component model fit results in a warm gas component of temperature comparable to that from a one-component model, and a hot gas component with extremely large uncertainties on the modeled temperature, thus favoring a one-component warm gas model. Figure \ref{fig:pdrtpy} shows the one-component model fit to the MIR H$_2$ lines (S1)-(S5) (filled black circles) with a warm gas at a temperature of 570$\pm$60 K and column density $N(H_2)m_{H_2}$ of  1.4$\pm$0.5 $\times$10$^{18}$ cm$^{-2}$. We also verify that the derived model (orange straight line) is consistent with the N$_u$/g$_u$ and E$_u$/k values determined from the 3$\sigma$ upper-limit on the undetected H$_2$(S6) and H$_2$ (S7) lines (Figure \ref{fig:pdrtpy}, open circles). We estimate the warm molecular gas density of 0.022$\pm$0.008 M$_{\odot}$pc$^{-2}$ using $\Sigma_{M(H_2)} = N(H_2)m_{H_2}$, where $ m_{H_2}$ is the mean molecular weight (=2.016\footnote{https://webbook.nist.gov/cgi/cbook.cgi?ID=C1333740\&\\Type=JANAFG\&Table=on}) of an H$_2$ molecule. 

\indent To probe the ``cold" ($\sim$10-20 K) molecular gas component, we predict the CO(2-1) intensity from H$_2$(S1) intensity by using the scaling relation from \citet{Whitcomb2023}, and then predict CO(1-0) by assuming a R21 (CO(2-1)/CO(1-0)) = 0.7. Our choice of R21 is driven by the observations that higher values ($\sim$0.6-0.8) correspond to cold and dense gas, even though R21 is found to be as low as $\sim$0.2 \citep{Penaloza2017}. The conversion of CO(1-0) intensity into gas surface density relies on the conversion factor $\alpha_{CO}$, which varies with metallicity, star-formation rate, and stellar mass, and thus on the galaxy type and environment \citep{Schinnerer2024}. %There are only a few studies on $\alpha_{CO}$ for low-metallicity dwarf galaxies, because of the limited detection of CO within these galaxies. 
%There is also evidence of CO-dark molecular gas in the Milky Way \citep{Smith2014} and dwarf galaxies \citep{Madden2020} which also impacts the $\alpha_{CO}$ conversion factor. Accounting for CO-dark gas, we use the metallicity-dependent $\alpha_{CO}$ recipe from \citet{Madden2020} to estimate the total gas surface density of 1.66$\pm$0.09 M$_{\odot}$pc$^{-2}$. Ignoring CO-dark molecular gas, we estimate a cold molecular gas surface density of 0.043$\pm$0.002 M$_{\odot}$pc$^{-2}$ using the metallicity-dependent $\alpha_{CO}$ expression derived by \citet{Hunt2020}. 
Though there are metallicity-dependent prescriptions available for determining $\alpha_{CO}$, it is also argued that such prescriptions would break down below a 0.2Z$\odot$ because of significant stochasticity of $\alpha_{CO}$ at low metallicities \citep{Glover2016, Schinnerer2024}, and hence might not be applicable to Pox 186, which is 0.15 Z$\odot$ \citep{Kumari2024}. Therefore, following \citet{Elmegreen2018}, who studied Kiso 5639 with similar metallicity to Pox 186, we assume an $\alpha_{CO}$ = 100 M$_{\odot}$ (K km s$^{-1}$ pc$^2$)$^{-1}$ and estimate the cold molecular gas surface density of 0.072$\pm$0.004 M$_{\odot}$pc$^{-2}$.

%We find that the majority of molecular gas is ``warm" ($\sim$78\%), while $\sim$19\% resides in ``cold" phase and only $\sim$2\%  in ``hot" phase. These fractions are consistent with the results of \citet{Madden2020}, which show that the CO-dark gas (warm+hot components) accounts for $>$70\% of the total H$_2$ over a wide range of dwarf galaxy properties. The surface density of total (hot+warm+cold) molecular gas is $\sim$ 0.8 M$_{\odot}$ pc$^{-2}$ and lies well within the range of dwarf galaxies derived by the self-consistent models of \citet{Madden2020} accounting for CO-dark gas.

\indent Assuming that the molecular gas in Pox 186 is concentrated within a $\sim$1'' radius, which corresponds to $\sim$84 pc at the luminosity distance of 17.5 Mpc, we use the surface gas densities to estimate the molecular mass of warm gas $\sim$500M$_{\odot}$, and cold gas $\sim$1.6$\times$10$^3$M$_{\odot}$. The gas mass of cold H$_2$ is extremely low compared to that derived from ALMA, likely because the empirical relation for predicting CO intensity from H$_2$S(1) line is derived for brighter sources, and such a low cold molecular gas mass in Pox 186 might indicate that the relation is not valid for fainter sources. %We find that cold, warm, and hot molecular gas forms  $\sim$3.5$\times$10$^3$ M$_{\odot}$, $\sim$1.5$\times$10$^4$ M$_{\odot}$, and $\sim$4.6$\times$10$^2$ M$_{\odot}$, respectively. The gas mass of warm H$_2$ in Pox 186 is within the range of 10$^3$--10$^7$ M$_{\odot}$ found for blue compact dwarfs \citep{Hunt2010} and also of the SINGS galaxies \citep[10$^3$--10$^8$ M$_{\odot}$;][]{Roussel2007} . 

\indent \uline{ALMA:} Figure \ref{fig:alma_CO} shows the 1D ALMA spectrum of Pox 186 with a tentative detection of CO(2-1) extracted from a circular aperture of 1\arcsec radius. The FWHM of this line is about 10 times (Figure \ref{fig:alma_CO}) narrower when compared to the warm H$_2$ line profile (Figure \ref{fig:H2_norm}). The case here is similar to that of Mrk 71 for which an FWHM as low as 6 km s$^{-1}$ is found for the CO(2-1) line \citep{Oey2017}. A factor of a few has also been reported in \citet{Cormier2014}, where narrower CO velocity widths are associated with lines originating from a smaller region. Hence, such a small line width of CO(2-1) indicates that the cold molecular gas traced by CO(2-1) is significantly concentrated compared to the warm molecular gas within Pox 186. Figure \ref{fig:alma_CO} also shows a Gaussian fit to the CO(2-1) line, which results in an integral of 11.1$\pm$1.6 mJy km s$^{-1}$. Using the aperture-corrected ($\times$2) flux integral from the Gaussian fit, we estimate the molecular gas mass from the following equation from \citet{Watson2016},

\begin{dmath}
    \left(\frac{M_{H_2}}{M_{\odot}}\right) = 3.8 \times 10^3 \left(\frac{\alpha_{10}}{4.3 M_{\odot} pc^{-2} [K. km/s]^{-1}}\right)
    \\
    \left(\frac{R_{21/10}}{0.7}\right)^{-1} \left(\frac{\int S_{21} dv}{Jy km/s}\right) \left(\frac{D}{Mpc}\right)^2
\end{dmath}

where, we assume $\alpha_{10}$ = 100 M$_{\odot}$ pc$^{-2}$ [K. km/s]$^{-1}$, R$_{21/10}$ = 0.7, and D = 17.5 Mpc, the same as the values we assumed while doing the MIRI/MRS analysis. This gives us a molecular gas mass of 5.9$\times$10$^5$M$_{\odot}$.

\indent \uline{Herschel:} \cii~ is also proposed as a diagnostic for the molecular gas \citet{Zanella2018, Ramambason2024}. We use the \cii~ flux of Pox 186 from Herschel reported in \citet{Cormier2015} within the relation from \citet{Ramambason2024} to estimate the molecular gas mass of 1.3$\pm$0.3$\times$10$^6$ M$_{\odot}$. \citet{Cormier2015} extracted \cii~ flux of Pox 186 from a single spaxel ($\sim$9.4$\arcsec\times$9.4$\arcsec$) via PSF modeling, which corresponds to a molecular gas surface density of 2M$_{\odot}$pc$^{-2}$. 

\subsection{High Star-formation Efficiency}
\label{sec:SFE}
\begin{figure}
    \centering
    \includegraphics[width=0.45\textwidth]{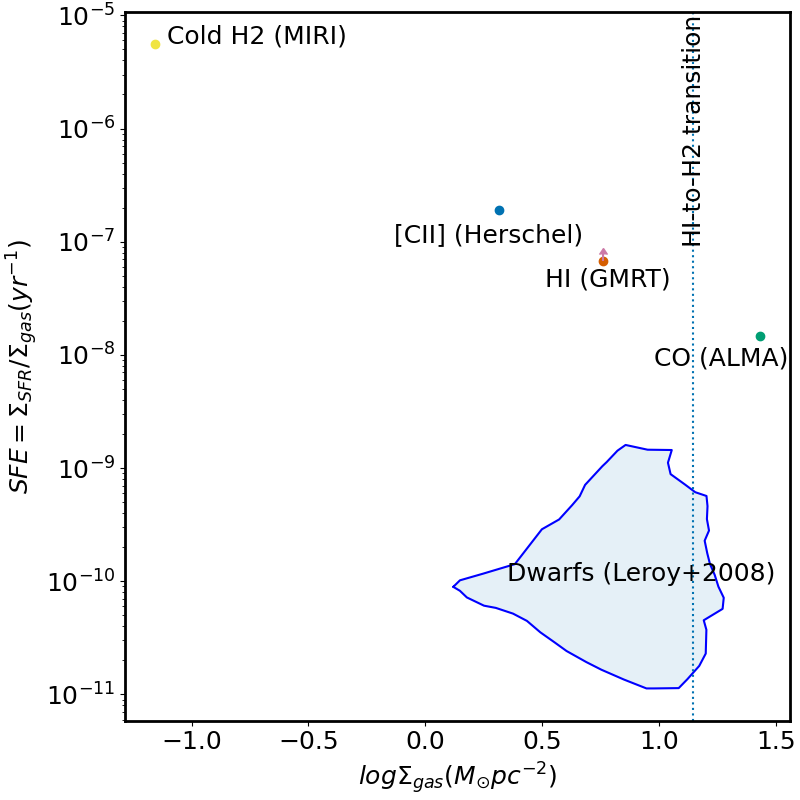}
    \caption{Variation of SFE with respect to the surface density of gas obtained from different tracers and observing facilities. All data points correspond to molecular gas surface density, with the exception of brown data with lower-limits corresponding to atomic gas surface density from H1 21cm from GMRT. The vertical dotted line corresponds to the surface gas density beyond which atomic gas converts into molecular gas. The light blue shaded region corresponds to the dwarfs in the local Universe.}
    \label{fig:sfe}
\end{figure}

Star-formation efficiency (SFE) quantifies how effectively different ISM components form stars. Though there are different mathematical definitions of SFE \citep[see e.g.,]{Leroy2008}, we assume SFE (yr$^{-1}$) = $\Sigma_{SFR}/\Sigma_{gas}$, where $\Sigma_{SFR}$ and $\Sigma_{gas}$ correspond to surface densities of star-formation rate (SFR) and gas, respectively. Figure \ref{fig:sfe} shows the variation of SFE with respect to the log $\Sigma_{gas}$ estimated from different tracers of neutral gas. To estimate the $\Sigma_{SFR}$ of Pox 186, we first estimated the H$\alpha$ luminosity from the reddening-corrected H$\alpha$ flux \citep[129$\pm$2$\times$10$^{-15}$ erg s$^{-1}$ cm$^{-2}$, ][]{Kumari2024} extracted from the HST/COS aperture, and used the metallicity-dependent SFR recipe from \citet{Ly2016} to measure SFR, and normalized it via the physical area of the COS aperture. We used the $\Sigma_{gas}$ estimated in the previous section \ref{sec:molecular} from MIRI, ALMA and Herschel, and the neutral atomic gas surface density of Pox 186 from \citet{Begum2005} based on Giant Metrewave Radio Telescope (GMRT) HI 21cm. We note that the physical areas used to obtain gas densities depend on the individual dataset and are different than the physical area used for estimating $\Sigma_{SFR}$. We find that the SFE of Pox 186 is significantly higher than the typical SFE of the local dwarf galaxy population from \citet{Leroy2008}. The lowest SFE (1.5$\times$10$^{-8}$) among all estimates done in this work is about an order of magnitude higher than the typical SFE of local dwarfs, and corresponds to a gas depletion time ($\tau_{dep}$ = 1/SFE) of 70 Myr, which is comparable to the gas depletion times of galaxies in the reionization era \citep{Vallini2024}.

%\subsection{Line Broadening}
%\label{sec:broad}

%Figure \ref{fig:H2_norm} shows the velocity line profiles of H$_2$ S(1)-S(5) transitions normalized by the peak of the line fluxes, showing that there is no line broadening with respect to UV O \textsc{iii}]1660 line, and is narrower than the UV C \textsc{iv} and C \textsc{iii} lines. 

\section{Decoding the Reionization era}
\label{sec:implications}
%\subsection{What is Pox 186?}
%\label{sec:doubt}

\subsection{Implications for interpreting the nature of high-z galaxies}
\label{sec:agn}
\indent  Emission line diagnostics across the electromagnetic spectrum (UV-optical-MIR) of Pox 186 clearly reveal a hard ionizing spectrum. The large values of EW(\ciii) in the UV,  EW(\oiii+\hb) in the optical, and the FIR \oiii/\cii\ line ratio of Pox 186 clearly place this unique local galaxy among the best analogs of reionization-era galaxies, whose ionizing spectra are hard as well, and for which the high-redshift community is actively investigating the dominant source of ionization -- star-formation versus AGN activity. Interestingly, UV emission lines (\ciii, \civ) diagnostics suggest that Pox 186 suggests AGN presence \citep[e.g.,][]{Nakajima2022} while the compactness of Pox 186 along with the presence of high ionization lines indicates star-formation \citep{Roberts-Borsani2026}. While the rest-frame MIR spectra of high-z galaxies are not accessible by the current observatories, extreme local analogs such as Pox 186 still provide a different angle to approach the much-investigated questions regarding the first stars and first galaxies. This work, which focused on MIR analysis of Pox 186, has further confirmed the hard ionizing spectrum that was already revealed in the UV-focused analysis \citep{Kumari2024}. The extreme MIR line ratios of Pox 186 could be well-reproduced by photoionization models incorporating the SED containing binary star populations from \texttt{BPASS} as discussed in Section \ref{sec:binaries}. Similarly, the MIR dust continuum of Pox 186 could be modeled by an SED of a young starburst as presented in Section \ref{sec:PAH}. Hence, detailed modeling of the MIR lines and continuum of Pox 186 provides strong evidence that the galaxy is dominated by star formation, and not an AGN. These findings for Pox 186 caution against using the UV diagnostics such as EW(\ciii) versus \civ/\ciii\ line ratios (Figure \ref{fig:intro_fig}) of high-z galaxies to establish the nature of high-z galaxies and categorize them as AGNs, because these diagnostics provide a rather incomplete picture. While EW(\ciii) undoubtedly calls for a source with a hard radiation field, other factors need to be considered while modeling the dominant ionization mechanism in these sources. This is because the equivalent width and the strengths of nebular emission lines (e.g., \ciii, \civ) are influenced by several factors such as density, temperature, geometry, nebular gas composition, and escape fraction of photons in addition to the ionizing spectrum \citep{Eldridge2020}.

\subsection{Implications for future observations}
\begin{figure}
    \centering
    \includegraphics[width=0.48\textwidth]{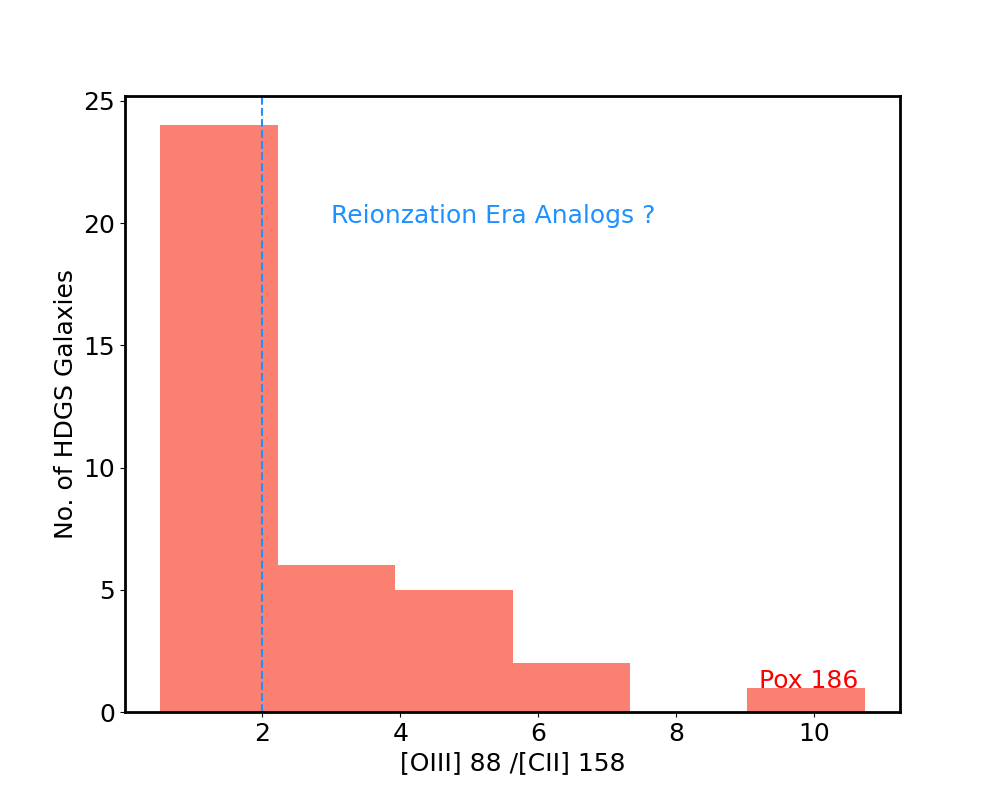}
    \caption{Distribution of HDGS galaxies based on their \oiii/\cii\ line ratios. The vertical dashed blue line indicates \oiii/\cii = 2, and dwarf galaxies with a relatively high \oiii/\cii\ line ratio above this value are potential analogs of reionization-era galaxies.}
    \label{fig:hist}
\end{figure}

\indent While discussing Pox 186, \citet{Kunth1988} had asked ``How common are such galaxies?" Pox 186 is the only local galaxy found so far with an \oiii/\cii\ line ratio $>$10, however, about 50\% of the HDGS sample shows a relatively high \oiii/\cii\ $>$2 line ratio (Figure \ref{fig:hist}), and are potential local ($<$200 Mpc) analogs of reionization-era galaxies. Characterizing these galaxies across the electromagnetic spectrum will provide a robust template collection for comparison with the high-z galaxies. Following this study on Pox 186 along with \citet{Kumari2024}, which have shown that a high \oiii/\cii\ line ratio is a good predictor of extreme UV, optical, and MIR properties and have high SFE, it would be useful to do such studies on the rest of the HDGS targets with relatively high \oiii/\cii. While bright dwarfs have been observed with HST and have UV spectroscopy \citep{Kumarihst2024, Berghst2019}, fainter targets still lack UV data. With HST nearing the mission end, it is now urgent to obtain UV spectroscopy of the fainter targets as well. Similarly, with JWST/MIRI, we can now obtain MIR spectra of these targets and probe ionization mechanisms in a statistically significant sample. In addition, the NIR spectrum of galaxies exhibits a plethora of spectral features characterizing dust, ionized and neutral gas composition \citep{Rodriguez-Ardila2005, Izotov2014, Tielens2008}. Similarly, the broadband X-ray imaging will help identify any potentially hidden AGN \citep{Maiolino2025} or ULX binaries \citep{Senchyna2020}. Hence, probing the NIR features via JWST/NIRSpec and X-ray imaging via Chandra for the HDGS-EoR analog sample (including Pox 186) will be useful and will provide insight into the properties for high-z galaxies. These observations will also guide the development of upcoming FIR mission concepts such as PRIMA \citep{Glenn2025} and Origins Space Telescope \citep{Leisawitz2021}. On one hand, the FIR facilities will provide FIR observations of local galaxies, thus potentially increasing the sample size of reionization-era analogs similar to Pox 186. On the other hand, these FIR observatories will provide the rest-frame NIR and MIR spectra of high-redshift targets, thus enabling MIR analysis similar to this one on Pox 186, and provide a more robust insight into the source of hard radiation prevalent in the distant Universe.

\section{Summary}
\label{sec:summary}
We use JWST/MIRI observations of Pox 186 to investigate the properties of this extreme local dwarf galaxy in the MIR wavelength range, which exhibits a variety of spectral features. The MIRI MRS allows us to make a spatially resolved data cube of Pox 186; however, the source remains unresolved. Hence, we extract a high signal-to-noise 1D spectrum of Pox 186 from the MRS data cube, study the global properties of this galaxy, and compare them with those of other HDGS dwarf galaxies using Spitzer MIR data. We also complement this analysis with the archival ALMA data. Our findings are summarized below:

\begin{enumerate}[a.]
    \item We probe the hardness of the radiation field in Pox 186 by using the MIR version of the softness diagram, based on the line ratios of \neiii/\neii~ and \siv/\siii, and by using $\eta\prime$, the observational quantity related to the theoretical softness parameter. Pox 186 shows an extremely low log $\eta\prime$ (=-0.82$\pm$0.01) indicative of a harder radiation field,  compared to other HDGS dwarf galaxies and normal star-forming galaxies. 
    
    \item We use the MIR diagnostic diagrams such as \oiv/\neiii~ versus \neiii/\neii~ and \siv/\neii~ versus \oiv/\neiii, along with the demarcation lines from \citet{Richardson2022} and \citet{Richardson2025} for separating dominant sources of ionization such as stars, ultraluminous X-ray sources, active galactic nuclei, and composites. Both of these diagnostic diagrams classify Pox 186 as star-forming. Our photoionization models provide evidence that the stripped binary stars are the main source of the hard radiation field.
    
    \item We use the software \texttt{cafe} to spectrally decompose the PAH features along with other features in the MIR spectrum of Pox 186. The SED of a 2 Myr old starburst is required to model the MIR continuum of Pox 186. We could detect only three of the most stable PAH features with an SNR$>$3, including PAH 12.0, PAH 13.6, and PAH 12.6 Complex. The non-detection of PAH features 11.3$\mu$m and 7.7$\mu$m along with a high \neiii/\neii\ ratio, indicates that these PAH molecules are destroyed by a hard radiation field.

    \item We use the software \texttt{pdrtpy} to model line fluxes of H$_2$ rotational lines, and find that a one-component gas model is favored over a two-component (warm and hot) gas model. We find that the warm gas is at a temperature of $\sim$570K, and has a surface gas density of 0.022 M$_{\odot}$pc$^{-2}$. 

    \item We also probe the cold gas mass by predicting the CO(2-1) from H$_2$ (S1) based on the empirical relation from \citet{Whitcomb2023}, and find the molecular gas mass to be about three orders of magnitude lower than that obtained by fitting a Gaussian line to CO(2-1) observations from ALMA. In either case, the SFE is $>$15$\times$ 10$^{-9}$ yr$^{-1}$, which is about 10 times larger than the local star-forming galaxies, and corresponds to a gas depletion time of $\lesssim$70 Myr, comparable to the EoR galaxies. 
    
\end{enumerate}

This work strongly suggests that binary stars are likely responsible for the hard ionizing radiation observed in Pox 186, and cautions against relying solely on UV diagnostics to determine the nature of high-z galaxies. In addition, this study demonstrated the need for a multi-wavelength study of a statistically significant sample of similar galaxies with large \oiii/\cii~ FIR line ratios to put better constraints on the existing models, and unravel the properties and nature of the dominant sources of ionization in reionization-era galaxies.    

%\clearpage

\begin{acknowledgments}
 We thank the JWST Instrument Development Teams and the instrument teams at the European Space Agency and the Space Telescope Science Institute. NK acknowledges discussions with Matilde Mingozzi, Oskar Arangue-Chong, Christophe Morisset, and Beena Meena during this work. J.A.-M. acknowledges support by grants PID2024-158856NA-I00 and PID2021-127718NB-100 from the Spanish Ministry of Science and Innovation/State Agency of Research MCIN/AEI/10.13039/501100011033 and by “ERDF A way of making Europe”. 
\end{acknowledgments}

\facilities{JWST(MIRI)}

%% Similar to \facility{}, there is the optional \software command to allow 
%% authors a place to specify which programs were used during the creation of 
%% the manuscript. Authors should list each code and include either a
%% citation or url to the code inside ()s when available.
\software{astropy \citep{Astropy2013}, cafe \citep{cafe2025}, carta \citep{Wang2026},  dust\_extinction \citep{Gordon2024}, pahfit \citep{Smith2007}, pdrtpy \citep{Pound2023}, 
          %Cloudy \citep{2013RMxAA..49..137F}, 
          %Source Extractor \citep{1996A&AS..117..393B}
          }

\appendix

\section{Line fluxes, ratios, PAH Equivalent Widths and abundances}
\label{app:line_pah}

Table \ref{tab:line} presents observed fluxes of emission lines  detected within 1D spectrum of Pox 186, and are in units of 10$^{-16}$erg s$^{-1}$cm$^{-2}$. Table \ref{tab:abun} shows the ionic and/or elemental abundances of available species in the MIR spectrum and the relative abundance with respect to oxygen for Pox 186. The ionic abundances are estimated by combining the MIR fluxes of collisional lines with the HI (7-6) recombination line and input them within the \texttt{pyneb} software. For estimating Ne elemental abundance, we have summed the ionic abundances, and for estimating Ar elemental abundance, we have used the ionization correction factor (ICF) from \citet{Kingsburgh1994}. No ICF prescription is available for Fe, and hence we do not estimate its elemental abundance. The relative abundance of elements with respect to oxygen is obtained by combining the elemental abundance of different species and gas-phase metallicity of Pox 186.

Table \ref{tab:H2_line} shows the observed surface brightnesses of H$_2$ rotational lines detected within 1D spectrum of the dwarf galaxy and are in units of 10$^{-6}$erg s$^{-1}$cm$^{-2}$steradian$^{-1}$. We report surface brightness rather than observed fluxes for H$_2$ rotational lines simply because we used the surface brightness while modeling these lines within \texttt{pdrtpy}. In Table \ref{tab:H2_line}, we also report the upper limits on the surface brightness of H$_2$ S(6) and H$_2$ S(7) lines which remain undetected in the 1D spectrum of Pox 186. We report upper limits as 3$\sigma$, with $\sigma$=rms$\times$FWHM$_{inst}$/$\sqrt{nres}$, where rms is the rms per wavelength channel, FWHM$_{inst}$ is the instrumental FWHM at the observed wavelength of the emission line, and nres is the number of resolution elements within FWHM$_{inst}$. We estimate FWHM$_{inst}$ from the resolving power reported in \citet{Argyriou2023}. We adopted nres=2 for both H$_2$ S(6) and H$_2$ S(7) lines.

Table \ref{tab:line_ratios} shows the relevant line ratios of Pox 186. Table \ref{tab:pah} shows the equivalent width of PAH features as determined by \texttt{CAFE}.

\indent Table \ref{tab:hdgs} presents $\eta\prime$ and line ratios used in computing the $\eta\prime$ for the HDGS sample, where they are estimated using the high-resolution Spitzer spectroscopic data.

\begin{table}
\centering
\caption{Observed fluxes of emission lines detected within 1D spectrum of Pox 186.}
\label{tab:line}
\begin{tabular}{ccc}
\hline
Line & rest-wavelength & Observed Fluxes \\
\hline
\arii & 6.985 & $0.43 \pm 0.05 $ \\
\ariii & 8.991 & $7.15 \pm 0.15 $ \\
%[ArIII] & 21.83 & $1.47 \pm 0.61 $ \\
%[ArV] & 7.9 & $0.00 \pm 0.00 $ \\
%[ArV] & 13.1 & $0.14 \pm 0.10 $ \\
\siv & 10.511 & $186.52 \pm 3.39 $ \\
\neii & 12.814 & $2.02 \pm 0.04 $ \\
%[NeV] & 14.32 & $0.00 \pm 0.00 $ \\
\neiii & 15.555 & $80.09 \pm 0.98 $ \\
\siii & 18.713 & $31.63 \pm 0.17 $ \\
%[NeV] & 24.31 & $3140.27 \pm nan $ \\
%[NeVI] & 7.65 & $0.00 \pm 0.00 $ \\
\oiv & 25.89 & $5.99 \pm 1.55 $ \\
%[FeII] & 25.9884 & $1.81 \pm 0.83 $ \\
%[PIII] & 17.885 & $0.30 \pm 0.10 $ \\
\feii & 5.34017 & $0.53 \pm 0.05 $ \\
HI(7-6) & 12.3719 & $2.67 \pm 0.14 $ \\
HI(8-6) & 7.50249 & $1.36 \pm 0.07 $ \\
\hline
\end{tabular}
\\
Notes: Units: 10$^{-16}$erg s$^{-1}$cm$^{-2}$.
\end{table}

\begin{table}
    \centering
    \caption{Abundances from MIR lines of Pox 186}
    \label{tab:abun}
    \begin{tabular}{lr}
    \hline
    Species & Values\\
    \hline
    12 + log(Ne$^+$/H$^+$) &  5.84 $\pm$ 0.03\\
12 + log(Ne$^{2+}$/H$^+$) &  7.07 $\pm$ 0.02\\
12 + log(Ne/H) &  7.1 $\pm$ 0.02\\
12 + log(Ar$^+$/H$^+$) &  4.87 $\pm$ 0.05\\
12 + log(Ar$^{2+}$/H$^+$) &  6.05 $\pm$ 0.03\\
12 + log(Ar/H) &  6.32 $\pm$ 0.03\\
12 + log(Fe$^+$/H$^+$) &  4.21 $\pm$ 0.07\\       
 \\
 
 Ne/O &  -0.77 $\pm$ 0.05\\
Ar/O &  -1.79 $\pm$ 0.05\\
 \hline 
    \end{tabular}    
\end{table}

\begin{table}
\centering
\caption{Observed surface brightnesses of H$_2$ rotational lines detected within 1D spectrum of Pox 186.}
\label{tab:H2_line}

\begin{tabular}{ccc}
\hline
H$_2$ line & Rest-wavelength & Surface Brightness \\
\hline
S(1) & 17.035 & $1.93 \pm 0.14 $ \\
S(2) & 12.279 & $1.74 \pm 0.10 $ \\
S(3) & 9.665 & $4.60 \pm 0.29 $ \\
S(4) & 8.025 & $1.26 \pm 0.18 $ \\
S(5) & 6.91 & $3.77 \pm 0.46 $ \\
S(6) & 6.109 & $<$0.19 \\
S(7) & 5.511 & $<$0.15 \\
\hline
\end{tabular}
\\
Notes: Units: 10$^{-6}$erg s$^{-1}$cm$^{-2}$steradian$^{-1}$. For S(6) and S(7), we report 3$\sigma$ upper-limits, where $\sigma$ is estimated as described in Appendix \ref{app:line_pah}.
\end{table}

\begin{table}
    \centering
    \caption{MIR emission line ratios of Pox 186 from MIRI/MRS 1D spectrum.}
    \label{tab:line_ratios}
    \begin{tabular}{lr}
    \hline
    Line rations & Values\\
    \hline
    
    log(\neiii/\neii)&  1.598$\pm$0.010\\
    log(\oiv/\neiii)&  -1.12$\pm$ 0.11\\
    log(\siv/\neii)& 1.966$\pm$0.012\\
    log(\siv/\siii) & 0.771$\pm$0.008\\ 
    \hline
    \end{tabular}
    
\end{table}

\begin{table}
\centering
\caption{Equivalent width of PAH features identified in 1D spectrum of Pox 186 by \texttt{CAFE}.}
\label{tab:pah}
\begin{tabular}{ccc}
\hline
PAH Features & Obs Strength ($\times$ 10$^{-18}$ W m$^{-2}$) & EW (micron) \\
\hline
PAH 12.0 &4.9 $\pm$ 0.9 & 0.64 $\pm$ 0.12 \\
PAH 12.6 C &4.8 $\pm$ 0.9 & 0.39 $\pm$ 0.07 \\
PAH 13.6 &4.4 $\pm$ 0.7 & 0.28 $\pm$ 0.04 \\
\hline
\end{tabular}
\\
Notes:C indicates complex.
\end{table}

\begin{table}
\centering
\caption{MIR line ratios of HDGS sample.}
\label{tab:hdgs}
\begin{tabular}{lccc}
\hline
Galaxy & \neiii/\neii & \siv/\siii & $\eta\prime$ \\
\hline
Haro3 & 2.8$\pm$0.23 & 0.81$\pm$0.07 & 0.29$\pm$0.04 \\
Haro11 & 3.43$\pm$0.18 & 0.93$\pm$0.05 & 0.27$\pm$0.02 \\
He2-10 & 0.41$\pm$0.02 & 0.12$\pm$0.01 & 0.3$\pm$0.03 \\
HS0822+3542 & $>$10.96 & 4.87$\pm$0.88 & $<$0.44 \\
HS1442+4250 & $>$7.51 & 4.86$\pm$1.71 & $<$0.65 \\
IIZw40 & 19.18$\pm$2.4 & 3.84$\pm$0.25 & 0.2$\pm$0.03 \\
IZw18 & $>$2.71 & 1.13$\pm$0.4 & $<$0.42 \\
Mrk153 & 7.98$\pm$1.63 & 1.69$\pm$0.47 & 0.21$\pm$0.07 \\
Mrk209 & 22.24$\pm$2.9 & 3.52$\pm$0.25 & 0.16$\pm$0.02 \\
Mrk930 & 5.38$\pm$0.37 & 1.47$\pm$0.09 & 0.27$\pm$0.03 \\
Mrk1450 & 10.14$\pm$0.87 & 1.66$\pm$0.06 & 0.16$\pm$0.02 \\
NGC1140 & 3.43$\pm$0.28 & 0.54$\pm$0.04 & 0.16$\pm$0.02 \\
NGC1569 & 10.62$\pm$1.03 & 1.89$\pm$0.08 & 0.18$\pm$0.02 \\
SBS0335-052 & 17.18$\pm$2.38 & 4.23$\pm$1.28 & 0.25$\pm$0.08 \\
SBS1159+545 & $<$1.36 & $<$1.21 & $<$0.89 \\
SBS1415+437 & 4.9$\pm$0.75 & 1.59$\pm$0.28 & 0.32$\pm$0.08 \\
Tol1214-277 & $>$3.57 & $>$2.96 & $<$0.83 \\
UM448 & 2.03$\pm$0.11 & 0.72$\pm$0.03 & 0.36$\pm$0.02 \\
UM461 & 18.42$\pm$4.61 & 5.36$\pm$0.5 & 0.29$\pm$0.08 \\
NGC625 & 4.49$\pm$0.21 & 1.3$\pm$0.05 & 0.29$\pm$0.02 \\
UM311 & 2.35$\pm$0.74 & 0.48$\pm$0.05 & 0.21$\pm$0.07 \\
VIIZw403 & 4.53$\pm$0.59 & 0.78$\pm$0.14 & 0.17$\pm$0.04 \\
\hline
\end{tabular}
\end{table}

\section{Issues with Models}
\label{app:models}

%We estimate ionization parameter from MIR line ratios using the calibrations from \citet{Kewley2019b} based on the density and pressure models from \citet{Kewley2019a} which are derived by combining \texttt{Starburst99} and \texttt{MAPPINGSv5.1} photoionization models. Using these calibrations, we find log U = -1.549 $\pm$ 0.011 from MIR \siv/\siii~ line ratio, and log U = -2.097 $\pm$ 0.015 from  MIR \neiii/\neii\ line ratio from the data presented in this work, and log U =-2.661$\pm$0.014 from the optical emission line ratio \siii/\sii\ presented in \citet{Kumari2024}, hence differing by $\sim$1 dex. By modelling the optical data of Pox 186 via \texttt{HCM-Teff} \citep{Perez-Montero2019}, \citet{Kumari2024} reports log U = -2.4$\pm$0.4. The differences in log U from different line ratios and modelling methods point toward the limitations in our theoretical understanding of the ISM physics. 

\begin{figure*}
    \centering
     \includegraphics[width=0.49\textwidth]{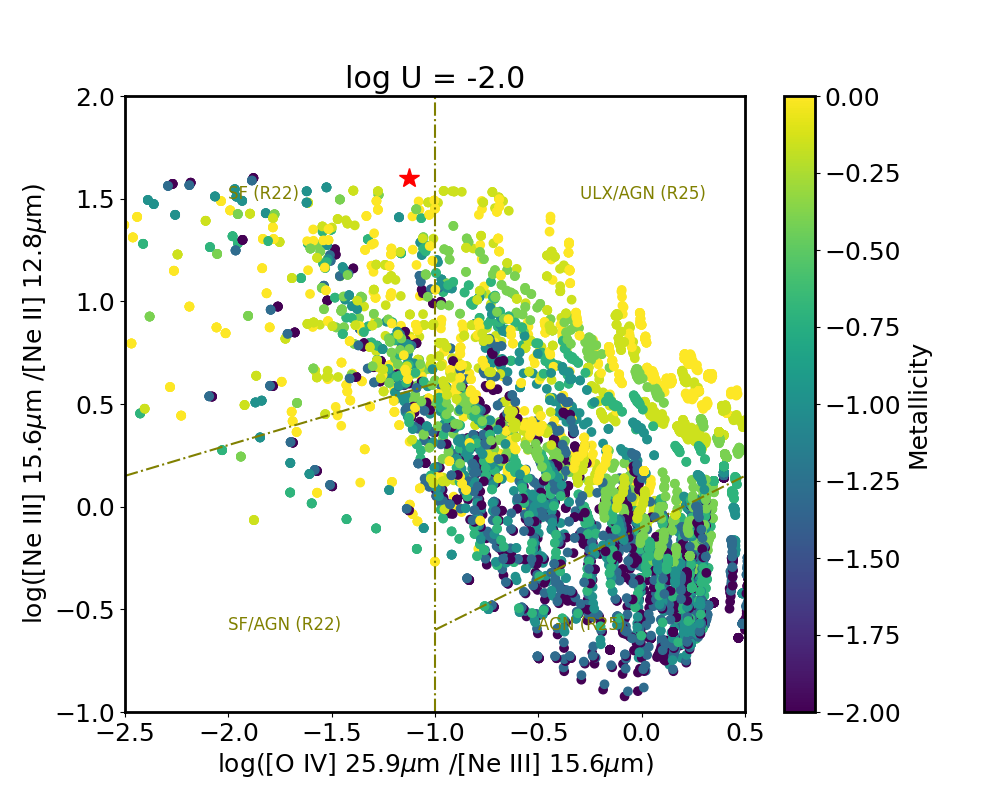}
    \includegraphics[width=0.49\textwidth]{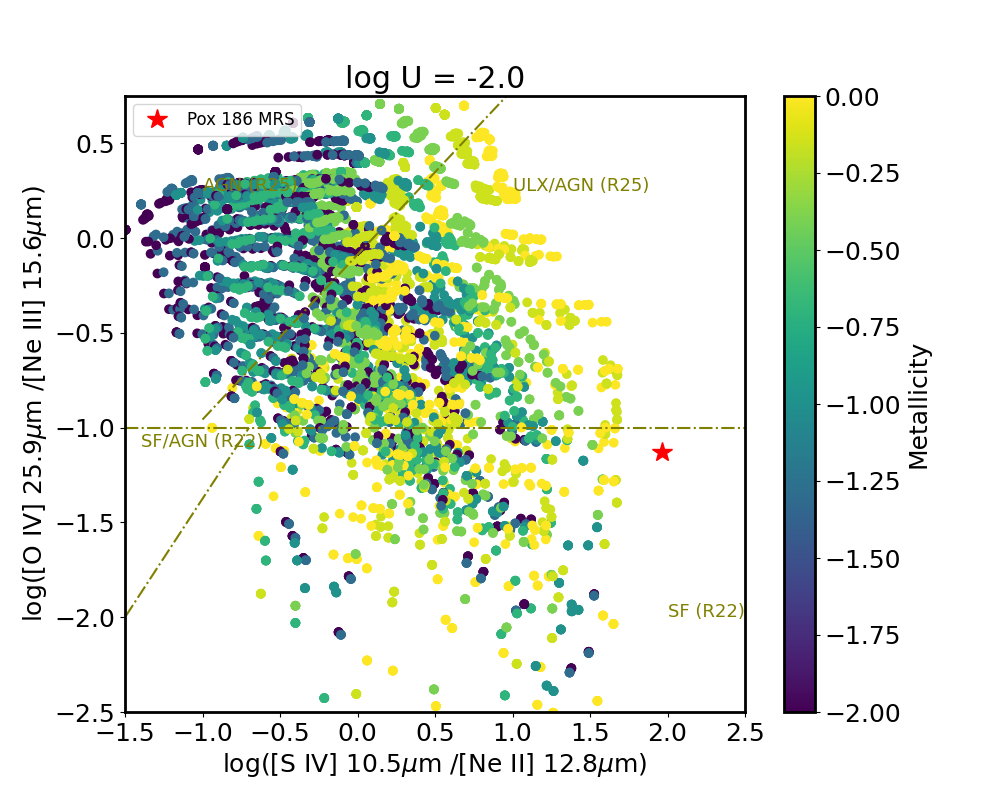}
    \includegraphics[width=0.49\textwidth]{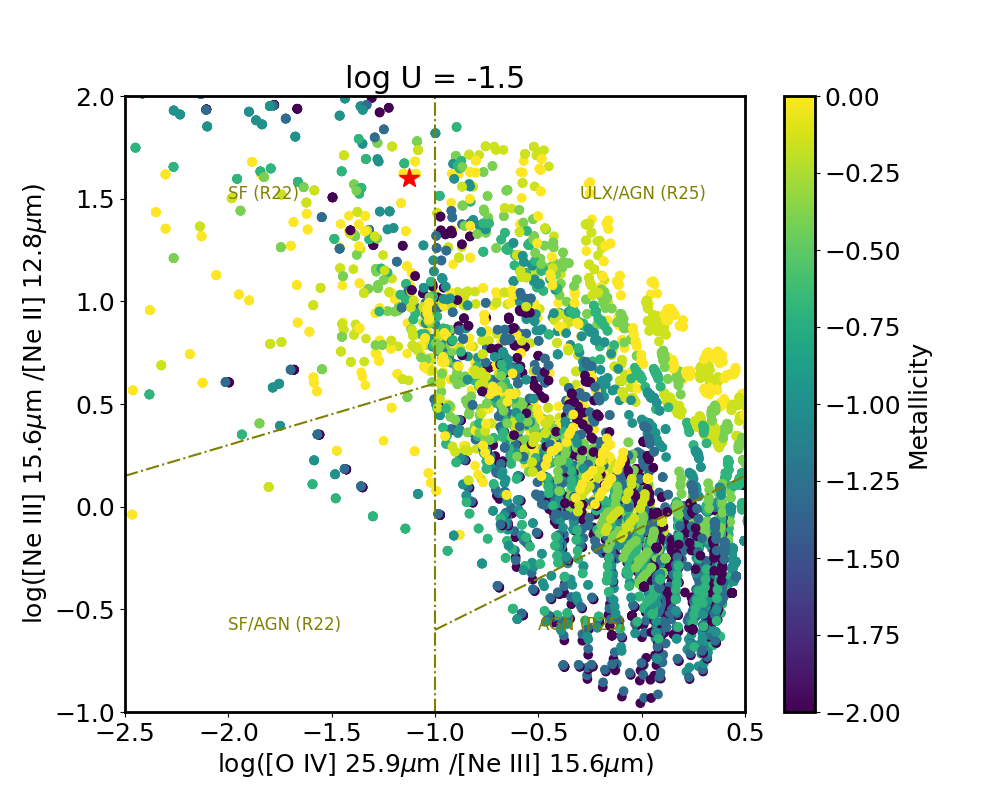}
    \includegraphics[width=0.49\textwidth]{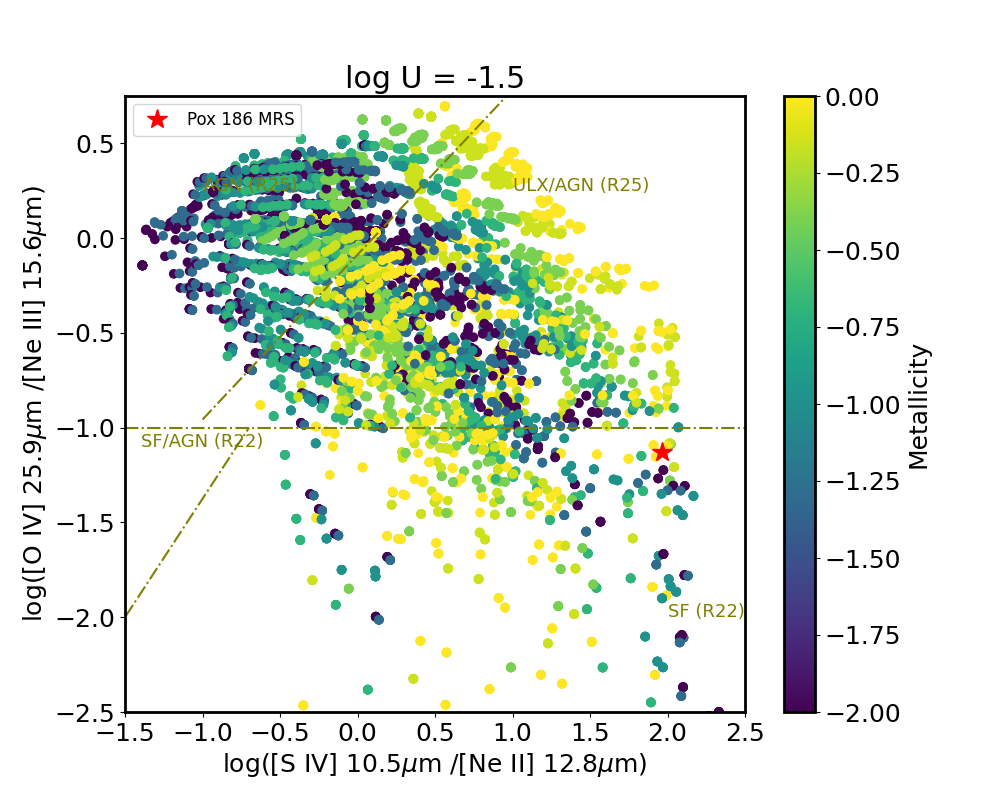}
    \caption{MIR line ratio models from \citet{Richardson2025} corresponding to log U = -1.5 (lower panel), and log U = -2.0 (upper panel), and color-coded with respect to metallicity relative to solar (log Z/Z$_{\odot}$) without taking into account the dust depletion, where log Z/Z$_{\odot}$ = 0 is Z$\odot$.}
    \label{fig:models}
\end{figure*}

Figure \ref{fig:models} shows the models from \citet{Richardson2025} on the line ratio diagrams of Pox 186. In particular, we have explored models corresponding to log U = -1.5 (lower panels), and log U =-2 (upper panels). The models corresponding to log U = -1.5 reproduce the line ratios of Pox 186, but the metallicities are relatively larger.

\section{AGN versus Starburst}
\label{app:groves}

\begin{figure}
    \centering
    \includegraphics[width=0.55\textwidth]{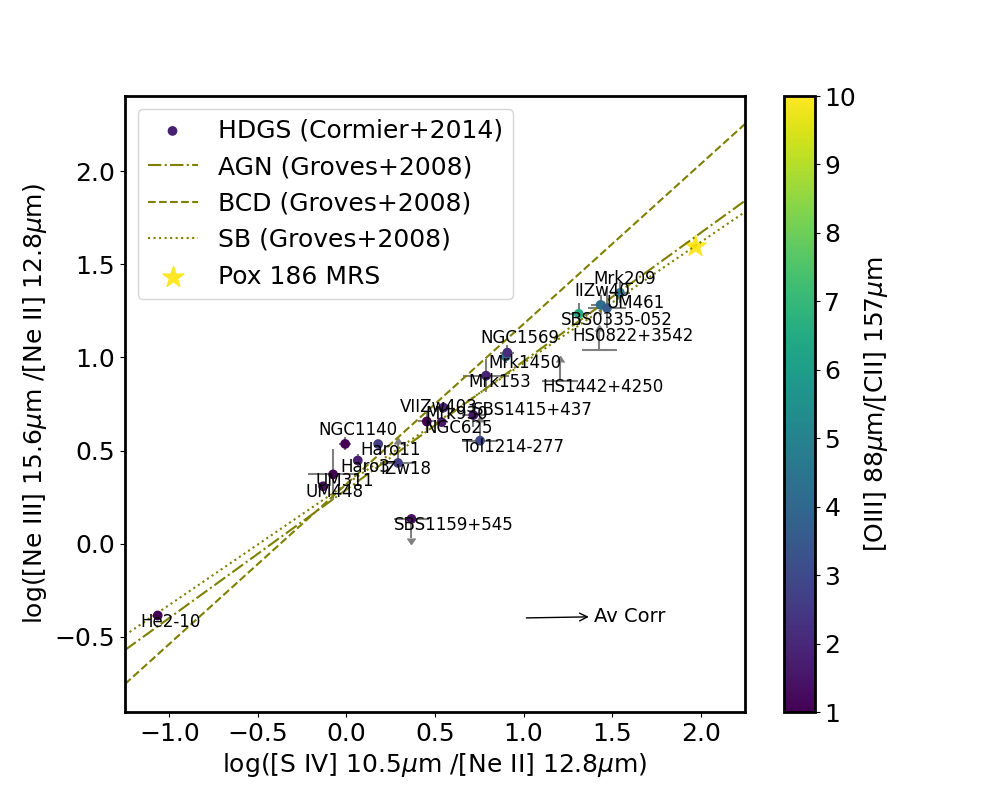}
    \caption{\neiii/\neii~ versus \siv/\neii~ of Pox 186 from MIRI/MRS data along with the dwarf galaxies from \citet{Cormier2015} based on Spitzer high-resolution data. The arrow in the bottom-right shows that the \siv/\neiii~ line ratios will increase when the line fluxes are corrected for dust extinction in the case of large A$_v$ as demonstrated in Figure \ref{fig:dust_WD01}.}
    \label{fig:Ne3Ne2_S4Ne2}
\end{figure}

Figure \ref{fig:Ne3Ne2_S4Ne2} shows the \neiii/\neii~ versus \siv/\neii~ of Pox 186 from MIRI/MRS data along with the dwarf galaxies from \citet{Cormier2015} based on Spitzer high-resolution data.

\section{SFE from different tracers}
\label{app:sfe}
Table \ref{tab:sfe} presents the SFE of Pox 186 estimated from different tracers of atomic and molecular gas, along with the observatories used for taking the gas data. The physical area used in estimating the neutral gas surface density varies from one data set to another, for example, circular apertures were used to estimate the gas density from ALMA and MIRI data, while rectangular apertures were used for Herschel and GMRT data.  
\begin{table}
\centering
\caption{SFE from different tracers}
\label{tab:sfe}
\begin{tabular}{ccccc}
\hline
Neutral Gas & Tracer & Facility & Aperture radius & SFE (yr$^{-1}$) \\
\hline
Molecular & H$_2$ & MIRI & $\sim$1$\arcsec$ & 5.6e-06 \\
Molecular & CO(2-1) & ALMA & $\sim$1$\arcsec$ & 1.5e-08 \\
Molecular & CII & Herschel & $\sim$9.4$\arcsec$x9.4$\arcsec$ & 1.9e-07 \\
Atomic & HI & Herschel & $\sim$7$\arcsec$x7$\arcsec$ & 6.8e-08 \\
\hline
\end{tabular}
\end{table}

\bibliography{biblio}{}

@ARTICLE{Maiolino2025,
       author = {{Maiolino}, Roberto and {Risaliti}, Guido and {Signorini}, Matilde and {Trefoloni}, Bartolomeo and {Juod{\v{z}}balis}, Ignas and {Scholtz}, Jan and {{\"U}bler}, Hannah and {D'Eugenio}, Francesco and {Carniani}, Stefano and {Fabian}, Andy and {Ji}, Xihan and {Mazzolari}, Giovanni and {Bertola}, Elena and {Brusa}, Marcella and {Bunker}, Andrew J. and {Charlot}, Stephane and {Comastri}, Andrea and {Cresci}, Giovanni and {DeCoursey}, Christa Noel and {Egami}, Eiichi and {Fiore}, Fabrizio and {Gilli}, Roberto and {Perna}, Michele and {Tacchella}, Sandro and {Venturi}, Giacomo},
        title = "{JWST meets Chandra: a large population of Compton thick, feedback-free, and intrinsically X-ray weak AGN, with a sprinkle of SNe}",
      journal = {\mnras},
         year = 2025,
        month = apr,
       volume = {538},
       number = {3},
        pages = {1921-1943},
          doi = {10.1093/mnras/staf359},
archivePrefix = {arXiv},
       eprint = {2405.00504},
 primaryClass = {astro-ph.GA},
       adsurl = {https://ui.adsabs.harvard.edu/abs/2025MNRAS.538.1921M}
}

@ARTICLE{Senchyna2020,
       author = {{Senchyna}, Peter and {Stark}, Daniel P. and {Mirocha}, Jordan and {Reines}, Amy E. and {Charlot}, St{\'e}phane and {Jones}, Tucker and {Mulchaey}, John S.},
        title = "{High-mass X-ray binaries in nearby metal-poor galaxies: on the contribution to nebular He II emission}",
      journal = {\mnras},
         year = 2020,
        month = may,
       volume = {494},
       number = {1},
        pages = {941-957},
          doi = {10.1093/mnras/staa586},
archivePrefix = {arXiv},
       eprint = {1909.10574},
 primaryClass = {astro-ph.GA},
       adsurl = {https://ui.adsabs.harvard.edu/abs/2020MNRAS.494..941S}
}

@ARTICLE{Maiolino2024,
       author = {{Maiolino}, Roberto and {{\"U}bler}, Hannah and {Perna}, Michele and {Scholtz}, Jan and {D'Eugenio}, Francesco and {Witten}, Callum and {Laporte}, Nicolas and {Witstok}, Joris and {Carniani}, Stefano and {Tacchella}, Sandro and {Baker}, William M. and {Arribas}, Santiago and {Nakajima}, Kimihiko and {Eisenstein}, Daniel J. and {Bunker}, Andrew J. and {Charlot}, St{\'e}phane and {Cresci}, Giovanni and {Curti}, Mirko and {Curtis-Lake}, Emma and {de Graaff}, Anna and {Egami}, Eiichi and {Ji}, Zhiyuan and {Johnson}, Benjamin D. and {Kumari}, Nimisha and {Looser}, Tobias J. and {Maseda}, Michael and {Nelson}, Erica and {Robertson}, Brant and {Rodr{\'\i}guez Del Pino}, Bruno and {Sandles}, Lester and {Simmonds}, Charlotte and {Smit}, Renske and {Sun}, Fengwu and {Venturi}, Giacomo and {Williams}, Christina C. and {Willmer}, Christopher N.~A.},
        title = "{JADES. Possible Population III signatures at z = 10.6 in the halo of GN-z11}",
      journal = {\aap},
         year = 2024,
        month = jul,
       volume = {687},
          eid = {A67},
        pages = {A67},
          doi = {10.1051/0004-6361/202347087},
archivePrefix = {arXiv},
       eprint = {2306.00953},
 primaryClass = {astro-ph.GA},
       adsurl = {https://ui.adsabs.harvard.edu/abs/2024A&A...687A..67M}
}

@ARTICLE{Roberts-Borsani2026,
       author = {{Roberts-Borsani}, Guido and {Oesch}, Pascal A. and {Ellis}, Richard and {Weibel}, Andrea and {Giovinazzo}, Emma and {Bouwens}, Rychard and {Dayal}, Pratika and {Fontana}, Adriano and {Heintz}, Kasper E. and {Matthee}, Jorryt and {Meyer}, Romain A. and {Pentericci}, Laura and {Shapley}, Alice and {Tacchella}, Sandro and {Treu}, Tommaso and {Walter}, Fabian and {Atek}, Hakim and {Bose}, Sownak and {Castellano}, Marco and {Fudamoto}, Yoshinobu and {Morishita}, Takahiro and {Naidu}, Rohan P. and {Sanders}, Ryan L. and {van der Wel}, Arjen},
        title = "{JWST spectroscopic insights into the diversity of galaxies in the first 500 Myr: short-lived snapshots along a common evolutionary pathway}",
      journal = {\mnras},
         year = 2026,
        month = may,
       volume = {548},
       number = {3},
          eid = {stag701},
        pages = {stag701},
          doi = {10.1093/mnras/stag701},
archivePrefix = {arXiv},
       eprint = {2508.21708},
 primaryClass = {astro-ph.GA},
       adsurl = {https://ui.adsabs.harvard.edu/abs/2026MNRAS.548ag701R}
}

@ARTICLE{Alvarez-Marquez2023,
       author = {{{\'A}lvarez-M{\'a}rquez}, J. and {Labiano}, A. and {Guillard}, P. and {Dicken}, D. and {Argyriou}, I. and {Patapis}, P. and {Law}, D.~R. and {Kavanagh}, P.~J. and {Larson}, K.~L. and {Gasman}, D. and {Mueller}, M. and {Alberts}, S. and {Brandl}, B.~R. and {Colina}, L. and {Garc{\'\i}a-Mar{\'\i}n}, M. and {Jones}, O.~C. and {Noriega-Crespo}, A. and {Shivaei}, I. and {Temim}, T. and {Wright}, G.~S.},
        title = "{Nuclear high-ionisation outflow in the Compton-thick AGN NGC 6552 as seen by the JWST mid-infrared instrument}",
      journal = {\aap},
         year = 2023,
        month = apr,
       volume = {672},
          eid = {A108},
        pages = {A108},
          doi = {10.1051/0004-6361/202244880},
archivePrefix = {arXiv},
       eprint = {2209.01695},
 primaryClass = {astro-ph.GA},
       adsurl = {https://ui.adsabs.harvard.edu/abs/2023A&A...672A.108A}
}

@ARTICLE{Alvarez-Marquez2024,
       author = {{{\'A}lvarez-M{\'a}rquez}, J. and {Colina}, L. and {Crespo G{\'o}mez}, A. and {Rinaldi}, P. and {Melinder}, J. and {{\"O}stlin}, G. and {Annunziatella}, M. and {Labiano}, A. and {Bik}, A. and {Bosman}, S. and {Greve}, T.~R. and {Wright}, G. and {Alonso-Herrero}, A. and {Boogaard}, L. and {Azollini}, R. and {Caputi}, K.~I. and {Costantin}, L. and {Eckart}, A. and {Garc{\'\i}a-Mar{\'\i}n}, M. and {Gillman}, S. and {Hjorth}, J. and {Iani}, E. and {Ilbert}, O. and {Jermann}, I. and {Langeroodi}, D. and {Meyer}, R. and {Pei{\ss}ker}, F. and {P{\'e}rez-Gonz{\'a}lez}, P. and {Pye}, J.~P. and {Tikkanen}, T. and {Topinka}, M. and {van der Werf}, P. and {Walter}, F. and {Henning}, Th. and {Ray}, T.},
        title = "{Spatially resolved H{\ensuremath{\alpha}} and ionizing photon production efficiency in the lensed galaxy MACS1149-JD1 at a redshift of 9.11}",
      journal = {\aap},
         year = 2024,
        month = jun,
       volume = {686},
          eid = {A85},
        pages = {A85},
          doi = {10.1051/0004-6361/202347946},
archivePrefix = {arXiv},
       eprint = {2309.06319},
 primaryClass = {astro-ph.GA},
       adsurl = {https://ui.adsabs.harvard.edu/abs/2024A&A...686A..85A}
}

@ARTICLE{Alvarez-Marquez2025,
       author = {{{\'A}lvarez-M{\'a}rquez}, J. and {Crespo G{\'o}mez}, A. and {Colina}, L. and {Langeroodi}, D. and {Marques-Chaves}, R. and {Prieto-Jim{\'e}nez}, C. and {Bik}, A. and {Alonso-Herrero}, A. and {Boogaard}, L. and {Costantin}, L. and {Garc{\'\i}a-Mar{\'\i}n}, M. and {Gillman}, S. and {Hjorth}, J. and {Iani}, E. and {Jermann}, I. and {Labiano}, A. and {Melinder}, J. and {Meyer}, R. and {{\"O}stlin}, G. and {P{\'e}rez-Gonz{\'a}lez}, P.~G. and {Rinaldi}, P. and {Walter}, F. and {van der Werf}, P. and {Wright}, G.},
        title = "{Insight into the starburst nature of Galaxy GN-z11 with JWST MIRI spectroscopy}",
      journal = {\aap},
         year = 2025,
        month = mar,
       volume = {695},
          eid = {A250},
        pages = {A250},
          doi = {10.1051/0004-6361/202451731},
archivePrefix = {arXiv},
       eprint = {2412.12826},
 primaryClass = {astro-ph.GA},
       adsurl = {https://ui.adsabs.harvard.edu/abs/2025A&A...695A.250A}
}

@ARTICLE{Alvarez-Marquez2026,
       author = {{{\'A}lvarez-M{\'a}rquez}, J. and {Colina}, L. and {Crespo-Gomez}, A. and {Kendrew}, S. and {Zavala}, J. and {Marques-Chaves}, R. and {Prieto-Jim{\'e}nez}, C. and {Abdurro'uf} and {Blanco-Prieto}, C. and {Boogaard}, L.~A. and {Castellano}, M. and {Fontana}, A. and {Fudamoto}, Y. and {Fujimoto}, S. and {Garc{\'\i}a-Mar{\'\i}n}, M. and {Harikane}, Y. and {Harish}, S. and {Hashimoto}, T. and {Hsiao}, T. and {Iani}, E. and {Inoue}, A.~K. and {Langeroodi}, D. and {Lin}, R. and {Melinder}, J. and {Napolitano}, L. and {Ostlin}, G. and {P{\'e}rez-Gonz{\'a}lez}, P.~G. and {Rinaldi}, P. and {Rodr{\'\i}guez Del Pino}, B. and {Santini}, P. and {Sugahara}, Y. and {Treu}, T. and {Varo-O'ferral}, A. and {Wright}, G.},
        title = "{PRISMS. UNCOVER-26185, a metal-poor SFG at z=10.05 with no evidence for a X-ray-luminous AGN}",
      journal = {arXiv e-prints},
         year = 2026,
        month = feb,
          eid = {arXiv:2602.02323},
        pages = {arXiv:2602.02323},
          doi = {10.48550/arXiv.2602.02323},
archivePrefix = {arXiv},
       eprint = {2602.02323},
 primaryClass = {astro-ph.GA},
       adsurl = {https://ui.adsabs.harvard.edu/abs/2026arXiv260202323A}
}

@ARTICLE{Kingsburgh1994,
       author = {{Kingsburgh}, R.~L. and {Barlow}, M.~J.},
        title = "{Elemental abundances for a sample of southern galactic planetary nebulae.}",
      journal = {\mnras},
         year = 1994,
        month = nov,
       volume = {271},
        pages = {257-299},
          doi = {10.1093/mnras/271.2.257},
       adsurl = {https://ui.adsabs.harvard.edu/abs/1994MNRAS.271..257K}
}

@MISC{Bergjwst2025,
       author = {{Berg}, Danielle and {James}, Bethan Lesley and {Amorin}, Ricardo and {Arellano Cordova}, Karla Ziboney and {Bolatto}, Alberto and {Brinchmann}, Jarle and {Carr}, Cody Andrew and {Chisholm}, John and {Dinerstein}, Harriet L. and {Erb}, Dawn K. and {Feltre}, Anna and {Fisher}, Deanne B. and {Hayes}, Matthew James and {Heckman}, Timothy M. and {Henry}, Alaina L. and {Hernandez}, Svea S. and {Hu}, Weida and {Hunt}, Leslie and {Kumari}, Nimisha and {Lai}, Thomas and {Leitherer}, Claus and {Martin}, Crystal Linn and {Martinez}, Zorayda and {Maseda}, Michael and {McQuinn}, Kristen B.~W. and {Mingozzi}, Matilde and {Parker}, Kaelee S. and {Pogge}, Richard W. and {Ravindranath}, Swara and {Rigby}, Jane R. and {Rogers}, Noah Sidney James and {Roy}, Namrata and {Sandstrom}, Karin Marie and {Scarlata}, Claudia and {Senchyna}, Peter and {Skillman}, Evan D. and {Smith}, JD and {Stark}, Daniel P. and {Strom}, Allison L. and {Wofford}, Aida and {Xu}, Xinfeng},
        title = "{The CLASSYIR Treasury: Unveiling the Cosmic Engines Powering Galaxies with JWST/MIRI}",
 howpublished = {JWST Proposal. Cycle 4, ID. \#7041},
         year = 2025,
        month = mar,
        pages = {7041},
       adsurl = {https://ui.adsabs.harvard.edu/abs/2025jwst.prop.7041B}
}

@ARTICLE{Leisawitz2021,
       author = {{Leisawitz}, David and {Amatucci}, Edward and {Allen}, Lynn and {Arenberg}, Jonathan and {Armus}, Lee and {Battersby}, Cara and {Bauer}, James and {Beaman}, Bobby G. and {Bell}, Ray and {Beltran}, Porfirio and {Benford}, Dominic and {Bergin}, Edward and {Bolognese}, Jeffrey and {Bradford}, Charles M. and {Bradley}, Damon and {Burgarella}, Denis and {Carey}, Sean and {Carter}, Ruth and {(Danny) Chi}, J.~D. and {Cooray}, Asantha and {Corsetti}, James and {D'Asto}, Thomas and {De Beck}, Elvire and {Denis}, Kevin and {Derkacz}, Christopher and {Dewell}, Larry and {DiPirro}, Michael and {Earle}, Cleland P. and {East}, Matthew and {Edgington}, Samantha and {Ennico}, Kimberly and {Fantano}, Louis and {Feller}, Greg and {Folta}, David and {Fortney}, Jonathan and {Gavares}, Benjamin J. and {Generie}, Joseph and {Gerin}, Maryvonne and {Granger}, Zachary and {Greene}, Thomas P. and {Griffiths}, Alex and {Harpole}, George and {Harvey}, Keith and {Helmich}, Frank and {Hilliard}, Lawrence and {Howard}, Joseph and {Jacoby}, Michael and {Jamil}, Anisa and {Jamison}, Tracee and {Kaltenegger}, Lisa and {Kataria}, Tiffany and {Knight}, John S. and {Knollenberg}, Perry and {Lawrence}, Charles and {Lightsey}, Paul and {Lipscy}, Sarah and {Mamajek}, Eric and {Martins}, Gregory and {Mather}, John C. and {Meixner}, Margaret and {Melnick}, Gary and {Milam}, Stefanie and {Mooney}, Ted and {Moseley}, Samuel H. and {Narayanan}, Desika and {Neff}, Susan and {Nguyen}, Thanh and {Nordt}, Alison and {Olson}, Jeffrey and {Padgett}, Deborah and {Petach}, Michael and {Petro}, Susanna and {Pohner}, John and {Pontoppidan}, Klaus and {Pope}, Alexandra and {Ramspacker}, Daniel and {Rao}, Alison and {Roellig}, Thomas and {Sakon}, Itsuki and {Sandin}, Carly and {Sandstrom}, Karin and {Scott}, Douglas and {Seals}, Len and {Sheth}, Kartik and {Sokolsky}, Lawrence M. and {Staguhn}, Johannes and {Steeves}, John and {Stevenson}, Kevin and {Stoneking}, Eric and {Su}, Kate and {Tajdaran}, Kiarash and {Tompkins}, Steven and {Vieira}, Joaquin and {Webster}, Cassandra and {Wiedner}, Martina C. and {Wright}, Edward L. and {Wu}, Chi and {Zmuidzinas}, Jonas},
        title = "{Origins Space Telescope: baseline mission concept}",
      journal = {Journal of Astronomical Telescopes, Instruments, and Systems},
         year = 2021,
        month = jan,
       volume = {7},
          eid = {011002},
        pages = {011002},
          doi = {10.1117/1.JATIS.7.1.011002},
       adsurl = {https://ui.adsabs.harvard.edu/abs/2021JATIS...7a1002L}
}

@ARTICLE{Gardner2023,
       author = {{Gardner}, Jonathan P. and {Mather}, John C. and {Abbott}, Randy and {Abell}, James S. and {Abernathy}, Mark and {Abney}, Faith E. and {Abraham}, John G. and {Abraham}, Roberto and {Abul-Huda}, Yasin M. and {Acton}, Scott and {Adams}, Cynthia K. and {Adams}, Evan and {Adler}, David S. and {Adriaensen}, Maarten and {Aguilar}, Jonathan Albert and {Ahmed}, Mansoor and {Ahmed}, Nasif S. and {Ahmed}, Tanjira and {Albat}, R{\"u}deger and {Albert}, Lo{\"\i}c and {Alberts}, Stacey and {Aldridge}, David and {Allen}, Mary Marsha and {Allen}, Shaune S. and {Altenburg}, Martin and {Altunc}, Serhat and {Alvarez}, Jose Lorenzo and {{\'A}lvarez-M{\'a}rquez}, Javier and {Alves de Oliveira}, Catarina and {Ambrose}, Leslie L. and {Anandakrishnan}, Satya M. and {Andersen}, Gregory C. and {Anderson}, Harry James and {Anderson}, Jay and {Anderson}, Kristen and {Anderson}, Sara M. and {Aprea}, Julio and {Archer}, Benita J. and {Arenberg}, Jonathan W. and {Argyriou}, Ioannis and {Arribas}, Santiago and {Artigau}, {\'E}tienne and {Arvai}, Amanda Rose and {Atcheson}, Paul and {Atkinson}, Charles B. and {Averbukh}, Jesse and {Aymergen}, Cagatay and {Bacinski}, John J. and {Baggett}, Wayne E. and {Bagnasco}, Giorgio and {Baker}, Lynn L. and {Balzano}, Vicki Ann and {Banks}, Kimberly A. and {Baran}, David A. and {Barker}, Elizabeth A. and {Barrett}, Larry K. and {Barringer}, Bruce O. and {Barto}, Allison and {Bast}, William and {Baudoz}, Pierre and {Baum}, Stefi and {Beatty}, Thomas G. and {Beaulieu}, Mathilde and {Bechtold}, Kathryn and {Beck}, Tracy and {Beddard}, Megan M. and {Beichman}, Charles and {Bellagama}, Larry and {Bely}, Pierre and {Berger}, Timothy W. and {Bergeron}, Louis E. and {Bernier}, Antoine-Darveau and {Bertch}, Maria D. and {Beskow}, Charlotte and {Betz}, Laura E. and {Biagetti}, Carl P. and {Birkmann}, Stephan and {Bjorklund}, Kurt F. and {Blackwood}, James D. and {Blazek}, Ronald Paul and {Blossfeld}, Stephen and {Bluth}, Marcel and {Boccaletti}, Anthony and {Boegner}, Jr., Martin E. and {Bohlin}, Ralph C. and {Boia}, John Joseph and {B{\"o}ker}, Torsten and {Bonaventura}, N. and {Bond}, Nicholas A. and {Bosley}, Kari Ann and {Boucarut}, Rene A. and {Bouchet}, Patrice and {Bouwman}, Jeroen and {Bower}, Gary and {Bowers}, Ariel S. and {Bowers}, Charles W. and {Boyce}, Leslye A. and {Boyer}, Christine T. and {Boyer}, Martha L. and {Boyer}, Michael and {Boyer}, Robert and {Bradley}, Larry D. and {Brady}, Gregory R. and {Brandl}, Bernhard R. and {Brannen}, Judith L. and {Breda}, David and {Bremmer}, Harold G. and {Brennan}, David and {Bresnahan}, Pamela A. and {Bright}, Stacey N. and {Broiles}, Brian J. and {Bromenschenkel}, Asa and {Brooks}, Brian H. and {Brooks}, Keira J. and {Brown}, Bob and {Brown}, Bruce and {Brown}, Thomas M. and {Bruce}, Barry W. and {Bryson}, Jonathan G. and {Bujanda}, Edwin D. and {Bullock}, Blake M. and {Bunker}, A.~J. and {Bureo}, Rafael and {Burt}, Irving J. and {Bush}, James Aaron and {Bushouse}, Howard A. and {Bussman}, Marie C. and {Cabaud}, Olivier and {Cale}, Steven and {Calhoon}, Charles D. and {Calvani}, Humberto and {Canipe}, Alicia M. and {Caputo}, Francis M. and {Cara}, Mihai and {Carey}, Larkin and {Case}, Michael Eli and {Cesari}, Thaddeus and {Cetorelli}, Lee D. and {Chance}, Don R. and {Chandler}, Lynn and {Chaney}, Dave and {Chapman}, George N. and {Charlot}, S. and {Chayer}, Pierre and {Cheezum}, Jeffrey I. and {Chen}, Bin and {Chen}, Christine H. and {Cherinka}, Brian and {Chichester}, Sarah C. and {Chilton}, Zachary S. and {Chittiraibalan}, Dharini and {Clampin}, Mark and {Clark}, Charles R. and {Clark}, Kerry W. and {Clark}, Stephanie M. and {Claybrooks}, Edward E. and {Cleveland}, Keith A. and {Cohen}, Andrew L. and {Cohen}, Lester M. and {Col{\'o}n}, Knicole D. and {Coleman}, Benee L. and {Colina}, Luis and {Comber}, Brian J. and {Comeau}, Thomas M. and {Comer}, Thomas and {Conde Reis}, Alain and {Connolly}, Dennis C. and {Conroy}, Kyle E. and {Contos}, Adam R. and {Contreras}, James and {Cook}, Neil J. and {Cooper}, James L. and {Cooper}, Rachel Aviva and {Correia}, Michael F. and {Correnti}, Matteo and {Cossou}, Christophe and {Costanza}, Brian F. and {Coulais}, Alain and {Cox}, Colin R. and {Coyle}, Ray T. and {Cracraft}, Misty M. and {Crew}, Keith A. and {Curtis}, Gary J. and {Cusveller}, Bianca and {Da Costa Maciel}, Cleyciane and {Dailey}, Christopher T. and {Daugeron}, Fr{\'e}d{\'e}ric and {Davidson}, Greg S. and {Davies}, James E. and {Davis}, Katherine Anne and {Davis}, Michael S. and {Day}, Ratna and {de Chambure}, Daniel and {de Jong}, Pauline and {De Marchi}, Guido and {Dean}, Bruce H. and {Decker}, John E. and {Delisa}, Amy S. and {Dell}, Lawrence C. and {Dellagatta}, Gail},
        title = "{The James Webb Space Telescope Mission}",
      journal = {\pasp},
         year = 2023,
        month = jun,
       volume = {135},
       number = {1048},
          eid = {068001},
        pages = {068001},
          doi = {10.1088/1538-3873/acd1b5},
archivePrefix = {arXiv},
       eprint = {2304.04869},
 primaryClass = {astro-ph.IM},
       adsurl = {https://ui.adsabs.harvard.edu/abs/2023PASP..135f8001G}
}

@ARTICLE{Glenn2025,
       author = {{Glenn}, Jason and {Meixner}, Margaret and {Bradford}, Charles M. and {Pontoppidan}, Klaus and {Pope}, Alexandra and {Kataria}, Tiffany and {Rocca}, Jennifer and {Luthman}, Elizabeth and {Armus}, Lee and {Baselmans}, Jochem and {Battersby}, Cara and {Bollato}, Alberto and {Burgarella}, Denis and {Chen}, Weibo and {Ciesla}, Laure and {Day}, Peter and {Di Giorgio}, Anna and {Dipirro}, Michael and {Dowell}, Charles Darren and {Echternach}, Pierre and {Essinger-Hileman}, Thomas and {Foote}, Marc and {Gruppioni}, Carlotta and {Hensley}, Brandon and {Henning}, Thomas and {Jellema}, Willem and {Johnson}, Matthew and {Kogut}, Alan and {Krause}, Oliver and {McGuire}, James and {Mills}, Elisabeth and {Moullet}, Arielle and {Rodgers}, Michael and {Sauvage}, Marc and {Smith}, John D. and {Somerville}, Rachel and {Staguhn}, Johannes and {Stevenson}, Thomas and {Tucker}, Carole and {Unwin}, Stephen and {Ziemer}, John and {Cannella}, Matthew and {Dissly}, Richard},
        title = "{PRIMA mission concept}",
      journal = {Journal of Astronomical Telescopes, Instruments, and Systems},
         year = 2025,
        month = jul,
       volume = {11},
          eid = {031628},
        pages = {031628},
          doi = {10.1117/1.JATIS.11.3.031628},
       adsurl = {https://ui.adsabs.harvard.edu/abs/2025JATIS..11c1628G}
}

@ARTICLE{Tielens2008,
       author = {{Tielens}, A.~G.~G.~M.},
        title = "{Interstellar polycyclic aromatic hydrocarbon molecules.}",
      journal = {\araa},
         year = 2008,
        month = sep,
       volume = {46},
        pages = {289-337},
          doi = {10.1146/annurev.astro.46.060407.145211},
       adsurl = {https://ui.adsabs.harvard.edu/abs/2008ARA&A..46..289T}
}

@ARTICLE{Rodriguez-Ardila2005,
       author = {{Rodr{\'\i}guez-Ardila}, A. and {Riffel}, R. and {Pastoriza}, M.~G.},
        title = "{Molecular hydrogen and [FeII] in active galactic nuclei - II. Results for Seyfert 2 galaxies}",
      journal = {\mnras},
         year = 2005,
        month = dec,
       volume = {364},
       number = {3},
        pages = {1041-1053},
          doi = {10.1111/j.1365-2966.2005.09638.x},
       adsurl = {https://ui.adsabs.harvard.edu/abs/2005MNRAS.364.1041R}
}

@ARTICLE{Izotov2014,
       author = {{Izotov}, Y.~I. and {Thuan}, T.~X. and {Guseva}, N.~G.},
        title = "{A new determination of the primordial He abundance using the He I {\ensuremath{\lambda}}10830 {\r{A}} emission line: cosmological implications}",
      journal = {\mnras},
         year = 2014,
        month = nov,
       volume = {445},
       number = {1},
        pages = {778-793},
          doi = {10.1093/mnras/stu1771},
archivePrefix = {arXiv},
       eprint = {1408.6953},
 primaryClass = {astro-ph.CO},
       adsurl = {https://ui.adsabs.harvard.edu/abs/2014MNRAS.445..778I}
}

@MISC{Berghst2019,
       author = {{Berg}, Danielle and {Chisholm}, John and {Heckman}, Timothy M. and {James}, Bethan Lesley and {Martin}, Crystal Linn and {Stark}, Daniel P. and {Aloisi}, Alessandra and {Amorin}, Ricardo and {Bayliss}, Matthew and {Bordoloi}, Rongmon and {Brinchmann}, Jarle and {Byler}, Nell and {Charlot}, Stephane and {Chevallard}, Jacopo and {Erb}, Dawn K. and {Feltre}, Anna and {Hayes}, Matthew James and {Henry}, Alaina L. and {Hernandez}, Svea S. and {Jaskot}, Anne and {Kewley}, Lisa and {Leitherer}, Claus and {Nanayakkara}, Themiya and {Ouchi}, Masami and {Pogge}, Richard W. and {Ravindranath}, Swara and {Rigby}, Jane R. and {Scarlata}, Claudia and {Senchyna}, Peter and {Skillman}, Evan D. and {Steidel}, Charles C. and {Strom}, Allison L. and {Wilkins}, Stephen Matthew and {Wofford}, Aida},
        title = "{The COS Legacy Archive Spectroscopic SurveY (CLASSY): A UV Treasury of Star-Forming Galaxies}",
 howpublished = {HST Proposal. Cycle 27, ID. \#15840},
         year = 2019,
        month = jun,
        pages = {15840},
       adsurl = {https://ui.adsabs.harvard.edu/abs/2019hst..prop15840B}
}

@software{Sutherland2018,
       author = {{Sutherland}, Ralph and {Dopita}, Mike and {Binette}, Luc and {Groves}, Brent},
        title = "{MAPPINGS V: Astrophysical plasma modeling code}",
 howpublished = {Astrophysics Source Code Library, record ascl:1807.005},
         year = 2018,
        month = jul,
          eid = {ascl:1807.005},
archivePrefix = {ascl},
       eprint = {1807.005},
       adsurl = {https://ui.adsabs.harvard.edu/abs/2018ascl.soft07005S}
}

@MISC{Kumarihst2024,
       author = {{Kumari}, Nimisha and {Leitherer}, Claus and {Smit}, Renske and {Witstok}, Joris},
        title = "{Decoding Reionization Era using UV and IR spectral features}",
 howpublished = {HST Proposal. Cycle 32, ID. \#17729},
         year = 2024,
        month = jul,
        pages = {17729},
       adsurl = {https://ui.adsabs.harvard.edu/abs/2024hst..prop17729K}
}

@ARTICLE{Madden2006,
       author = {{Madden}, S.~C. and {Galliano}, F. and {Jones}, A.~P. and {Sauvage}, M.},
        title = "{ISM properties in low-metallicity environments}",
      journal = {\aap},
         year = 2006,
        month = feb,
       volume = {446},
       number = {3},
        pages = {877-896},
          doi = {10.1051/0004-6361:20053890},
archivePrefix = {arXiv},
       eprint = {astro-ph/0510086},
 primaryClass = {astro-ph},
       adsurl = {https://ui.adsabs.harvard.edu/abs/2006A&A...446..877M}
}

@ARTICLE{Rodighiero2026,
       author = {{Rodighiero}, Giulia and {Ferrara}, Andrea and {Catone}, Michele and {Napolitano}, Lorenzo and {Cassata}, Paolo and {Gandolfi}, Giovanni and {Merlin}, Emiliano and {Grazian}, Andrea and {Renzini}, Alvio and {Bisigello}, Laura and {Castellano}, Marco and {P{\'e}rez-Gonz{\'a}lez}, Pablo G. and {P{\'e}rez-D{\'\i}az}, Borja and {Iani}, Edoardo and {Gruppioni}, Carlotta and {Finkelstein}, Steven L. and {Koekemoer}, Anton M. and {Bianchetti}, Alessandro and {Sinigaglia}, Francesco},
        title = "{EGS-z11-R0: a red, dust-rich galaxy at Cosmic Dawn}",
      journal = {arXiv e-prints},
         year = 2026,
        month = mar,
          eid = {arXiv:2603.15841},
        pages = {arXiv:2603.15841},
          doi = {10.48550/arXiv.2603.15841},
archivePrefix = {arXiv},
       eprint = {2603.15841},
 primaryClass = {astro-ph.GA},
       adsurl = {https://ui.adsabs.harvard.edu/abs/2026arXiv260315841R}
}

@ARTICLE{deGraff2025,
       author = {{de Graaff}, Anna and {Brammer}, Gabriel and {Weibel}, Andrea and {Lewis}, Zach and {Maseda}, Michael V. and {Oesch}, Pascal A. and {Bezanson}, Rachel and {Boogaard}, Leindert A. and {Cleri}, Nikko J. and {Cooper}, Olivia R. and {Gottumukkala}, Rashmi and {Greene}, Jenny E. and {Hirschmann}, Michaela and {Hviding}, Raphael E. and {Katz}, Harley and {Labb{\'e}}, Ivo and {Leja}, Joel and {Matthee}, Jorryt and {McConachie}, Ian and {Miller}, Tim B. and {Naidu}, Rohan P. and {Price}, Sedona H. and {Rix}, Hans-Walter and {Setton}, David J. and {Suess}, Katherine A. and {Wang}, Bingjie and {Whitaker}, Katherine E. and {Williams}, Christina C.},
        title = "{RUBIES: A complete census of the bright and red distant Universe with JWST/NIRSpec}",
      journal = {\aap},
         year = 2025,
        month = may,
       volume = {697},
          eid = {A189},
        pages = {A189},
          doi = {10.1051/0004-6361/202452186},
archivePrefix = {arXiv},
       eprint = {2409.05948},
 primaryClass = {astro-ph.GA},
       adsurl = {https://ui.adsabs.harvard.edu/abs/2025A&A...697A.189D}
}

@ARTICLE{Fabian2026,
       author = {{Fabian}, A.~C. and {Jiang}, J. and {Baker}, W.~M. and {Maiolino}, R. and {Ji}, X. and {Juod{\v{z}}balis}, I. and {Scholtz}, J.},
        title = "{The possible accretion discs of GN-z11 at redshift z = 10.6, MoM-z14 at z = 14.44, and other high-redshift objects}",
      journal = {\mnras},
         year = 2026,
        month = apr,
       volume = {547},
       number = {3},
          eid = {stag379},
        pages = {stag379},
          doi = {10.1093/mnras/stag379},
archivePrefix = {arXiv},
       eprint = {2509.05459},
 primaryClass = {astro-ph.GA},
       adsurl = {https://ui.adsabs.harvard.edu/abs/2026MNRAS.547ag379F}
}

@ARTICLE{Kovacs2024,
       author = {{Kov{\'a}cs}, Orsolya E. and {Bogd{\'a}n}, {\'A}kos and {Natarajan}, Priyamvada and {Werner}, Norbert and {Azadi}, Mojegan and {Volonteri}, Marta and {Tremblay}, Grant R. and {Chadayammuri}, Urmila and {Forman}, William R. and {Jones}, Christine and {Kraft}, Ralph P.},
        title = "{A Candidate Supermassive Black Hole in a Gravitationally Lensed Galaxy at Z {\ensuremath{\approx}} 10}",
      journal = {\apjl},
         year = 2024,
        month = apr,
       volume = {965},
       number = {2},
          eid = {L21},
        pages = {L21},
          doi = {10.3847/2041-8213/ad391f},
archivePrefix = {arXiv},
       eprint = {2403.14745},
 primaryClass = {astro-ph.GA},
       adsurl = {https://ui.adsabs.harvard.edu/abs/2024ApJ...965L..21K}
}

@ARTICLE{Kunth1988,
       author = {{Kunth}, D. and {Maurogordato}, S. and {Vigroux}, L.},
        title = "{CCD observations of blue compact galaxies : a mixed bag of morphological types.}",
      journal = {\aap},
         year = 1988,
        month = oct,
       volume = {204},
        pages = {10-20},
       adsurl = {https://ui.adsabs.harvard.edu/abs/1988A&A...204...10K}
}

@ARTICLE{SmithNathan2014,
       author = {{Smith}, Nathan},
        title = "{Mass Loss: Its Effect on the Evolution and Fate of High-Mass Stars}",
      journal = {\araa},
         year = 2014,
        month = aug,
       volume = {52},
        pages = {487-528},
          doi = {10.1146/annurev-astro-081913-040025},
archivePrefix = {arXiv},
       eprint = {1402.1237},
 primaryClass = {astro-ph.SR},
       adsurl = {https://ui.adsabs.harvard.edu/abs/2014ARA&A..52..487S}
}

@ARTICLE{Grasha2021,
       author = {{Grasha}, K. and {Roy}, A. and {Sutherland}, R.~S. and {Kewley}, L.~J.},
        title = "{Stromlo Stellar Tracks: Non-solar-scaled Abundances for Massive Stars}",
      journal = {\apj},
         year = 2021,
        month = feb,
       volume = {908},
       number = {2},
          eid = {241},
        pages = {241},
          doi = {10.3847/1538-4357/abd6bf},
archivePrefix = {arXiv},
       eprint = {2101.01197},
 primaryClass = {astro-ph.GA},
       adsurl = {https://ui.adsabs.harvard.edu/abs/2021ApJ...908..241G}
}

@ARTICLE{Hovis-Afflerbach2025,
       author = {{Hovis-Afflerbach}, B. and {G{\"o}tberg}, Y. and {Schootemeijer}, A. and {Klencki}, J. and {Strom}, A.~L. and {Ludwig}, B.~A. and {Drout}, M.~R.},
        title = "{The mass distribution of stars stripped in binaries: The effect of metallicity}",
      journal = {\aap},
         year = 2025,
        month = may,
       volume = {697},
          eid = {A239},
        pages = {A239},
          doi = {10.1051/0004-6361/202453185},
archivePrefix = {arXiv},
       eprint = {2412.05356},
 primaryClass = {astro-ph.SR},
       adsurl = {https://ui.adsabs.harvard.edu/abs/2025A&A...697A.239H}
}

@ARTICLE{Gotberg2019,
       author = {{G{\"o}tberg}, Y. and {de Mink}, S.~E. and {Groh}, J.~H. and {Leitherer}, C. and {Norman}, C.},
        title = "{The impact of stars stripped in binaries on the integrated spectra of stellar populations}",
      journal = {\aap},
         year = 2019,
        month = sep,
       volume = {629},
          eid = {A134},
        pages = {A134},
          doi = {10.1051/0004-6361/201834525},
archivePrefix = {arXiv},
       eprint = {1908.06102},
 primaryClass = {astro-ph.GA},
       adsurl = {https://ui.adsabs.harvard.edu/abs/2019A&A...629A.134G}
}

@ARTICLE{Gotberg2017,
       author = {{G{\"o}tberg}, Y. and {de Mink}, S.~E. and {Groh}, J.~H.},
        title = "{Ionizing spectra of stars that lose their envelope through interaction with a binary companion: role of metallicity}",
      journal = {\aap},
         year = 2017,
        month = nov,
       volume = {608},
          eid = {A11},
        pages = {A11},
          doi = {10.1051/0004-6361/201730472},
archivePrefix = {arXiv},
       eprint = {1701.07439},
 primaryClass = {astro-ph.SR},
       adsurl = {https://ui.adsabs.harvard.edu/abs/2017A&A...608A..11G}
}

@INPROCEEDINGS{Polles2023,
       author = {{Polles}, Fiorella},
        title = "{The electron density stratification in IC10: a low-metallicity galaxy}",
    booktitle = {American Astronomical Society Meeting Abstracts \#241},
         year = 2023,
       series = {American Astronomical Society Meeting Abstracts},
       volume = {241},
        month = jan,
          eid = {235.04},
        pages = {235.04},
       adsurl = {https://ui.adsabs.harvard.edu/abs/2023AAS...24123504P}
}

@ARTICLE{Topping2024,
       author = {{Topping}, Michael W. and {Stark}, Daniel P. and {Senchyna}, Peter and {Plat}, Adele and {Zitrin}, Adi and {Endsley}, Ryan and {Charlot}, St{\'e}phane and {Furtak}, Lukas J. and {Maseda}, Michael V. and {Smit}, Renske and {Mainali}, Ramesh and {Chevallard}, Jacopo and {Molyneux}, Stephen and {Rigby}, Jane R.},
        title = "{Metal-poor star formation at z > 6 with JWST: new insight into hard radiation fields and nitrogen enrichment on 20 pc scales}",
      journal = {\mnras},
         year = 2024,
        month = apr,
       volume = {529},
       number = {4},
        pages = {3301-3322},
          doi = {10.1093/mnras/stae682},
archivePrefix = {arXiv},
       eprint = {2401.08764},
 primaryClass = {astro-ph.GA},
       adsurl = {https://ui.adsabs.harvard.edu/abs/2024MNRAS.529.3301T}
}

@ARTICLE{Nakajima2022,
       author = {{Nakajima}, K. and {Maiolino}, R.},
        title = "{Diagnostics for PopIII galaxies and direct collapse black holes in the early universe}",
      journal = {\mnras},
         year = 2022,
        month = jul,
       volume = {513},
       number = {4},
        pages = {5134-5147},
          doi = {10.1093/mnras/stac1242},
archivePrefix = {arXiv},
       eprint = {2204.11870},
 primaryClass = {astro-ph.GA},
       adsurl = {https://ui.adsabs.harvard.edu/abs/2022MNRAS.513.5134N}
}

@ARTICLE{Naidu2026,
       author = {{Naidu}, Rohan P. and {Oesch}, Pascal A. and {Brammer}, Gabriel and {Weibel}, Andrea and {Li}, Yijia and {Matthee}, Jorryt and {Chisolm}, John and {Pollock}, Clara L. and {Heintz}, Kasper E. and {Johnson}, Benjamin D. and {Shen}, Xuejian and {Hviding}, Raphael E. and {Leja}, Joel and {Tacchella}, Sandro and {Ganguly}, Arpita and {Witten}, Callum and {Atek}, Hakim and {Belli}, Siro and {Bose}, Sownak and {Bouwens}, Rychard and {Dayal}, Pratika and {Decarli}, Roberto and {de Graaff}, Anna and {Fudamoto}, Yoshinobu and {Giovinazzo}, Emma and {Greene}, Jenny E. and {Illingworth}, Garth and {Inoue}, Akio K. and {Kane}, Sarah G. and {Labbe}, Ivo and {Leonova}, Ecaterina and {Marques-Chaves}, Rui and {Meyer}, Roman A. and {Nelson}, Erica J. and {Roberts-Borsani}, Guido and {Schaerer}, Daniel and {Simcoe}, Robert A. and {Stefanon}, Mauro and {Sugahara}, Yuma and {Toft}, Sune and {van der Wel}, Arjen and {van Dokkum}, Pieter and {Walter}, Fabian and {Watson}, Darrach and {Weaver}, John R. and {Whitaker}, Katherine E.},
        title = "{A Cosmic Miracle: A Remarkably Luminous Galaxy at zspec = 14.44 Confirmed with JWST}",
      journal = {The Open Journal of Astrophysics},
         year = 2026,
        month = jan,
       volume = {9},
        pages = {56033},
          doi = {10.33232/001c.156033},
archivePrefix = {arXiv},
       eprint = {2505.11263},
 primaryClass = {astro-ph.GA},
       adsurl = {https://ui.adsabs.harvard.edu/abs/2026OJAp....956033N}
}

@ARTICLE{Treiber2025,
       author = {{Treiber}, Helena and {Greene}, Jenny E. and {Weaver}, John R. and {Miller}, Tim B. and {Furtak}, Lukas J. and {Setton}, David J. and {Wang}, Bingjie and {de Graaff}, Anna and {Bezanson}, Rachel and {Brammer}, Gabriel and {Cutler}, Sam E. and {Dayal}, Pratika and {Feldmann}, Robert and {Fujimoto}, Seiji and {Goulding}, Andy D. and {Kokorev}, Vasily and {Labbe}, Ivo and {Leja}, Joel and {Marchesini}, Danilo and {Nanayakkara}, Themiya and {Nelson}, Erica and {Pan}, Richard and {Price}, Sedona H. and {Siegel}, Jared and {Suess}, Katherine A. and {Whitaker}, Katherine E.},
        title = "{UNCOVERing the High-redshift AGN Population among Extreme UV Line Emitters}",
      journal = {\apj},
         year = 2025,
        month = may,
       volume = {984},
       number = {1},
          eid = {93},
        pages = {93},
          doi = {10.3847/1538-4357/adc38f},
archivePrefix = {arXiv},
       eprint = {2409.12232},
 primaryClass = {astro-ph.GA},
       adsurl = {https://ui.adsabs.harvard.edu/abs/2025ApJ...984...93T}
}

@ARTICLE{Arevalo-Gonzalez2026,
       author = {{Arevalo-Gonzalez}, F. and {Tripodi}, R. and {Llerena}, M. and {Pentericci}, L. and {Plat}, A. and {Barro}, G. and {Amor{\'\i}n}, R.~O. and {Backhaus}, B. and {Calabr{\`o}}, A. and {Cleri}, N.~J. and {Dickinson}, M. and {Dunlop}, J.~S. and {Finkelstein}, S.~L. and {Giavalisco}, M. and {Hirschmann}, M. and {Kartaltepe}, J. and {Koekemoer}, A.~M. and {Lucas}, R.~A. and {Napolitano}, L. and {Piconcelli}, E. and {Taylor}, A.~J. and {Tombesi}, F. and {Trump}, J.~R. and {Wang}, X.},
        title = "{The AGN nature of strong C III] emitters in the early Universe with JWST}",
      journal = {\aap},
         year = 2026,
        month = may,
       volume = {709},
          eid = {A46},
        pages = {A46},
          doi = {10.1051/0004-6361/202558652},
archivePrefix = {arXiv},
       eprint = {2512.16365},
 primaryClass = {astro-ph.GA},
       adsurl = {https://ui.adsabs.harvard.edu/abs/2026A&A...709A..46A}
}

@ARTICLE{Curti2025,
       author = {{Curti}, Mirko and {Witstok}, Joris and {Jakobsen}, Peter and {Kobayashi}, Chiaki and {Curtis-Lake}, Emma and {Hainline}, Kevin and {Ji}, Xihan and {D'Eugenio}, Francesco and {Chevallard}, Jacopo and {Maiolino}, Roberto and {Scholtz}, Jan and {Carniani}, Stefano and {Arribas}, Santiago and {Baker}, William M. and {Bhatawdekar}, Rachana and {Boyett}, Kristan and {Bunker}, Andrew J. and {Cameron}, Alex and {Cargile}, Phillip A. and {Charlot}, St{\'e}phane and {Eisenstein}, Daniel J. and {Ji}, Zhiyuan and {Johnson}, Benjamin D. and {Kumari}, Nimisha and {Maseda}, Michael V. and {Robertson}, Brant and {Silcock}, Maddie S. and {Tacchella}, Sandro and {{\"U}bler}, Hannah and {Venturi}, Giacomo and {Williams}, Christina C. and {Willmer}, Christopher N.~A. and {Willott}, Chris},
        title = "{JADES: The star formation and chemical enrichment history of a luminous galaxy at z {\ensuremath{\sim}} 9.43 probed by ultra-deep JWST/NIRSpec spectroscopy}",
      journal = {\aap},
         year = 2025,
        month = may,
       volume = {697},
          eid = {A89},
        pages = {A89},
          doi = {10.1051/0004-6361/202451410},
archivePrefix = {arXiv},
       eprint = {2407.02575},
 primaryClass = {astro-ph.GA},
       adsurl = {https://ui.adsabs.harvard.edu/abs/2025A&A...697A..89C}
}

@ARTICLE{Tang2026,
       author = {{Tang}, Mengtao and {Stark}, Daniel P. and {Mason}, Charlotte A. and {Gelli}, Viola and {Chen}, Zuyi and {Topping}, Michael W.},
        title = "{The JWST Spectroscopic Properties of Galaxies at z = 9{\ensuremath{-}}14}",
      journal = {\apj},
         year = 2026,
        month = apr,
       volume = {1001},
       number = {1},
          eid = {38},
        pages = {38},
          doi = {10.3847/1538-4357/ae4edb},
archivePrefix = {arXiv},
       eprint = {2507.08245},
 primaryClass = {astro-ph.GA},
       adsurl = {https://ui.adsabs.harvard.edu/abs/2026ApJ..1001...38T}
}

@ARTICLE{Napolitano2025,
       author = {{Napolitano}, L. and {Castellano}, M. and {Pentericci}, L. and {Arrabal Haro}, P. and {Fontana}, A. and {Treu}, T. and {Bergamini}, P. and {Calabr{\`o}}, A. and {Mascia}, S. and {Morishita}, T. and {Roberts-Borsani}, G. and {Santini}, P. and {Vanzella}, E. and {Vulcani}, B. and {Zakharova}, D. and {Bakx}, T. and {Dickinson}, M. and {Grillo}, C. and {Leethochawalit}, N. and {Llerena}, M. and {Merlin}, E. and {Paris}, D. and {Rojas-Ruiz}, S. and {Rosati}, P. and {Wang}, X. and {Yoon}, I. and {Zavala}, J.},
        title = "{Seven wonders of Cosmic Dawn: JWST confirms a high abundance of galaxies and AGN at z ≃ 9─11 in the GLASS field}",
      journal = {\aap},
         year = 2025,
        month = jan,
       volume = {693},
          eid = {A50},
        pages = {A50},
          doi = {10.1051/0004-6361/202452090},
archivePrefix = {arXiv},
       eprint = {2410.10967},
 primaryClass = {astro-ph.GA},
       adsurl = {https://ui.adsabs.harvard.edu/abs/2025A&A...693A..50N}
}

@ARTICLE{Chen2026,
       author = {{Chen}, Zuyi and {Stark}, Daniel P. and {Mason}, Charlotte A. and {Senchyna}, Peter and {Tang}, Mengtao and {Keerthi Vasan G.}, C. and {Whitler}, Lily and {Plat}, Adele and {Gelli}, Viola},
        title = "{SPURS: Massive Stars, Dense Gas, and Ly$α$ Escape in GN-z11 at $z = 10.6$}",
      journal = {arXiv e-prints},
         year = 2026,
        month = aug,
          eid = {arXiv:2608.12699},
        pages = {arXiv:2608.12699},
          doi = {10.48550/arXiv.2608.12699},
archivePrefix = {arXiv},
       eprint = {2608.12699},
 primaryClass = {astro-ph.GA},
       adsurl = {https://ui.adsabs.harvard.edu/abs/2026arXiv260812699C}
}

@ARTICLE{Valle-Espinosa2026,
       author = {{del Valle-Espinosa}, Macarena G. and {Mingozzi}, Matilde and {James}, Bethan and {S{\'a}nchez-Janssen}, Rub{\'e}n and {Fern{\'a}ndez-Ontiveros}, Juan Antonio and {Vaught}, Ryan J. and {Amor{\'\i}n}, Ricardo O. and {Hunt}, Leslie and {Aloisi}, Alessandra and {Arellano-C{\'o}rdova}, Karla Z. and {Berg}, Danielle A. and {Chisholm}, John and {Hayes}, Matthew and {Hernandez}, Svea and {Hirschauer}, Alec S. and {Jones}, Logan and {Martin}, Crystal L. and {Vallini}, Livia and {Xu}, Xinfeng},
        title = "{JWST/MIRI-MRS View of the Metal-poor Galaxy CGCG 007-025: The Spatial Location of Polycyclic Aromatic Hydrocarbons and Very Highly Ionized Gas}",
      journal = {\apjl},
         year = 2026,
        month = feb,
       volume = {998},
       number = {2},
          eid = {L45},
        pages = {L45},
          doi = {10.3847/2041-8213/ae43ed},
archivePrefix = {arXiv},
       eprint = {2510.11581},
 primaryClass = {astro-ph.GA},
       adsurl = {https://ui.adsabs.harvard.edu/abs/2026ApJ...998L..45D}
}

@ARTICLE{Allamandola1989,
       author = {{Allamandola}, L.~J. and {Tielens}, A.~G.~G.~M. and {Barker}, J.~R.},
        title = "{Interstellar Polycyclic Aromatic Hydrocarbons: The Infrared Emission Bands, the Excitation/Emission Mechanism, and the Astrophysical Implications}",
      journal = {\apjs},
         year = 1989,
        month = dec,
       volume = {71},
        pages = {733},
          doi = {10.1086/191396},
       adsurl = {https://ui.adsabs.harvard.edu/abs/1989ApJS...71..733A}
}

@ARTICLE{Marshall2007,
       author = {{Marshall}, J.~A. and {Herter}, T.~L. and {Armus}, L. and {Charmandaris}, V. and {Spoon}, H.~W.~W. and {Bernard-Salas}, J. and {Houck}, J.~R.},
        title = "{Decomposing Dusty Galaxies. I. Multicomponent Spectral Energy Distribution Fitting}",
      journal = {\apj},
         year = 2007,
        month = nov,
       volume = {670},
       number = {1},
        pages = {129-155},
          doi = {10.1086/521588},
archivePrefix = {arXiv},
       eprint = {0707.2962},
 primaryClass = {astro-ph},
       adsurl = {https://ui.adsabs.harvard.edu/abs/2007ApJ...670..129M}
}

@ARTICLE{Vallini2024,
       author = {{Vallini}, Livia and {Witstok}, Joris and {Sommovigo}, Laura and {Pallottini}, Andrea and {Ferrara}, Andrea and {Carniani}, Stefano and {Kohandel}, Mahsa and {Smit}, Renske and {Gallerani}, Simona and {Gruppioni}, Carlotta},
        title = "{Spatially resolved Kennicutt-Schmidt relation at z {\ensuremath{\approx}} 7 and its connection with the interstellar medium properties}",
      journal = {\mnras},
         year = 2024,
        month = jan,
       volume = {527},
       number = {1},
        pages = {10-22},
          doi = {10.1093/mnras/stad3150},
archivePrefix = {arXiv},
       eprint = {2309.07957},
 primaryClass = {astro-ph.GA},
       adsurl = {https://ui.adsabs.harvard.edu/abs/2024MNRAS.527...10V}
}

@ARTICLE{Zanella2018,
       author = {{Zanella}, A. and {Daddi}, E. and {Magdis}, G. and {Diaz Santos}, T. and {Cormier}, D. and {Liu}, D. and {Cibinel}, A. and {Gobat}, R. and {Dickinson}, M. and {Sargent}, M. and {Popping}, G. and {Madden}, S.~C. and {Bethermin}, M. and {Hughes}, T.~M. and {Valentino}, F. and {Rujopakarn}, W. and {Pannella}, M. and {Bournaud}, F. and {Walter}, F. and {Wang}, T. and {Elbaz}, D. and {Coogan}, R.~T.},
        title = "{The [C II] emission as a molecular gas mass tracer in galaxies at low and high redshifts}",
      journal = {\mnras},
         year = 2018,
        month = dec,
       volume = {481},
       number = {2},
        pages = {1976-1999},
          doi = {10.1093/mnras/sty2394},
archivePrefix = {arXiv},
       eprint = {1808.10331},
 primaryClass = {astro-ph.GA},
       adsurl = {https://ui.adsabs.harvard.edu/abs/2018MNRAS.481.1976Z}
}

@ARTICLE{Glover2016,
       author = {{Glover}, Simon C.~O. and {Clark}, Paul C.},
        title = "{Is atomic carbon a good tracer of molecular gas in metal-poor galaxies?}",
      journal = {\mnras},
         year = 2016,
        month = mar,
       volume = {456},
       number = {4},
        pages = {3596-3609},
          doi = {10.1093/mnras/stv2863},
archivePrefix = {arXiv},
       eprint = {1509.01939},
 primaryClass = {astro-ph.GA},
       adsurl = {https://ui.adsabs.harvard.edu/abs/2016MNRAS.456.3596G}
}

@ARTICLE{Rich2023,
       author = {{Rich}, J. and {Aalto}, S. and {Evans}, A.~S. and {Charmandaris}, V. and {Privon}, G.~C. and {Lai}, T. and {Inami}, H. and {Linden}, S. and {Armus}, L. and {Diaz-Santos}, T. and {Appleton}, P. and {Barcos-Mu{\~n}oz}, L. and {B{\"o}ker}, T. and {Larson}, K.~L. and {Law}, D.~R. and {Malkan}, M.~A. and {Medling}, A.~M. and {Song}, Y. and {U}, V. and {van der Werf}, P. and {Bohn}, T. and {Brown}, M.~J.~I. and {Finnerty}, L. and {Hayward}, C. and {Howell}, J. and {Iwasawa}, K. and {Kemper}, F. and {Marshall}, J. and {Mazzarella}, J.~M. and {McKinney}, J. and {Muller-Sanchez}, F. and {Murphy}, E.~J. and {Sanders}, D. and {Soifer}, B.~T. and {Stierwalt}, S. and {Surace}, J.},
        title = "{GOALS-JWST: Pulling Back the Curtain on the AGN and Star Formation in VV 114}",
      journal = {\apjl},
         year = 2023,
        month = feb,
       volume = {944},
       number = {2},
          eid = {L50},
        pages = {L50},
          doi = {10.3847/2041-8213/acb2b8},
archivePrefix = {arXiv},
       eprint = {2301.02338},
 primaryClass = {astro-ph.GA},
       adsurl = {https://ui.adsabs.harvard.edu/abs/2023ApJ...944L..50R}
}

@ARTICLE{U2022,
       author = {{U}, Vivian and {Lai}, Thomas and {Bianchin}, Marina and {Remigio}, Raymond P. and {Armus}, Lee and {Larson}, Kirsten L. and {D{\'\i}az-Santos}, Tanio and {Evans}, Aaron and {Stierwalt}, Sabrina and {Law}, David R. and {Malkan}, Matthew A. and {Linden}, Sean and {Song}, Yiqing and {van der Werf}, Paul P. and {Gao}, Tianmu and {Privon}, George C. and {Medling}, Anne M. and {Barcos-Mu{\~n}oz}, Loreto and {Hayward}, Christopher C. and {Inami}, Hanae and {Rich}, Jeff and {Aalto}, Susanne and {Appleton}, Philip and {Bohn}, Thomas and {B{\"o}ker}, Torsten and {Brown}, Michael J.~I. and {Charmandaris}, Vassilis and {Finnerty}, Luke and {Howell}, Justin and {Iwasawa}, Kazushi and {Kemper}, Francisca and {Marshall}, Jason and {Mazzarella}, Joseph M. and {McKinney}, Jed and {Muller-Sanchez}, Francisco and {Murphy}, Eric J. and {Sanders}, David and {Surace}, Jason},
        title = "{GOALS-JWST: Resolving the Circumnuclear Gas Dynamics in NGC 7469 in the Mid-infrared}",
      journal = {\apjl},
         year = 2022,
        month = nov,
       volume = {940},
       number = {1},
          eid = {L5},
        pages = {L5},
          doi = {10.3847/2041-8213/ac961c},
archivePrefix = {arXiv},
       eprint = {2209.01210},
 primaryClass = {astro-ph.GA},
       adsurl = {https://ui.adsabs.harvard.edu/abs/2022ApJ...940L...5U}
}

@ARTICLE{Schouws2026,
       author = {{Schouws}, Sander and {Bouwens}, Rychard J. and {Algera}, Hiddo and {Smit}, Renske and {Kumari}, Nimisha and {Rowland}, Lucie E. and {van Leeuwen}, Ivana and {Sommovigo}, Laura and {Ferrara}, Andrea and {Oesch}, Pascal A. and {Ormerod}, Katherine and {Stefanon}, Mauro and {Herard-Demanche}, Thomas and {Hodge}, Jacqueline and {Fudamoto}, Yoshinobu and {R{\"o}ttgering}, Huub and {van der Werf}, Paul},
        title = "{Deep Constraints on [CII]158$\mu$m in JADES-GS-z14-0: Further Evidence for a Galaxy with Low Gas Content at z=14.2}",
      journal = {arXiv e-prints},
         year = 2025,
        month = feb,
          eid = {arXiv:2502.01610},
        pages = {arXiv:2502.01610},
          doi = {10.48550/arXiv.2502.01610},
archivePrefix = {arXiv},
       eprint = {2502.01610},
 primaryClass = {astro-ph.GA},
       adsurl = {https://ui.adsabs.harvard.edu/abs/2025arXiv250201610S}
}

@ARTICLE{Schouws2025,
       author = {{Schouws}, Sander and {Bouwens}, Rychard J. and {Ormerod}, Katherine and {Smit}, Renske and {Algera}, Hiddo and {Sommovigo}, Laura and {Hodge}, Jacqueline and {Ferrara}, Andrea and {Oesch}, Pascal A. and {Rowland}, Lucie E. and {van Leeuwen}, Ivana and {Stefanon}, Mauro and {Herard-Demanche}, Thomas and {Fudamoto}, Yoshinobu and {R{\"o}ttgering}, Huub and {van der Werf}, Paul},
        title = "{Detection of [O III]$_{88 {\ensuremath{\mu}}m}$ in JADES-GS-z14-0 at z = 14.1793}",
      journal = {\apj},
         year = 2025,
        month = jul,
       volume = {988},
       number = {1},
          eid = {19},
        pages = {19},
          doi = {10.3847/1538-4357/adbf1b},
archivePrefix = {arXiv},
       eprint = {2409.20549},
 primaryClass = {astro-ph.GA},
       adsurl = {https://ui.adsabs.harvard.edu/abs/2025ApJ...988...19S}
}

@ARTICLE{Zavala2024,
       author = {{Zavala}, Jorge A. and {Bakx}, Tom and {Mitsuhashi}, Ikki and {Castellano}, Marco and {Calabro}, Antonello and {Akins}, Hollis and {Buat}, Veronique and {Casey}, Caitlin M. and {Fernandez-Arenas}, David and {Franco}, Maximilien and {Fontana}, Adriano and {Hatsukade}, Bunyo and {Ho}, Luis C. and {Ikeda}, Ryota and {Kartaltepe}, Jeyhan and {Koekemoer}, Anton M. and {McKinney}, Jed and {Napolitano}, Lorenzo and {P{\'e}rez-Gonz{\'a}lez}, Pablo G. and {Santini}, Paola and {Serjeant}, Stephen and {Terlevich}, Elena and {Terlevich}, Roberto and {Yung}, L.~Y. Aaron},
        title = "{ALMA Detection of [O III] 88 {\ensuremath{\mu}}m at z = 12.33: Exploring the Nature and Evolution of GHZ2 as a Massive Compact Stellar System}",
      journal = {\apjl},
         year = 2024,
        month = dec,
       volume = {977},
       number = {1},
          eid = {L9},
        pages = {L9},
          doi = {10.3847/2041-8213/ad8f38},
archivePrefix = {arXiv},
       eprint = {2411.03593},
 primaryClass = {astro-ph.GA},
       adsurl = {https://ui.adsabs.harvard.edu/abs/2024ApJ...977L...9Z}
}

@article{Smit2026,
   author = "Smit, Renske and Bowler, Rebecca A.A.",
   title = "The ALMA View of High-Redshift Galaxy Formation",
   journal = "Annual Review of Astronomy and Astrophysics",
   issn = "0066-4146",
   year = "2026",
   publisher = "Annual Reviews",
   url = "https://www.annualreviews.org/content/journals/10.1146/annurev-astro-052722-104242",
   doi = "https://doi.org/10.1146/annurev-astro-052722-104242"
   
}

@ARTICLE{Begum2005,
       author = {{Begum}, Ayesha and {Chengalur}, Jayaram N.},
        title = "{A search for HI in some peculiar faint dwarf galaxies}",
      journal = {\mnras},
         year = 2005,
        month = sep,
       volume = {362},
       number = {2},
        pages = {609-611},
          doi = {10.1111/j.1365-2966.2005.09342.x},
archivePrefix = {arXiv},
       eprint = {astro-ph/0506667},
 primaryClass = {astro-ph},
       adsurl = {https://ui.adsabs.harvard.edu/abs/2005MNRAS.362..609B}
}

@ARTICLE{Ramambason2024,
       author = {{Ramambason}, L. and {Lebouteiller}, V. and {Madden}, S.~C. and {Galliano}, F. and {Richardson}, C.~T. and {Saintonge}, A. and {De Looze}, I. and {Chevance}, M. and {Abel}, N.~P. and {Hernandez}, S. and {Braine}, J.},
        title = "{Modeling the molecular gas content and CO-to-H$_{2}$ conversion factors in low-metallicity star-forming dwarf galaxies}",
      journal = {\aap},
         year = 2024,
        month = jan,
       volume = {681},
          eid = {A14},
        pages = {A14},
          doi = {10.1051/0004-6361/202347280},
archivePrefix = {arXiv},
       eprint = {2306.14881},
 primaryClass = {astro-ph.GA},
       adsurl = {https://ui.adsabs.harvard.edu/abs/2024A&A...681A..14R}
}

@ARTICLE{Mingozzi2025,
       author = {{Mingozzi}, Matilde and {del Valle-Espinosa}, Macarena G. and {James}, Bethan L. and {Rickards Vaught}, Ryan J. and {Hayes}, Matthew and {Amor{\'\i}n}, Ricardo O. and {Leitherer}, Claus and {Aloisi}, Alessandra and {Hunt}, Leslie and {Law}, David and {Richardson}, Chris T. and {Pidgeon}, Aidan and {Arellano-C{\'o}rdova}, Karla Z. and {Berg}, Danielle A. and {Chisholm}, John and {Hernandez}, Svea and {Jones}, Logan and {Kumari}, Nimisha and {Martin}, Crystal L. and {Ravindranath}, Swara and {Vallini}, Livia and {Xu}, Xinfeng},
        title = "{Exploring the Mysterious High-ionization Source Powering [Ne V] in High-z Analog SBS0335-052 E with JWST/MIRI}",
      journal = {\apj},
         year = 2025,
        month = jun,
       volume = {985},
       number = {2},
          eid = {253},
        pages = {253},
          doi = {10.3847/1538-4357/adc996},
archivePrefix = {arXiv},
       eprint = {2502.07662},
 primaryClass = {astro-ph.GA},
       adsurl = {https://ui.adsabs.harvard.edu/abs/2025ApJ...985..253M}
}

@ARTICLE{Oey2017,
       author = {{Oey}, M.~S. and {Herrera}, C.~N. and {Silich}, Sergiy and {Reiter}, Megan and {James}, Bethan L. and {Jaskot}, A.~E. and {Micheva}, Genoveva},
        title = "{Dense CO in Mrk 71-A: Superwind Suppressed in a Young Super Star Cluster}",
      journal = {\apjl},
         year = 2017,
        month = nov,
       volume = {849},
       number = {1},
          eid = {L1},
        pages = {L1},
          doi = {10.3847/2041-8213/aa9215},
archivePrefix = {arXiv},
       eprint = {1710.03261},
 primaryClass = {astro-ph.GA},
       adsurl = {https://ui.adsabs.harvard.edu/abs/2017ApJ...849L...1O}
}

@ARTICLE{Galliano2008,
       author = {{Galliano}, Fr{\'e}d{\'e}ric and {Madden}, Suzanne C. and {Tielens}, Alexander G.~G.~M. and {Peeters}, Els and {Jones}, Anthony P.},
        title = "{Variations of the Mid-IR Aromatic Features inside and among Galaxies}",
      journal = {\apj},
         year = 2008,
        month = may,
       volume = {679},
       number = {1},
        pages = {310-345},
          doi = {10.1086/587051},
archivePrefix = {arXiv},
       eprint = {0801.4955},
 primaryClass = {astro-ph},
       adsurl = {https://ui.adsabs.harvard.edu/abs/2008ApJ...679..310G}
}

@ARTICLE{Sturm2000,
       author = {{Sturm}, E. and {Lutz}, D. and {Tran}, D. and {Feuchtgruber}, H. and {Genzel}, R. and {Kunze}, D. and {Moorwood}, A.~F.~M. and {Thornley}, M.~D.},
        title = "{ISO-SWS spectra of galaxies: Continuum and features}",
      journal = {\aap},
         year = 2000,
        month = jun,
       volume = {358},
        pages = {481-493},
          doi = {10.48550/arXiv.astro-ph/0002195},
archivePrefix = {arXiv},
       eprint = {astro-ph/0002195},
 primaryClass = {astro-ph},
       adsurl = {https://ui.adsabs.harvard.edu/abs/2000A&A...358..481S}
}

@ARTICLE{Soifer1987,
       author = {{Soifer}, B.~T. and {Neugebauer}, G. and {Houck}, J.~R.},
        title = "{The IRAS view of the extragalactic sky.}",
      journal = {\araa},
         year = 1987,
        month = jan,
       volume = {25},
        pages = {187-230},
          doi = {10.1146/annurev.aa.25.090187.001155},
       adsurl = {https://ui.adsabs.harvard.edu/abs/1987ARA&A..25..187S}
}

@ARTICLE{Leitherer1995,
       author = {{Leitherer}, Claus and {Ferguson}, Henry C. and {Heckman}, Timothy M. and {Lowenthal}, James D.},
        title = "{The Lyman Continuum in Starburst Galaxies Observed with the Hopkins Ultraviolet Telescope}",
      journal = {\apjl},
         year = 1995,
        month = nov,
       volume = {454},
        pages = {L19},
          doi = {10.1086/309760},
       adsurl = {https://ui.adsabs.harvard.edu/abs/1995ApJ...454L..19L}
}

@ARTICLE{Jaskot2025,
       author = {{Jaskot}, Anne E.},
        title = "{Ionizing Radiation Escape from Low-Redshift Galaxies and Its Connection to Cosmic Reionization}",
      journal = {\araa},
         year = 2025,
        month = aug,
       volume = {63},
       number = {1},
        pages = {45-82},
          doi = {10.1146/annurev-astro-111324-074935},
archivePrefix = {arXiv},
       eprint = {2508.18411},
 primaryClass = {astro-ph.GA},
       adsurl = {https://ui.adsabs.harvard.edu/abs/2025ARA&A..63...45J}
}

@ARTICLE{Mingozzi2022,
       author = {{Mingozzi}, Matilde and {James}, Bethan L. and {Arellano-C{\'o}rdova}, Karla Z. and {Berg}, Danielle A. and {Senchyna}, Peter and {Chisholm}, John and {Brinchmann}, Jarle and {Aloisi}, Alessandra and {Amor{\'\i}n}, Ricardo O. and {Charlot}, St{\'e}phane and {Feltre}, Anna and {Hayes}, Matthew and {Heckman}, Timothy and {Henry}, Alaina and {Hernandez}, Svea and {Kumari}, Nimisha and {Leitherer}, Claus and {Llerena}, Mario and {Martin}, Crystal L. and {Nanayakkara}, Themiya and {Ravindranath}, Swara and {Skillman}, Evan D. and {Sugahara}, Yuma and {Wofford}, Aida and {Xu}, Xinfeng},
        title = "{CLASSY IV. Exploring UV Diagnostics of the Interstellar Medium in Local High-z Analogs at the Dawn of the JWST Era}",
      journal = {\apj},
         year = 2022,
        month = nov,
       volume = {939},
       number = {2},
          eid = {110},
        pages = {110},
          doi = {10.3847/1538-4357/ac952c},
archivePrefix = {arXiv},
       eprint = {2209.09047},
 primaryClass = {astro-ph.GA},
       adsurl = {https://ui.adsabs.harvard.edu/abs/2022ApJ...939..110M}
}

@ARTICLE{Izotov2021,
       author = {{Izotov}, Y.~I. and {Guseva}, N.~G. and {Fricke}, K.~J. and {Henkel}, C. and {Schaerer}, D. and {Thuan}, T.~X.},
        title = "{Low-redshift compact star-forming galaxies as analogues of high-redshift star-forming galaxies}",
      journal = {\aap},
         year = 2021,
        month = feb,
       volume = {646},
          eid = {A138},
        pages = {A138},
          doi = {10.1051/0004-6361/202039772},
archivePrefix = {arXiv},
       eprint = {2103.01505},
 primaryClass = {astro-ph.GA},
       adsurl = {https://ui.adsabs.harvard.edu/abs/2021A&A...646A.138I}
}

@ARTICLE{Kunth2000,
       author = {{Kunth}, D. and {{\"O}stlin}, G.},
        title = "{The most metal-poor galaxies}",
      journal = {\aapr},
         year = 2000,
        month = jan,
       volume = {10},
        pages = {1-79},
          doi = {10.1007/s001590000005},
archivePrefix = {arXiv},
       eprint = {astro-ph/9911094},
 primaryClass = {astro-ph},
       adsurl = {https://ui.adsabs.harvard.edu/abs/2000A&ARv..10....1K}
}

@ARTICLE{Bian2016,
       author = {{Bian}, Fuyan and {Kewley}, Lisa J. and {Dopita}, Michael A. and {Juneau}, Stephanie},
        title = "{Local Analogs for High-redshift Galaxies: Resembling the Physical Conditions of the Interstellar Medium in High-redshift Galaxies}",
      journal = {\apj},
         year = 2016,
        month = may,
       volume = {822},
       number = {2},
          eid = {62},
        pages = {62},
          doi = {10.3847/0004-637X/822/2/62},
archivePrefix = {arXiv},
       eprint = {1603.05275},
 primaryClass = {astro-ph.GA},
       adsurl = {https://ui.adsabs.harvard.edu/abs/2016ApJ...822...62B}
}

@ARTICLE{Heckman2005,
       author = {{Heckman}, Timothy M. and {Hoopes}, Charles G. and {Seibert}, Mark and {Martin}, D. Christopher and {Salim}, Samir and {Rich}, R. Michael and {Kauffmann}, Guinevere and {Charlot}, Stephane and {Barlow}, Tom A. and {Bianchi}, Luciana and {Byun}, Yong-Ik and {Donas}, Jose and {Forster}, Karl and {Friedman}, Peter G. and {Jelinsky}, Patrick N. and {Lee}, Young-Wook and {Madore}, Barry F. and {Malina}, Roger F. and {Milliard}, Bruno and {Morrissey}, Patrick F. and {Neff}, Susan G. and {Schiminovich}, David and {Siegmund}, Oswald H.~W. and {Small}, Todd and {Szalay}, Alex S. and {Welsh}, Barry Y. and {Wyder}, Ted K.},
        title = "{The Properties of Ultraviolet-luminous Galaxies at the Current Epoch}",
      journal = {\apjl},
         year = 2005,
        month = jan,
       volume = {619},
       number = {1},
        pages = {L35-L38},
          doi = {10.1086/425979},
archivePrefix = {arXiv},
       eprint = {astro-ph/0412577},
 primaryClass = {astro-ph},
       adsurl = {https://ui.adsabs.harvard.edu/abs/2005ApJ...619L..35H}
}

@ARTICLE{Borthakur2014,
       author = {{Borthakur}, Sanchayeeta and {Heckman}, Timothy M. and {Leitherer}, Claus and {Overzier}, Roderik A.},
        title = "{A local clue to the reionization of the universe}",
      journal = {Science},
         year = 2014,
        month = oct,
       volume = {346},
       number = {6206},
        pages = {216-219},
          doi = {10.1126/science.1254214},
archivePrefix = {arXiv},
       eprint = {1410.3511},
 primaryClass = {astro-ph.GA},
       adsurl = {https://ui.adsabs.harvard.edu/abs/2014Sci...346..216B}
}

@ARTICLE{Garcia-Bernete2024,
       author = {{Garc{\'\i}a-Bernete}, I. and {Alonso-Herrero}, A. and {Rigopoulou}, D. and {Pereira-Santaella}, M. and {Shimizu}, T. and {Davies}, R. and {Donnan}, F.~R. and {Roche}, P.~F. and {Gonz{\'a}lez-Mart{\'\i}n}, O. and {Ramos Almeida}, C. and {Bellocchi}, E. and {Boorman}, P. and {Combes}, F. and {Efstathiou}, A. and {Esparza-Arredondo}, D. and {Garc{\'\i}a-Burillo}, S. and {Gonz{\'a}lez-Alfonso}, E. and {Hicks}, E.~K.~S. and {H{\"o}nig}, S. and {Labiano}, A. and {Levenson}, N.~A. and {L{\'o}pez-Rodr{\'\i}guez}, E. and {Ricci}, C. and {Packham}, C. and {Rouan}, D. and {Stalevski}, M. and {Ward}, M.~J.},
        title = "{The Galaxy Activity, Torus, and Outflow Survey (GATOS). III. Revealing the inner icy structure in local active galactic nuclei}",
      journal = {\aap},
         year = 2024,
        month = jan,
       volume = {681},
          eid = {L7},
        pages = {L7},
          doi = {10.1051/0004-6361/202348266},
archivePrefix = {arXiv},
       eprint = {2310.09093},
 primaryClass = {astro-ph.GA},
       adsurl = {https://ui.adsabs.harvard.edu/abs/2024A&A...681L...7G}
}

@INPROCEEDINGS{Hermosa2025,
       author = {{Hermosa Mu{\~n}oz}, L. and {Alonso-Herrero}, A. and {MICONIC Team}},
        title = "{MIri Characterization Of Nearby Infrared Galaxy Centers (MICONIC): First results of the JWST MIRI/MRS GTO program}",
    booktitle = {Highlights of Spanish Astrophysics XII},
         year = 2025,
       editor = {{Manteiga}, M. and {Gonz{\'a}lez-Galindo}, F. and {Labiano-Ortega}, A. and {Mart{\'\i}nez-Gonz{\'a}lez}, M.~J. and {Rea}, N. and {Romero-G{\'o}mez}, M. and {Ulla-Miguel}, A. and {Yepes}, G. and {Rodr{\'\i}guez-L{\'o}pez}, C. and {G{\'o}mez-Garc{\'\i}a}, A. and {Dafonte}, C.},
        month = may,
        pages = {87},
       adsurl = {https://ui.adsabs.harvard.edu/abs/2025hsa..conf...87H}
}

@ARTICLE{Williams2024,
       author = {{Williams}, Thomas G. and {Lee}, Janice C. and {Larson}, Kirsten L. and {Leroy}, Adam K. and {Sandstrom}, Karin and {Schinnerer}, Eva and {Thilker}, David A. and {Belfiore}, Francesco and {Egorov}, Oleg V. and {Rosolowsky}, Erik and {Sutter}, Jessica and {DePasquale}, Joseph and {Pagan}, Alyssa and {Berger}, Travis A. and {Anand}, Gagandeep S. and {Barnes}, Ashley T. and {Bigiel}, Frank and {Boquien}, M{\'e}d{\'e}ric and {Cao}, Yixian and {Chastenet}, J{\'e}r{\'e}my and {Chevance}, M{\'e}lanie and {Chown}, Ryan and {Dale}, Daniel A. and {Deger}, Sinan and {Eibensteiner}, Cosima and {Emsellem}, Eric and {Faesi}, Christopher M. and {Glover}, Simon C.~O. and {Grasha}, Kathryn and {Hannon}, Stephen and {Hassani}, Hamid and {Henshaw}, Jonathan D. and {Jim{\'e}nez-Donaire}, Mar{\'\i}a J. and {Kim}, Jaeyeon and {Klessen}, Ralf S. and {Koch}, Eric W. and {Li}, Jing and {Liu}, Daizhong and {Meidt}, Sharon E. and {M{\'e}ndez-Delgado}, J. Eduardo and {Murphy}, Eric J. and {Neumann}, Justus and {Neumann}, Lukas and {Neumayer}, Nadine and {Oakes}, Elias K. and {Pathak}, Debosmita and {Pety}, J{\'e}r{\^o}me and {Pinna}, Francesca and {Querejeta}, Miguel and {Ramambason}, Lise and {Romanelli}, Andrea and {Sormani}, Mattia C. and {Stuber}, Sophia K. and {Sun}, Jiayi and {Teng}, Yu-Hsuan and {Usero}, Antonio and {Watkins}, Elizabeth J. and {Weinbeck}, Tony D.},
        title = "{PHANGS-JWST: Data-processing Pipeline and First Full Public Data Release}",
      journal = {\apjs},
         year = 2024,
        month = jul,
       volume = {273},
       number = {1},
          eid = {13},
        pages = {13},
          doi = {10.3847/1538-4365/ad4be5},
archivePrefix = {arXiv},
       eprint = {2401.15142},
 primaryClass = {astro-ph.GA},
       adsurl = {https://ui.adsabs.harvard.edu/abs/2024ApJS..273...13W}
}

@ARTICLE{Telford2024,
       author = {{Telford}, O. Grace and {Sandstrom}, Karin M. and {McQuinn}, Kristen B.~W. and {Glover}, Simon C.~O. and {Tarantino}, Elizabeth J. and {Bolatto}, Alberto D. and {Rickards Vaught}, Ryan J.},
        title = "{Molecular Hydrogen in the Extremely Metal-Poor, Star-Forming Galaxy Leo P}",
      journal = {arXiv e-prints},
         year = 2024,
        month = oct,
          eid = {arXiv:2410.21368},
        pages = {arXiv:2410.21368},
          doi = {10.48550/arXiv.2410.21368},
archivePrefix = {arXiv},
       eprint = {2410.21368},
 primaryClass = {astro-ph.GA},
       adsurl = {https://ui.adsabs.harvard.edu/abs/2024arXiv241021368T}
}

@ARTICLE{Curtis-Lake2023,
       author = {{Curtis-Lake}, Emma and {Carniani}, Stefano and {Cameron}, Alex and {Charlot}, Stephane and {Jakobsen}, Peter and {Maiolino}, Roberto and {Bunker}, Andrew and {Witstok}, Joris and {Smit}, Renske and {Chevallard}, Jacopo and {Willott}, Chris and {Ferruit}, Pierre and {Arribas}, Santiago and {Bonaventura}, Nina and {Curti}, Mirko and {D'Eugenio}, Francesco and {Franx}, Marijn and {Giardino}, Giovanna and {Looser}, Tobias J. and {L{\"u}tzgendorf}, Nora and {Maseda}, Michael V. and {Rawle}, Tim and {Rix}, Hans-Walter and {Rodr{\'\i}guez del Pino}, Bruno and {{\"U}bler}, Hannah and {Sirianni}, Marco and {Dressler}, Alan and {Egami}, Eiichi and {Eisenstein}, Daniel J. and {Endsley}, Ryan and {Hainline}, Kevin and {Hausen}, Ryan and {Johnson}, Benjamin D. and {Rieke}, Marcia and {Robertson}, Brant and {Shivaei}, Irene and {Stark}, Daniel P. and {Tacchella}, Sandro and {Williams}, Christina C. and {Willmer}, Christopher N.~A. and {Bhatawdekar}, Rachana and {Bowler}, Rebecca and {Boyett}, Kristan and {Chen}, Zuyi and {de Graaff}, Anna and {Helton}, Jakob M. and {Hviding}, Raphael E. and {Jones}, Gareth C. and {Kumari}, Nimisha and {Lyu}, Jianwei and {Nelson}, Erica and {Perna}, Michele and {Sandles}, Lester and {Saxena}, Aayush and {Suess}, Katherine A. and {Sun}, Fengwu and {Topping}, Michael W. and {Wallace}, Imaan E.~B. and {Whitler}, Lily},
        title = "{Spectroscopic confirmation of four metal-poor galaxies at z = 10.3-13.2}",
      journal = {Nature Astronomy},
         year = 2023,
        month = may,
       volume = {7},
        pages = {622-632},
          doi = {10.1038/s41550-023-01918-w},
archivePrefix = {arXiv},
       eprint = {2212.04568},
 primaryClass = {astro-ph.GA},
       adsurl = {https://ui.adsabs.harvard.edu/abs/2023NatAs...7..622C}
}

@ARTICLE{Witstok2026,
       author = {{Witstok}, Joris and {Smit}, Renske and {Baker}, William M. and {Rinaldi}, Pierluigi and {Hainline}, Kevin N. and {Algera}, Hiddo S.~B. and {Arribas}, Santiago and {Bakx}, Tom J.~L.~C. and {Bunker}, Andrew J. and {Carniani}, Stefano and {Charlot}, St{\'e}phane and {Chevallard}, Jacopo and {Curti}, Mirko and {Curtis-Lake}, Emma and {Eisenstein}, Daniel J. and {Heintz}, Kasper E. and {Helton}, Jakob M. and {Jones}, Gareth C. and {Maiolino}, Roberto and {Maseda}, Michael V. and {P{\'e}rez-Gonz{\'a}lez}, Pablo G. and {Pollock}, Clara L. and {Robertson}, Brant E. and {Saxena}, Aayush and {Scholtz}, Jan and {Shivaei}, Irene and {Sun}, Fengwu and {Tacchella}, Sandro and {{\"U}bler}, Hannah and {Watson}, Darach and {Willott}, Chris J. and {Wu}, Zihao},
        title = "{On the origins of oxygen: ALMA and JWST characterise the multi-phase, metal-enriched, star-bursting medium within a 'normal' z > 11 galaxy}",
      journal = {The Open Journal of Astrophysics},
         year = 2026,
        month = jan,
       volume = {9},
        pages = {55261},
          doi = {10.33232/001c.155261},
archivePrefix = {arXiv},
       eprint = {2507.22888},
 primaryClass = {astro-ph.GA},
       adsurl = {https://ui.adsabs.harvard.edu/abs/2026OJAp....955261W}
}

@ARTICLE{Carniani2025,
       author = {{Carniani}, Stefano and {D'Eugenio}, Francesco and {Ji}, Xihan and {Parlanti}, Eleonora and {Scholtz}, Jan and {Sun}, Fengwu and {Venturi}, Giacomo and {Bakx}, Tom J.~L.~C. and {Curti}, Mirko and {Maiolino}, Roberto and {Tacchella}, Sandro and {Zavala}, Jorge A. and {Hainline}, Kevin and {Witstok}, Joris and {Johnson}, Benjamin D. and {Alberts}, Stacey and {Bunker}, Andrew J. and {Charlot}, St{\'e}phane and {Eisenstein}, Daniel J. and {Helton}, Jakob M. and {Jakobsen}, Peter and {Kumari}, Nimisha and {Robertson}, Brant and {Saxena}, Aayush and {{\"U}bler}, Hannah and {Williams}, Christina C. and {Willmer}, Christopher N.~A. and {Willott}, Chris},
        title = "{The eventful life of a luminous galaxy at z = 14: metal enrichment, feedback, and low gas fraction?}",
      journal = {\aap},
         year = 2025,
        month = apr,
       volume = {696},
          eid = {A87},
        pages = {A87},
          doi = {10.1051/0004-6361/202452451},
archivePrefix = {arXiv},
       eprint = {2409.20533},
 primaryClass = {astro-ph.GA},
       adsurl = {https://ui.adsabs.harvard.edu/abs/2025A&A...696A..87C}
}

@INPROCEEDINGS{Stark2026,
       author = {{Stark}, Daniel P. and {Topping}, Michael W. and {Endsley}, Ryan and {Tang}, Mengtao},
        title = "{Observations of the first galaxies in the Era of JWST}",
    booktitle = {Encyclopedia of Astrophysics, Volume 4},
         year = 2026,
       volume = {4},
        month = jan,
        pages = {453-499},
          doi = {10.1016/B978-0-443-21439-4.00128-0},
archivePrefix = {arXiv},
       eprint = {2501.17078},
 primaryClass = {astro-ph.GA},
       adsurl = {https://ui.adsabs.harvard.edu/abs/2026enap....4..453S}
}

@ARTICLE{Leroy2008,
       author = {{Leroy}, Adam K. and {Walter}, Fabian and {Brinks}, Elias and {Bigiel}, Frank and {de Blok}, W.~J.~G. and {Madore}, Barry and {Thornley}, M.~D.},
        title = "{The Star Formation Efficiency in Nearby Galaxies: Measuring Where Gas Forms Stars Effectively}",
      journal = {\aj},
         year = 2008,
        month = dec,
       volume = {136},
       number = {6},
        pages = {2782-2845},
          doi = {10.1088/0004-6256/136/6/2782},
archivePrefix = {arXiv},
       eprint = {0810.2556},
 primaryClass = {astro-ph},
       adsurl = {https://ui.adsabs.harvard.edu/abs/2008AJ....136.2782L}
}

@ARTICLE{Watson2016,
       author = {{Watson}, Linda C. and {Koda}, Jin},
        title = "{Molecular Gas in the Outskirts of Galaxies}",
      journal = {arXiv e-prints},
         year = 2016,
        month = dec,
          eid = {arXiv:1612.05275},
        pages = {arXiv:1612.05275},
          doi = {10.48550/arXiv.1612.05275},
archivePrefix = {arXiv},
       eprint = {1612.05275},
 primaryClass = {astro-ph.GA},
       adsurl = {https://ui.adsabs.harvard.edu/abs/2016arXiv161205275W}
}

@ARTICLE{Groves2008,
       author = {{Groves}, Brent and {Nefs}, Bas and {Brandl}, Bernhard},
        title = "{The mid-infrared [SIV]/[NeII] versus [NeIII]/[NeII] correlation}",
      journal = {\mnras},
         year = 2008,
        month = nov,
       volume = {391},
       number = {1},
        pages = {L113-L116},
          doi = {10.1111/j.1745-3933.2008.00568.x},
archivePrefix = {arXiv},
       eprint = {0810.0010},
 primaryClass = {astro-ph},
       adsurl = {https://ui.adsabs.harvard.edu/abs/2008MNRAS.391L.113G}
}

@ARTICLE{Perez-Montero2025,
       author = {{P{\'e}rez-Montero}, E. and {Fern{\'a}ndez-Ontiveros}, J.~A. and {P{\'e}rez-D{\'\i}az}, B. and {V{\'\i}lchez}, J.~M. and {Amor{\'\i}n}, R.},
        title = "{Exploring the hardness of the ionizing radiation with the infrared softness diagram: II. Bimodal distributions in both the ionizing continuum slope and the excitation in active galactic nuclei}",
      journal = {\aap},
         year = 2025,
        month = apr,
       volume = {696},
          eid = {A229},
        pages = {A229},
          doi = {10.1051/0004-6361/202453276},
archivePrefix = {arXiv},
       eprint = {2503.09267},
 primaryClass = {astro-ph.GA},
       adsurl = {https://ui.adsabs.harvard.edu/abs/2025A&A...696A.229P}
}

@ARTICLE{Wang2026,
       author = {{Wang}, Kuo-Song and {Simmonds}, Robert and {Comrie}, Angus and {Hwang}, Yu-Hsuan and {Pi{\'n}ska}, Adrianna and {Harris}, Pamela and {Moraghan}, Anthony and {Pang}, Qi and {Raul-Omar}, Carli and {Hou}, Kuan-Chou and {Aikema}, David and {Chiang}, Cheng-Chin and {Chang}, Tien-Hao and {Hsu}, Shou-Chieh and {Lin}, Ming-Yi and {Gao}, Zhen-Kai and {Huang}, Po-Sheng and {Hibbard}, John and {Ott}, Juergen and {Collier}, Jordan and {Stoehr}, Felix and {Raba}, Ryan and {Kirkham}, Kechil and {Rosolowsky}, Erik and {Kern}, Jeff and {Lee}, Chin-Fei and {Taylor}, Russ and {CARTA Collaboration}},
        title = "{CARTA{\textemdash}Cube Analysis and Rendering Tool for Astronomy: A Tool for Big Imaging Data}",
      journal = {\pasp},
         year = 2026,
        month = feb,
       volume = {138},
       number = {2},
          eid = {024506},
        pages = {024506},
          doi = {10.1088/1538-3873/ae3eb4},
       adsurl = {https://ui.adsabs.harvard.edu/abs/2026PASP..138b4506W}
}

@ARTICLE{Richardson2025,
       author = {{Richardson}, Chris T. and {Wels}, Jordan and {Garofali}, Kristen and {Levanti}, Julianna M. and {Lebouteiller}, Vianney and {Lehmer}, Bret and {Basu-Zych}, Antara and {Berg}, Danielle and {Bellovary}, Jillian M. and {Chisholm}, John and {Kannappan}, Sheila J. and {Lambrides}, Erini and {Polimera}, Mugdha S. and {Ramambason}, Lise and {Varese}, Maxime and {Vivona}, Thomas},
        title = "{Emission-line Diagnostics for IMBHs in Dwarf Galaxies: Accounting for Black Hole Seeding and Ultraluminous X-Ray Source Excitation}",
      journal = {\apj},
         year = 2025,
        month = nov,
       volume = {993},
       number = {1},
          eid = {154},
        pages = {154},
          doi = {10.3847/1538-4357/adf215},
archivePrefix = {arXiv},
       eprint = {2505.07749},
 primaryClass = {astro-ph.GA},
       adsurl = {https://ui.adsabs.harvard.edu/abs/2025ApJ...993..154R}
}

@ARTICLE{Richardson2022,
       author = {{Richardson}, Chris T. and {Simpson}, Connor and {Polimera}, Mugdha S. and {Kannappan}, Sheila J. and {Bellovary}, Jillian M. and {Greene}, Christopher and {Jenkins}, Sam},
        title = "{Optical and JWST Mid-IR Emission Line Diagnostics for Simultaneous IMBH and Stellar Excitation in z 0 Dwarf Galaxies}",
      journal = {\apj},
         year = 2022,
        month = mar,
       volume = {927},
       number = {2},
          eid = {165},
        pages = {165},
          doi = {10.3847/1538-4357/ac510c},
archivePrefix = {arXiv},
       eprint = {2202.01330},
 primaryClass = {astro-ph.GA},
       adsurl = {https://ui.adsabs.harvard.edu/abs/2022ApJ...927..165R}
}

@ARTICLE{Bisigello2023,
       author = {{Bisigello}, L. and {Gandolfi}, G. and {Grazian}, A. and {Rodighiero}, G. and {Costantin}, L. and {Cooray}, A.~R. and {Feltre}, A. and {Gruppioni}, C. and {Hathi}, N.~P. and {Holwerda}, B.~W. and {Koekemoer}, A.~M. and {Lucas}, R.~A. and {Newman}, J.~A. and {P{\'e}rez-Gonz{\'a}lez}, P.~G. and {Yung}, L.~Y.~A. and {de la Vega}, A. and {Arrabal Haro}, P. and {Bagley}, M.~B. and {Dickinson}, M. and {Finkelstein}, S.~L. and {Kartaltepe}, J.~S. and {Papovich}, C. and {Pirzkal}, N. and {Wilkins}, S.},
        title = "{Delving deep: A population of extremely dusty dwarfs observed by JWST}",
      journal = {\aap},
         year = 2023,
        month = aug,
       volume = {676},
          eid = {A76},
        pages = {A76},
          doi = {10.1051/0004-6361/202346219},
archivePrefix = {arXiv},
       eprint = {2302.12270},
 primaryClass = {astro-ph.GA},
       adsurl = {https://ui.adsabs.harvard.edu/abs/2023A&A...676A..76B}
}

@ARTICLE{Gordon2023,
       author = {{Gordon}, Karl D. and {Clayton}, Geoffrey C. and {Decleir}, Marjorie and {Fitzpatrick}, E.~L. and {Massa}, Derck and {Misselt}, Karl A. and {Tollerud}, Erik J.},
        title = "{One Relation for All Wavelengths: The Far-ultraviolet to Mid-infrared Milky Way Spectroscopic R(V)-dependent Dust Extinction Relationship}",
      journal = {\apj},
         year = 2023,
        month = jun,
       volume = {950},
       number = {2},
          eid = {86},
        pages = {86},
          doi = {10.3847/1538-4357/accb59},
archivePrefix = {arXiv},
       eprint = {2304.01991},
 primaryClass = {astro-ph.GA},
       adsurl = {https://ui.adsabs.harvard.edu/abs/2023ApJ...950...86G}
}

@ARTICLE{Weingartner2001,
       author = {{Weingartner}, Joseph C. and {Draine}, B.~T.},
        title = "{Electron-Ion Recombination on Grains and Polycyclic Aromatic Hydrocarbons}",
      journal = {\apj},
         year = 2001,
        month = dec,
       volume = {563},
       number = {2},
        pages = {842-852},
          doi = {10.1086/324035},
archivePrefix = {arXiv},
       eprint = {astro-ph/0105237},
 primaryClass = {astro-ph},
       adsurl = {https://ui.adsabs.harvard.edu/abs/2001ApJ...563..842W}
}

@article{Gordon2024, doi = {10.21105/joss.07023}, url = {https://doi.org/10.21105/joss.07023}, year = {2024}, publisher = {The Open Journal}, volume = {9}, number = {100}, pages = {7023}, author = {Gordon, Karl D.}, title = {dust_extinction: Interstellar Dust Extinction Models}, journal = {Journal of Open Source Software} }

@ARTICLE{Smith2007,
       author = {{Smith}, J.~D.~T. and {Draine}, B.~T. and {Dale}, D.~A. and {Moustakas}, J. and {Kennicutt}, Jr., R.~C. and {Helou}, G. and {Armus}, L. and {Roussel}, H. and {Sheth}, K. and {Bendo}, G.~J. and {Buckalew}, B.~A. and {Calzetti}, D. and {Engelbracht}, C.~W. and {Gordon}, K.~D. and {Hollenbach}, D.~J. and {Li}, A. and {Malhotra}, S. and {Murphy}, E.~J. and {Walter}, F.},
        title = "{The Mid-Infrared Spectrum of Star-forming Galaxies: Global Properties of Polycyclic Aromatic Hydrocarbon Emission}",
      journal = {\apj},
         year = 2007,
        month = feb,
       volume = {656},
       number = {2},
        pages = {770-791},
          doi = {10.1086/510549},
archivePrefix = {arXiv},
       eprint = {astro-ph/0610913},
 primaryClass = {astro-ph},
       adsurl = {https://ui.adsabs.harvard.edu/abs/2007ApJ...656..770S}
}

@software{cafe2025,
author = {{Diaz-Santos}, Tanio and {Lai}, Thomas S. -Y. and {Finnerty}, Luke and {Privon}, George and {Bonfini}, Paolo and {Larson}, Kirsten and {Marshall}, Jason and {Armus}, Lee and {Charmandaris}, Vassilis},
title = "{CAFE: Continuum And Feature Extraction tool}",
howpublished = {Astrophysics Source Code Library, record ascl:2501.001},
year = 2025,
month = jan,
eid = {ascl:2501.001},
adsurl = {https://ui.adsabs.harvard.edu/abs/2025ascl.soft01001D}
}
\bibliographystyle{aasjournalv7}

\end{document}